\documentclass[sn-mathphys]{sn-jnl}

\usepackage{hyperref}

\jyear{2023}%

\theoremstyle{thmstyleone}%
\theoremstyle{thmstyletwo}%

\theoremstyle{thmstylethree}%

\begin{document}

\title[Observations of disks in multiple systems]{Observations of planet forming disks in multiple stellar systems}


\author*[1,2,3]{\fnm{Alice} \sur{Zurlo}}\email{alice.zurlo@mail.udp.cl}

\author[4]{\fnm{Raffaele} \sur{Gratton}}
\equalcont{These authors contributed equally to this work.}

\author[5,3]{\fnm{Sebasti\'an} \sur{P\'erez}} 
\equalcont{These authors contributed equally to this work.}

\author[1,3]{\fnm{Lucas} \sur{Cieza}} 
\equalcont{These authors contributed equally to this work.}


\affil*[1]{Instituto de Estudios Astrof\'isicos, Facultad de Ingenier\'ia y Ciencias, Universidad Diego Portales, Av. Ejercito 441, Santiago, Chile}

\affil[2]{Escuela de Ingenier\'ia Industrial, Facultad de Ingenier\'ia y Ciencias, Universidad Diego Portales, Av. Ejercito 441, Santiago, Chile}

\affil[3]{Millennium Nucleus on Young Exoplanets and their Moons (YEMS)}

\affil[4]{\orgdiv{Osservatorio Astronomico di Padova}, \orgname{INAF}, \orgaddress{\street{Vicolo dell'Osservatorio 5}, \city{Padova}, \country{Italia}}}

\affil[5]{Departamento de Física, Universidad de Santiago de Chile, Av. Victor Jara 3659, Santiago, Chile}



\abstract{The demographic of circumstellar disks, the birthplaces of planets, is diverse and rich in disks featuring rings, gaps, spirals, filaments, and arcs. Many studies revealing these disk structures have focused on objects around single stars and disks in isolation. The scenario is more complex if binarity or multiplicity is involved; most stars are part of multiple systems in crowded star-forming regions. How does the presence of one or more stellar companions affect the shape and size of the circumstellar disks? Here we review the landscape of results from optical, infrared, and (sub-) millimeter observations of the effects of multiplicity on protoplanetary disks, emphasizing the demographic studies of nearby molecular clouds and the high-resolution studies of multiple disk systems. }




\maketitle

\section{Introduction}\label{sec1}

\subsection{Motivation}

Protoplanetary disks  are the cribs of planets. As of today, around 5300 extrasolar planets have been detected with different techniques. The diverse population of exoplanets brings particular attention to the environment where planets form, to better understand why we find so many different system architectures. 

Since a large fraction of stars forms in multiple systems \cite{2013ARA&A..51..269D}, studying how (sub)stellar companions affect the properties and evolution of protoplanetary disks is critical for our understanding of planet formation.  Such a line of research is complicated by two important factors: 1) multiplicity and disk properties are both a function of stellar mass and age, and 2) significant incompleteness and biases still exist in the characterization of disks and multiple systems. In particular, multiplicity increases with stellar mass and decreases with time.   The overall binary fraction of OB stars is close to 100$\%$ but 
decreases to  25$\%$ for M-type stars \cite{2013ARA&A..51..269D}.  Similarly, the fraction of visual companions for solar mass (0.7–1.5 M$_\odot$) stars decreases from 20$\%$ in the pre-main-sequence stage to 10$\%$ in the field \cite{2013ARA&A..51..269D}.  

Regarding incompleteness and biases, it is important to keep in mind that the census of tight companions to young stellar objects is highly incomplete, especially for objects with large mass ratios.   The distance to nearby star-forming regions (d $>$ 140 pc) and the diffraction limit of current telescopes operating at near-IR wavelengths (0.05$"$) implies that only visual companions with separations beyond  8 au can be readily identified. The scarcity of dedicated spectroscopic surveys of pre-main sequence stars also results in an incomplete census of spectroscopic binaries. Furthermore, most disks in nearby star-forming regions still remain unresolved (see, e.g., \cite{2019A&A...626A..80C, 2021MNRAS.500.4878L}, Miley et al., in prep.), limiting the number of disk sizes available in the literature. 

Even with all the limitations discussed above, important observational trends have already been identified in the sense that protoplanetary disks in binary systems with separations $<$100 au are less frequent \cite{2009ApJ...696L..84C,2012ApJ...757..141K} and less massive \cite{2005ApJ...631.1134A,2017ApJ...851...83C} than those observed around single stars. On the other hand, very tight (a $<<$  1 au) companions seem to have a much weaker effect on disks since massive circumbinary disks  can be found around spectroscopic binaries \cite{2012ApJ...751..115H}.  These basic trends can be explained by tidal truncation.  Stars in a binary system are expected to truncate each other´s outer disks at 0.3-0-5 the orbital separation, depending on the mass ratio \cite{1977MNRAS.181..441P}.   Similarly, and depending on eccentricity, the truncation radius of a circumbinary disk is expected to occur at 1.8 to 2.6 times the separation of the binary \cite{1994ApJ...421..651A}. { The truncation radius can be affected by the disk misalignment:  \cite{2015ApJ...800...96L} and \cite{2015MNRAS.452.2396M} found that misaligned disks in binary systems can be more extended compared to coplanar binary disks.}


\subsection{Observational primer}

This review focuses on the landscape of recent observations aimed at completing the picture of demographics of stellar and disk multiplicity in nearby star-forming regions, as well as observations aimed at resolving and characterizing the structures around multiple disk systems. In this review, we acknowledge that there are at least five observational tracers (largely dependent on wavelength) that are used to study multiple stellar and disk systems: direct, scattered, and polarized light imaging using optical and infrared high-contrast imaging cameras, as well as, thermal continuum and molecular line emission using radio interferometers. 

This review highlights the work on visual binaries using near-infrared adaptive optics (AO) in nearby star-forming regions (Ophiuchus, Lupus, and Taurus) and disk studies of the same regions using the Atacama Large Millimeter/submillimeter Array (ALMA). We also discuss the results on several systems observed in scattered and polarized light where dramatic effects can be seen in the morphologies of interacting protoplanetary disks. In the manuscript, we refer to ``companion" as both stellar and sub-stellar objects physically bound to a star.


\section{Insights from demographics}

To better understand the effect of the stellar multiplicity on protoplanetary disks it is convenient to have the measurements of the mass and radius of the disk. With ALMA, it is now possible to measure with extreme accuracy the size of the circumstellar disks, even in the case of low-mass or compact objects. { It is important to mention that the disk radii reported in this Section have been measured in the dust size, which is affected by the radial drift. This mechanism has different effects in single and multiple systems, as discussed in \cite{2023EPJP..138...25Z} and reference therein.} We report here the results on three different star-forming regions (SFR): Ophiuchus, Lupus, and Taurus.

\subsection{Multiplicity in Ophiuchus} 

Ophiuchus is a star-forming region at a distance of 139 pc \cite{2019A&A...626A..80C} and ages from 2 to 5 Myr \cite{2008hsf2.book..351W}. In Ophiuchus 289 circumstellar disks have been identified. All of them have been observed with ALMA during the survey Ophiuchus Disk Survey Employing ALMA (ODISEA; \cite{2019MNRAS.482..698C}). These observations permitted the accurate measurement of the size and luminosity of each disk, allowing the estimation of their masses. \cite{2020MNRAS.496.5089Z} used the parameters measured from the ALMA data to compare the disks around isolated stars to the ones in multiple systems. 

In the analysis presented by \cite{2020MNRAS.496.5089Z} multiple systems have been identified via NIR imaging. This means that only the visual binaries are considered in the study. They collected NIR imaging from the archive and from dedicated surveys to detect (sub)stellar companions for each disk in Ophiuchus. As a result, 20 new binaries and one triple system were discovered. Considering the already-known multiple systems, the visual binaries frequency in Ophiuchus is 18\%. Considering the resolution and contrast limits of the instruments implied in the analysis, this translates to the rate of companions with separation from 9 to 1200 au, and flux ratios from 0.01 to 1 {(allowing the detection of brown dwarfs of a few jupiter masses)}. To make a comparison with other SFR, the companion-star frequency for a separation range of $\sim$17 to 780 au, for the Coronae Australis (CrA) region is 36.2 $\pm$ 8.8 \% \cite{2008A&A...488..997K}. In Taurus-Auriga, this frequency is almost identical, 37.4 $\pm$ 4.6 \% \cite{1993A&A...278..129L, 1998A&A...331..977K}.

\begin{figure}[h]%
\centering
\includegraphics[width=0.48\textwidth]{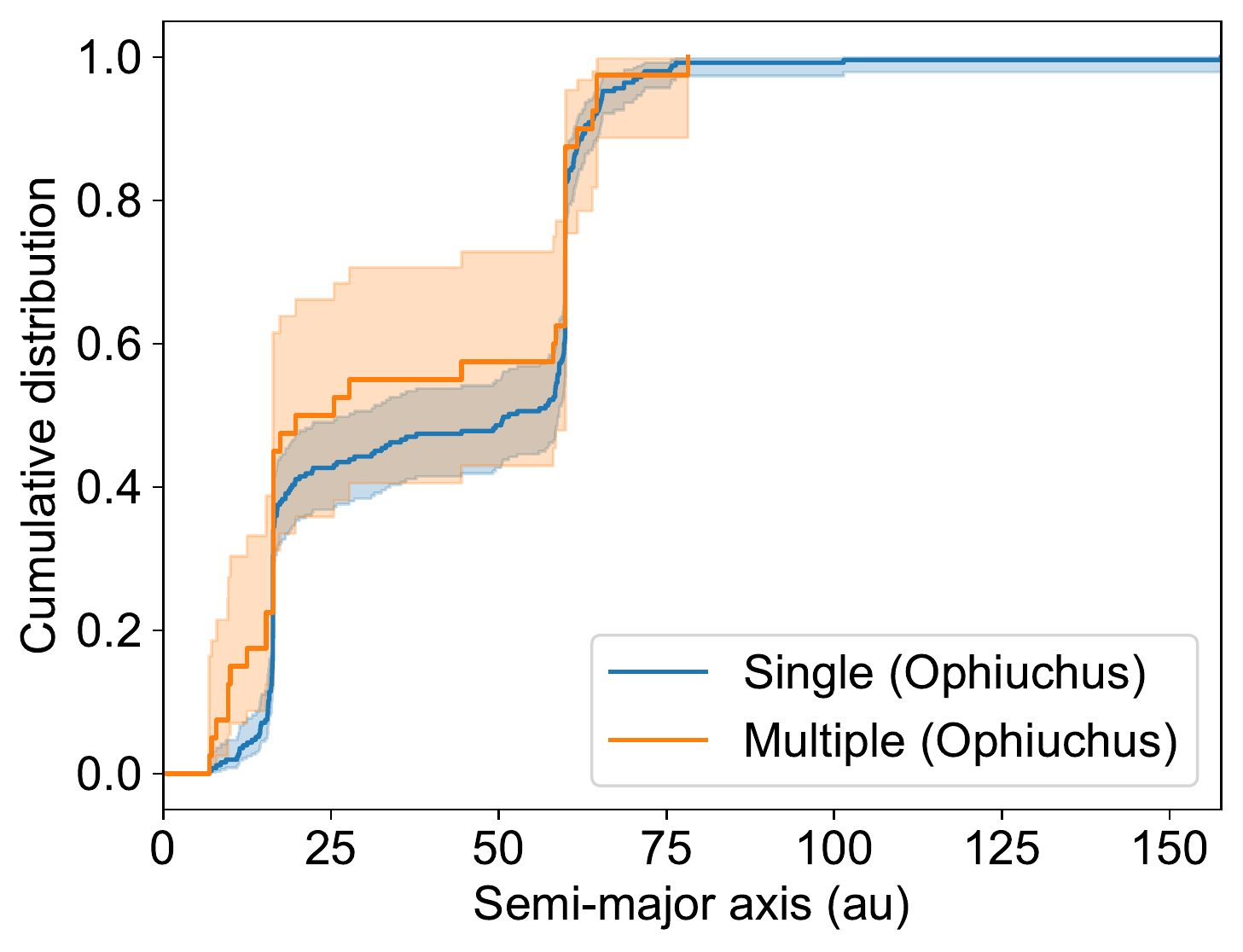}
\includegraphics[width=0.48\textwidth]{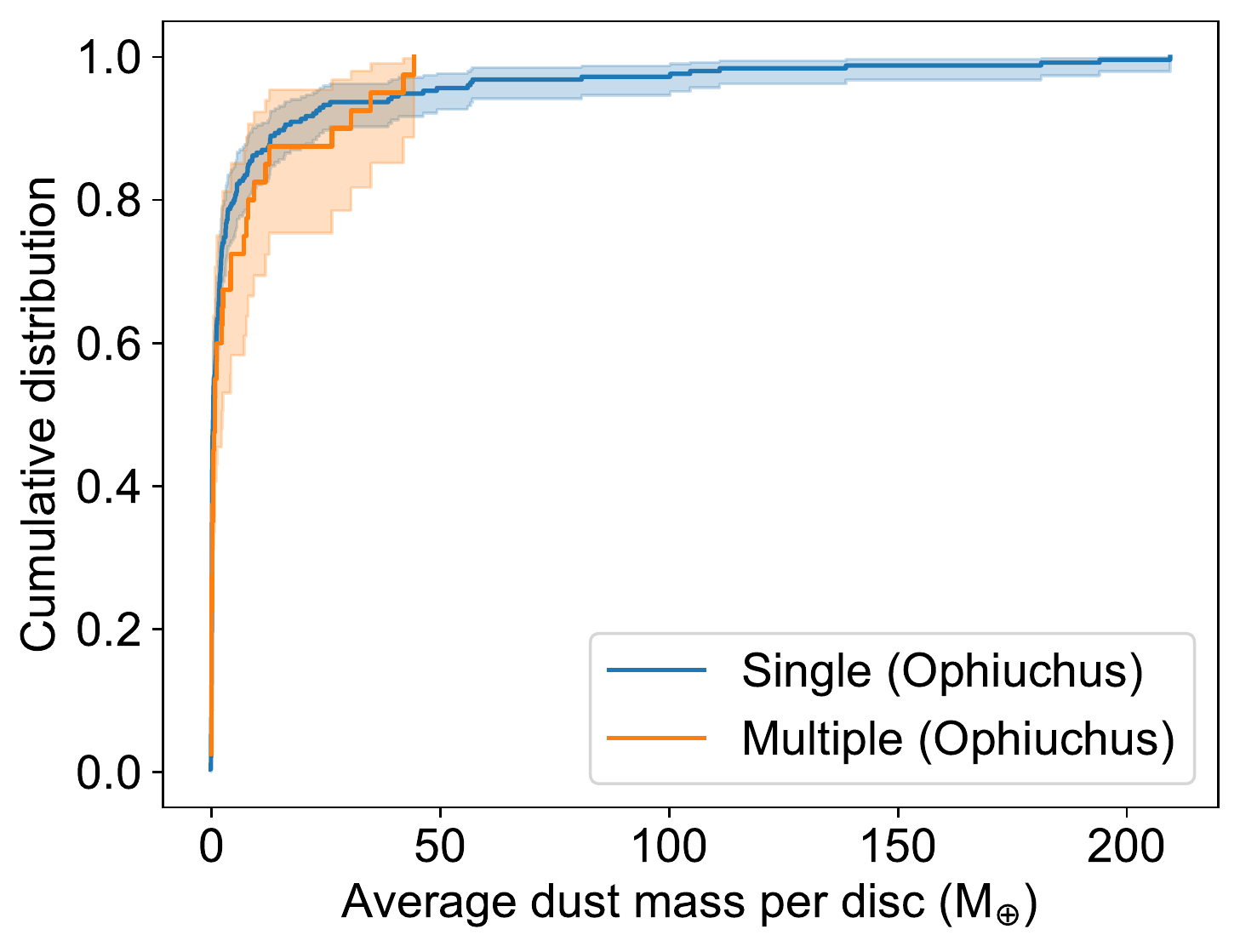}
\caption{Cumulative distribution of the projected semi-major axis and masses of the disks in Ophiuchus as measured in the ALMA data. The shaded area represents the lower and upper confidence intervals for the cumulative density. Adapted from \cite{2020MNRAS.496.5089Z}.}\label{fig_oph}
\end{figure}

From the same study, we can derive the comparison between the size and mass of disks around single stars and multiple systems. The size of the disks around single stars is systematically higher than the one around stars with companions, which are more compact. The semi-major axis of disks around single stars can extend up to 150 au. The extension of disks around stars with companions is half this value (see Fig.~\ref{fig_oph}, left). In the same line, massive disks are found only around single stars. The distribution of masses is similar until 50 M$_{\oplus}$ for disks around single and multiple systems, but masses from 50 up to 200 are found only around isolated stars (see Fig.~\ref{fig_oph}, right).

\subsection{Multiplicity in Lupus}

The Lupus SFR is another nearby molecular cloud subject of numerous studies. Its median distance is 158 pc \cite{2018A&A...616A...1G} { and age from 1 to 3 Myr \cite{2020A&A...643A.148G}}. { This region was also targeted by ALMA,} in particular, 125 objects including the Class II and III in the Lupus I-IV clouds were part of two surveys presented in \cite{2016ApJ...828...46A, 2018ApJ...859...21A, 2021MNRAS.500.4878L}. The ALMA observations allowed the measurement of the size and masses of these 125 protoplanetary disks. As for Ophiuchus, the same analysis can be conducted for Lupus, selecting the disks around isolated stars from the ones that are part of multiple systems. For that, \cite{2021MNRAS.501.2305Z} collected the NIR imaging for each of the disks observed with ALMA to identify any (sub)stellar visual companion. Some of the objects had NIR imaging available in the archive, but the majority of them have been observed with VLT/NACO for the first time. Among the 105 objects observable in the NIR \cite{2021MNRAS.501.2305Z} identified 13 multiple systems: 11 binaries and 2 triple systems. Considering the distance of Lupus, the survey is sensitive to visual binaries with projected separations between 20–4800 au. 

\begin{figure}[h]%
\centering
\includegraphics[width=0.48\textwidth]{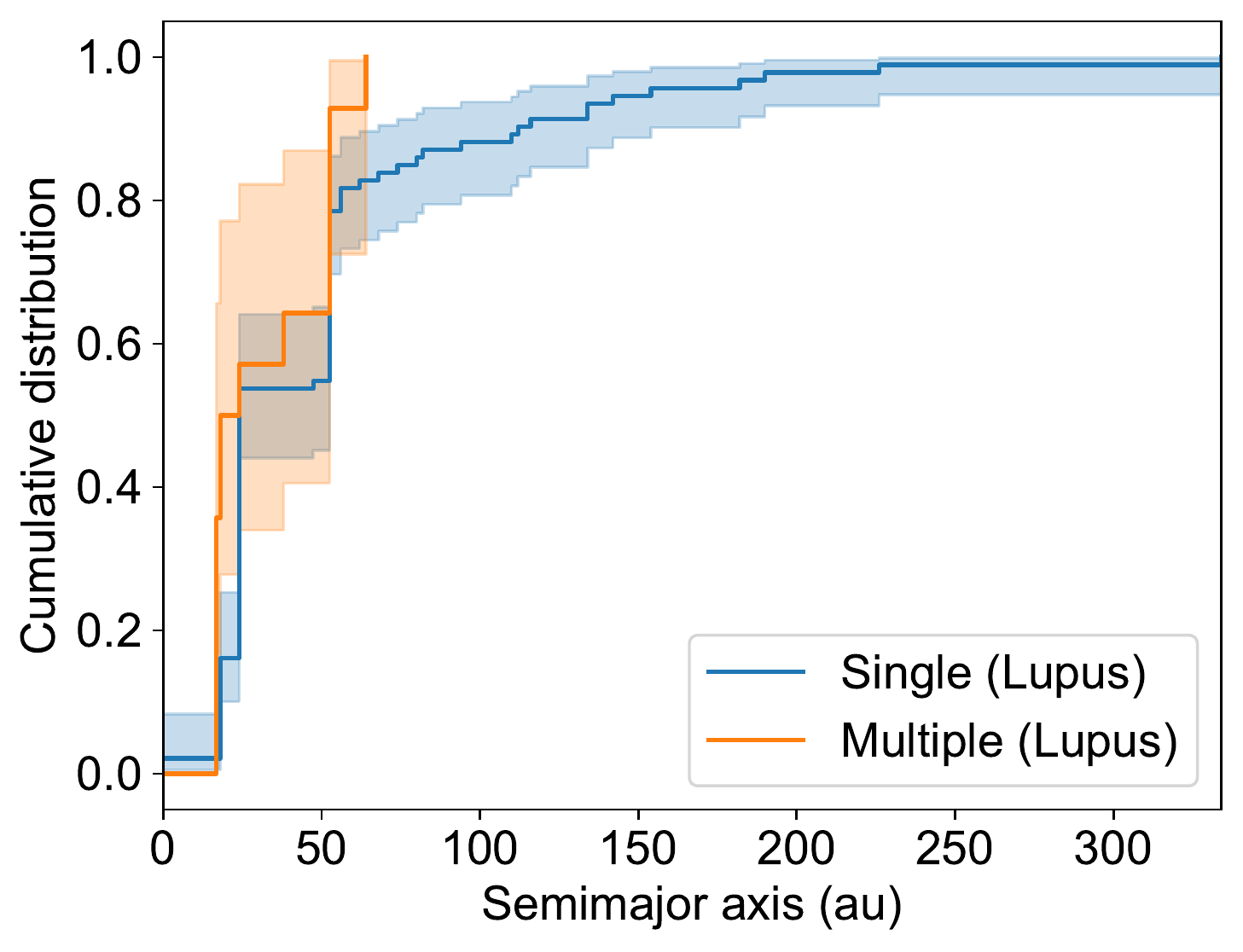}
\includegraphics[width=0.48\textwidth]{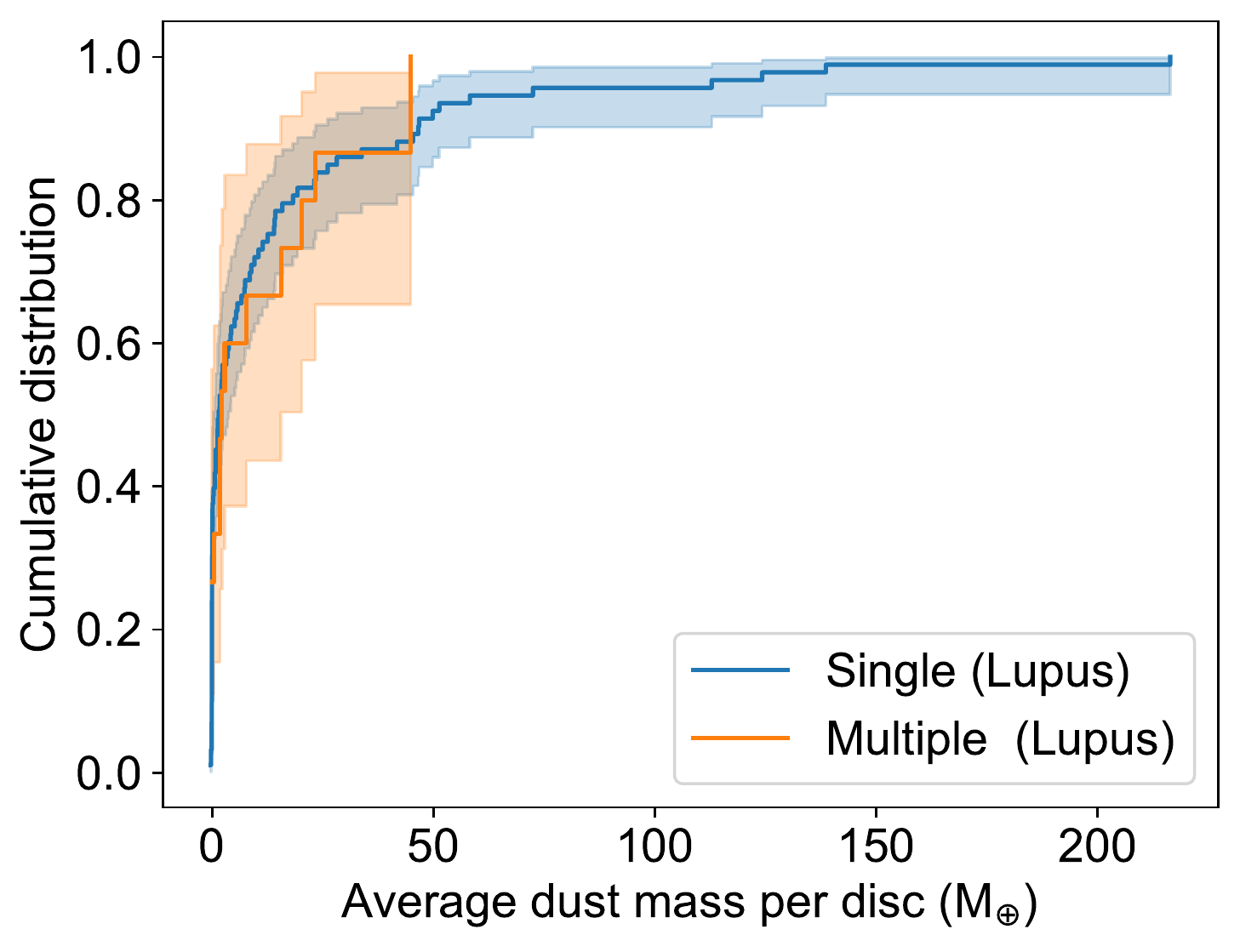}
\caption{Cumulative distribution of the projected semi-major axis and masses of the disks in Lupus as measured in the ALMA data. The shaded area represents the lower and upper confidence intervals for the cumulative density. Adapted from \cite{2021MNRAS.501.2305Z}.}\label{fig_lup}
\end{figure}

As for Lupus, it is possible to compare the distribution of size and mass of the disks for single objects vs multiple systems. The disks found in multiple systems are barely resolved by ALMA, meaning that they are very compact (a $<$ 70 au). The masses are also very different, while massive disks of 200 M$_{\oplus}$ can be found around isolated stars, multiple systems host disks with a maximum mass of 50 M$_{\oplus}$. The cumulative distributions of the size and mass of the disks in Lupus are shown in Fig.~\ref{fig_lup}.

\subsection{Multiplicity in Taurus} 

\begin{figure}
\centering
\includegraphics[width=0.5\textwidth]{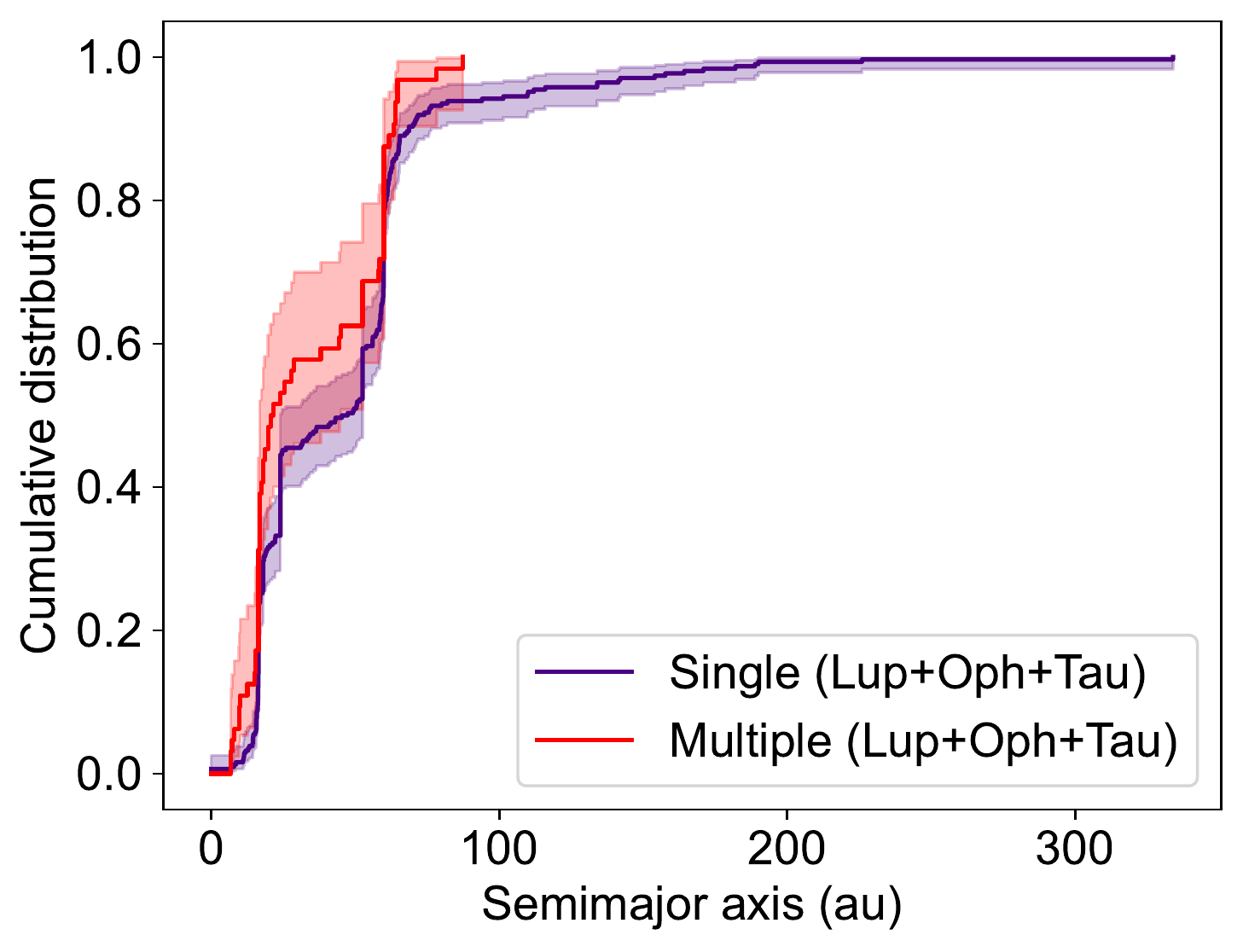}
\caption{Cumulative distribution of the projected semi-major axis of the disks as measured in the ALMA data. The shaded area represents the lower and upper confidence intervals for the cumulative density. All the disks in the Lupus, Ophiuchus, and Taurus regions are included.}\label{fig_a_tot}
\end{figure}

\begin{figure}
\centering
\includegraphics[width=0.5\textwidth]{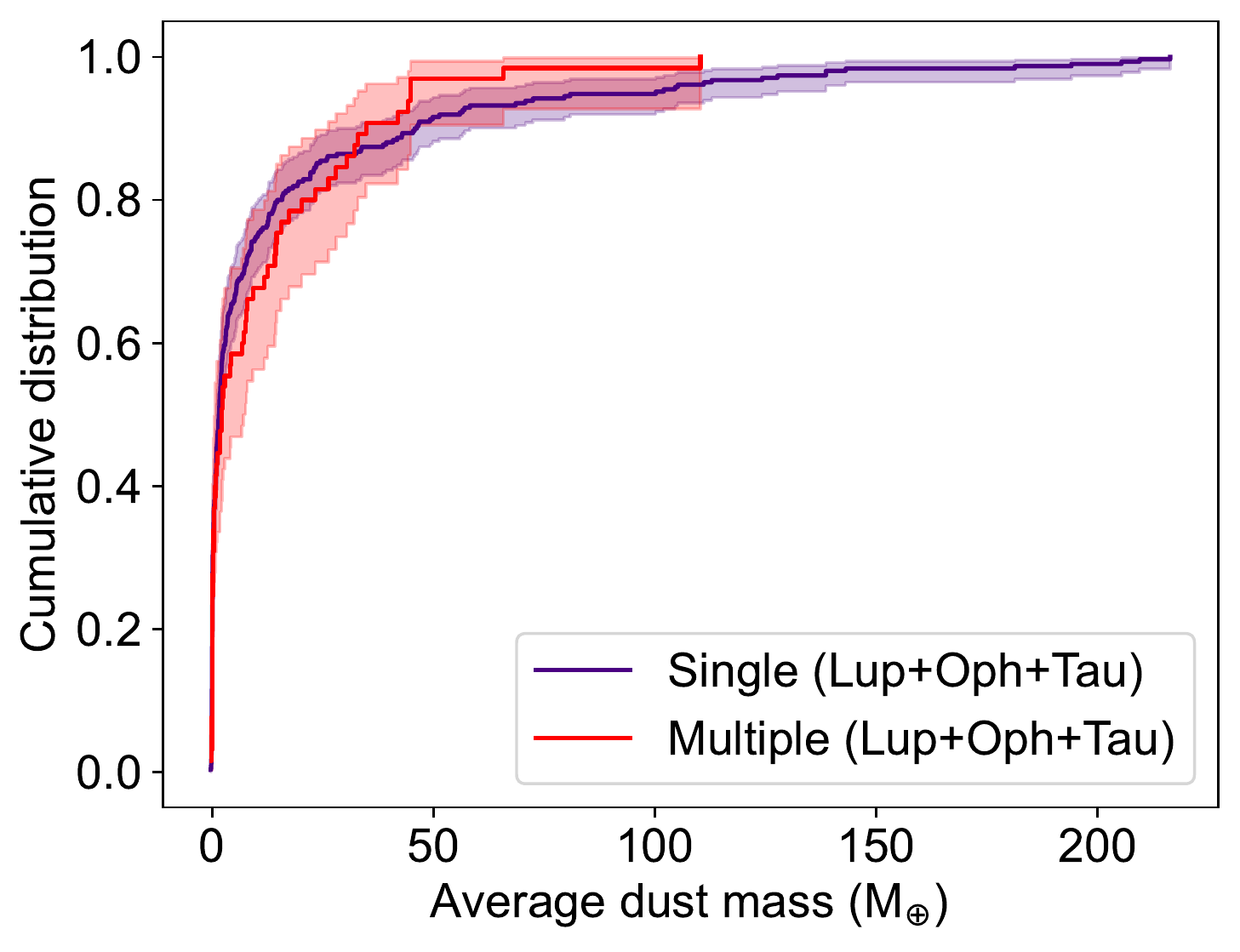}
\caption{Distribution of the average mass per component for single disks vs disks with stellar companions. The sample includes all the disks in the Ophiuchus, Lupus, and Taurus regions.}\label{fig_mass_tot}
\end{figure}

The Taurus-Auriga SFR is composed of various clusters with a depth of about 25 pc \cite{2019A&A...630A.137G}. The median distance is found at a distance of 140 pc from us, { and age $\sim$1--3 Myr \cite{2023AJ....165...37L}}. It is one of the most extensively studied star-forming regions, as it represents the perfect laboratory to study young stellar objects. Some of the members of Taurus were observed during a survey of disk structures in ALMA Band 6. The observations included 32 targets with spectral type earlier than M3. The sample and survey are described in \cite{2019ApJ...882...49L}. From this survey, the disk size and masses were measured, as for the case of Ophiuchus and Lupus. Among the 32 objects, ten are identified as multiple systems. \cite{2019A&A...628A..95M} presented the comparison of the disks around single stars vs the ones in multiple systems. Two systems are triple, the rest are binaries. With a similar analysis of \cite{2020MNRAS.496.5089Z,2021MNRAS.501.2305Z} but with a smaller sample, we calculated the cumulative distribution of the projected semi-major axis and mass for all the disks included in the Taurus sample. 
Even if the sample is small, if combined with the other two SFRs, it can help increase the statistics about the differences among disks around isolated stars and multiple systems. In Fig~\ref{fig_a_tot} the cumulative distribution of the projected semi-major axis for the whole sample is shown. In the plot the difference in size between disks around single and multiple stars is clear. Disks around single stars can be three times more extended in size than disks in multiple systems. 
{ We also report that a similar analysis has been presented in \cite{2022A&A...662A.121R}, in this case, the authors compare the gas (not the dust as in this work) radii of the Taurus disks.} In the case of the mass per disk, disks around single stars can be twice more massive than disks around multiple systems (Fig~\ref{fig_mass_tot}). 

\begin{figure}
\centering
\includegraphics[width=0.5\textwidth]{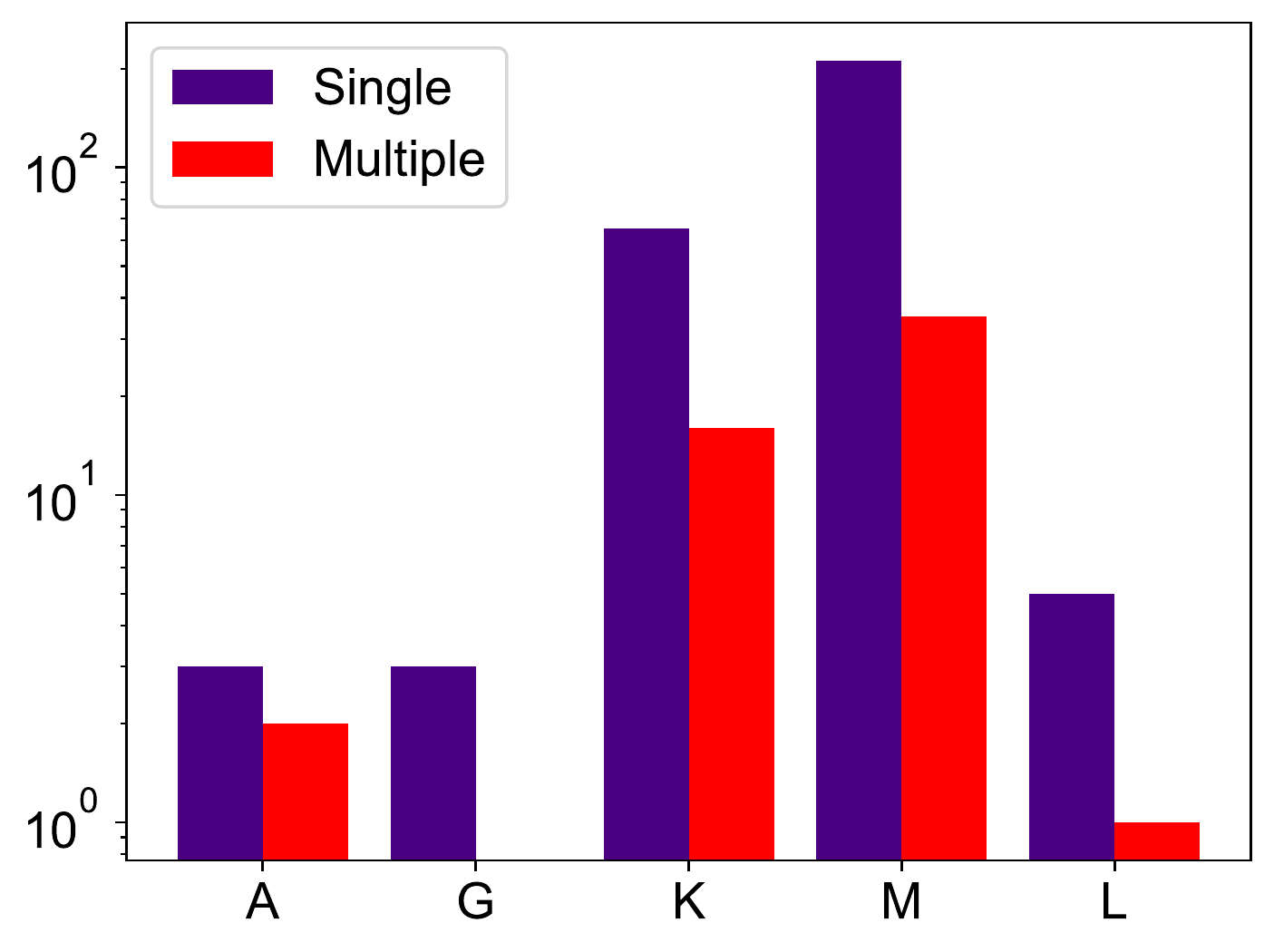}
\caption{Histogram showing the distribution of the spectral type of the disk-bearing stars for the isolated systems vs multiple systems. The scale is logarithmic. The sample includes all the disks in the Ophiuchus, Lupus, and Taurus regions. }\label{figsp}
\end{figure}

Considering all the disks of the three SFRs together, we investigate the spectral type of the primary star. As shown in Fig.~\ref{figsp}, the majority of the stars in the samples are M-type, but companions can be found around all the spectral types.

\section{Resolved observations of young multiple systems}
\label{sec:resobs}
\subsection{A word about formation pathways}

There are several ideas for forming multiple stellar systems, and they usually rely on either fragmentation processes or dynamical interactions. These ideas can be listed as four main mechanisms: filament fragmentation, core fragmentation, disk fragmentation, and dynamical captures. Each of these processes requires different spatial and temporal scales and a different number of events of gravitational collapse. Fragmentation applies mainly to self-gravitating objects that evolve following the development of dense structures that collapse under their own weight. Filament and core fragmentation operate at fractions of parsec scales, while disk fragmentation is more local to the circumstellar environment with sizes from hundreds of au down to a few au. In terms of timescales, the fragmentation processes are relatively fast, lasting for a fraction of a mega year. At the same time, dynamical capture can be slow and dependent on the presence of gas for friction to reduce orbital separations, making this process environment-dependent. { See review by \cite{2022arXiv220310066O}, and references therein, for more details on each evolutionary pathway.}

Multiple star formation is hard to model as the overall process is likely a combination of the mechanisms mentioned above, each regulated by a not fully understood interplay between gravity, thermal pressure, turbulence, and magnetic fields. Detailed models for each formation mechanism provide predictions that can be tested against observations. We describe some of these observations in the { following sections}.

\subsection{High-contrast imaging observations}

Optical high-contrast imaging observations can be a powerful method to study the interaction of the disks in multiple systems. In the era of VLT/SPHERE \cite{2019A&A...631A.155B}, Gemini/GPI \cite{2014PNAS..11112661M} and SUBARU/HiCIAO \cite{2010SPIE.7735E..30S}, many faint substructures of protoplanetary disks have been revealed for the first time. During the last years, these instruments have targeted numerous multiple systems. In some cases, all the components of the multiple systems host a protoplanetary disk, which can lead to the study of the mutual interaction of dust and gas of these disks. 

Recent observations of multiple systems during the Disk Evolution Study Through Imaging of Nearby Young Stars (DESTINY) survey \cite{2020A&A...642A.119G,2021ApJ...908L..25G} with SPHERE were presented in \cite{2023A&A...672A.145Z}. Three multiple systems were observed in polarized light: CHX 22, S CrA, and HP Cha. The first system, CHX 22 is composed of a weak T Tauri star with a 0.24 arcsec separated companion \cite{2012A&A...540A..46D}. The two components of the system are 16 and 9 Myr old \cite{2012A&A...540A..46D}, and they are part of the Chamaeleon I cloud. The second system, S CrA, is also composed of T Tauri stars, very young ($\sim$ 1 Myr), located in the Corona Australis SFR \cite{2018A&A...614A.117G}. The last system, HP Cha, is a G-type WTTS also located in Chamaeleon I SFR, with a binary system separated by 2.4 arcsec \cite{1988A&A...207...46C, 2013A&A...554A..43D, 2013A&A...557A..80S}. The triple system has an age of 1.8 Myr \cite{2013A&A...554A..43D}. 
\begin{figure}[h!]%
\centering
\includegraphics[width=0.95\textwidth]{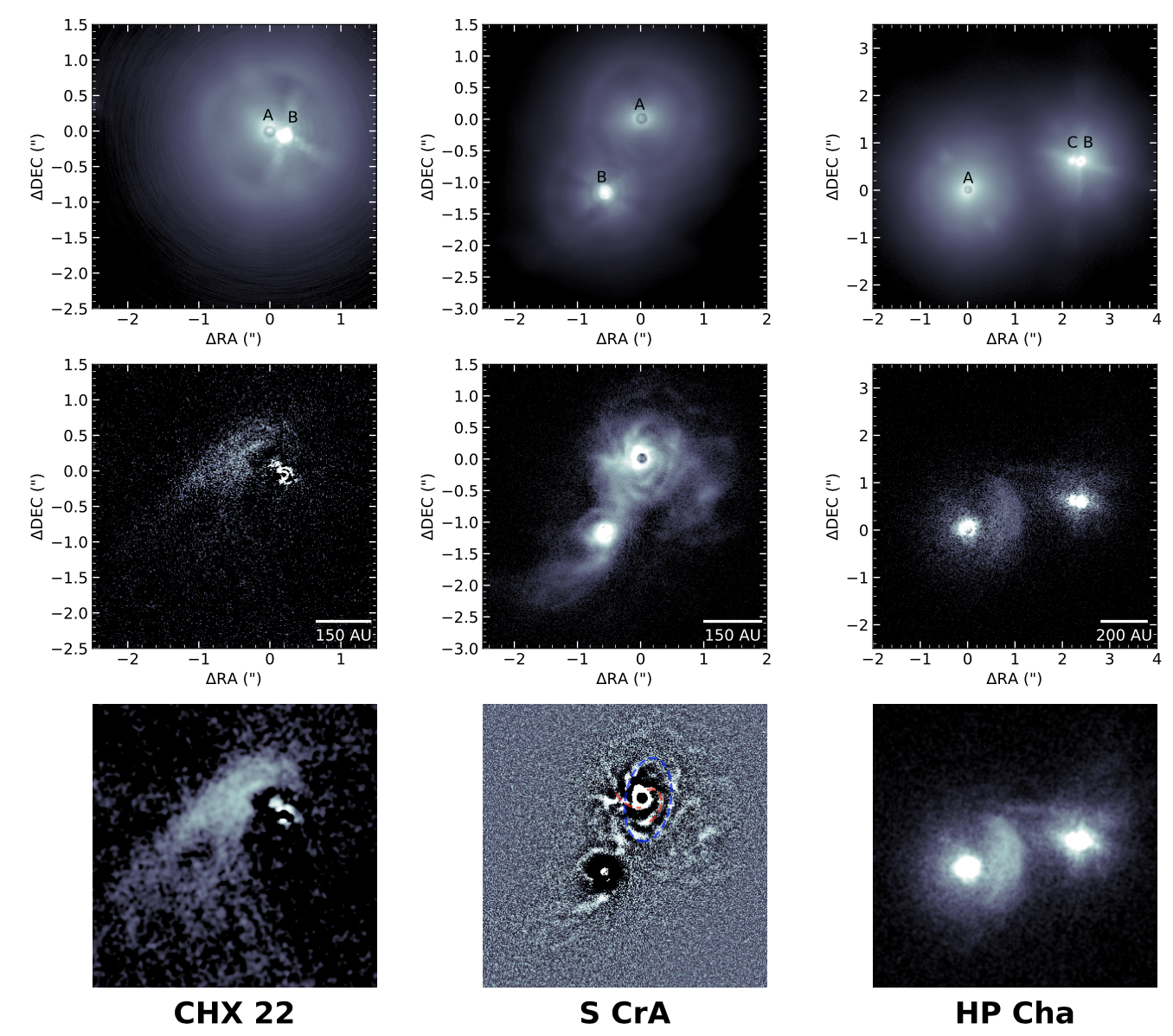}
\caption{Total intensity (top), polarized intensity( middle), and smoothed polarized intensity (bottom panels) of the three systems CHX 22, S CrA, and HP Cha as seen with SPHERE. The components of the multiple systems are indicated with letters. Image from \cite{2023A&A...672A.145Z}. }\label{sphere}
\end{figure}

The three systems have been observed with the SPHERE instrument in polarimetric mode, H band (see, e.g., \cite{2020A&A...633A..63D}). This mode allows to detect the scattered light emission from the disks in these multiple systems. The final images of each system are shown in Fig.~\ref{sphere}. In CHX 22, which is a close binary, the dust surrounds both companions, but a long dust tail is seen towards the South. The circumstellar material makes an arc-shaped feature which may be produced more likely by the interaction with the surrounding cloud of the Chamaeleon molecular cloud, more than the result of a stellar fly-by {\cite{2019A&A...628A..20D}}. The second system, S CrA, clearly shows two individual disks: the circumstellar disk around the primary (North) shows a gap and a spiral feature. The spiral feature has a pitch angle of $\sim$50 deg, which is compatible with the perturbation by the secondary. On the other hand, the ring structure cannot be explained by the presence of the secondary, and it is more likely carved by a substellar companion. The last system is a triple system, HP Cha. In the mm only the disk around the primary component is seen, which appears to be a transition disk with a ring \cite{2020ApJ...892..111F}. But in scattered light, the gas is detected around all the components. The material is flowing from the primary onto the secondary system. The disk around the primary shows a regular shape, which means that the interaction with the disk of the companions was not violent. The material around the BC component may be shaped by the extreme eccentricity of their orbit. 

In this example of three multiple systems where all the components show circumstellar material, it is seen how the separation of the companion from the host star plays an important role in the interaction of the different disks. In the first case, where the components of the system are close (a=46 au), the primordial disk has been significantly truncated by the binary interaction. In the second case, where the binary is more separated (a=322 au), the interaction between the two disks is evident: the spiral arm identified in the polarized images is compatible with the presence of the binary companion. The last case represents an example where the very wide orbit of the binary (a$\sim$500 au) makes the interaction non-destructive: the shape of the disk around the primary is regular.  

\begin{figure}[h]
\centering
\includegraphics[width=0.9\textwidth]{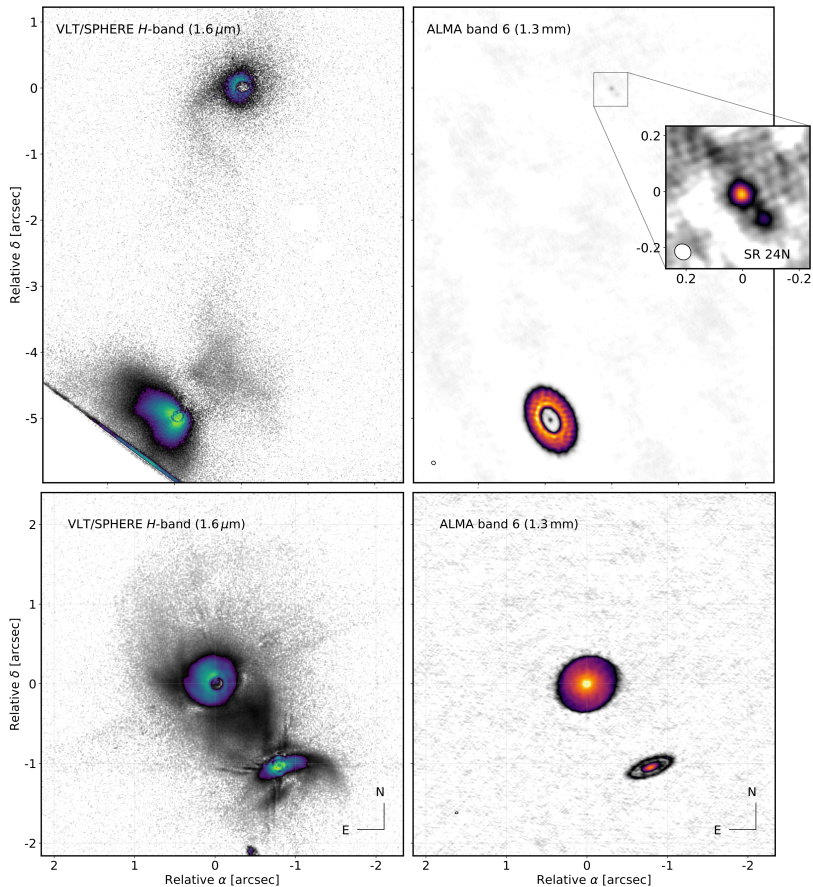}
\vspace*{-.5cm}
\caption{Gallery showing the comparison between high-resolution observations at 1.6 $\mu$m with VLT/SPHERE (left) and 1.3 millimeters with ALMA (right) for two interacting multiple systems: SR24 (top) and AS205 (bottom). All observations have a resolution of about 50 mas. Light that is scattered off small dust grains is shown with a green color map while the thermal continuum emission from sub-millimeter grains is shown as a heat color map. Noise below a 3-sigma threshold has been clipped out. Figure adapted from \cite{2023MNRAS.518.5620W} with permission from the author.}\label{fig:interacting}
\end{figure}

Last but not least, a notable example of an observation of interacting protoplanetary disks can be seen in the images of the SR 24 multiple system. SR 24 is a hierarchical triple system located in Ophiuchus and consists of a southern disk (primary) and a circumbinary disk to the north (SR24 Na and Nb). An image obtained with the SUBARU/HiCIAO AO camera in $H$-band (1.6 $\mu$m) shows a bright bridge of reflected light connecting both disks  and a surprisingly long tidal arm extending from the southern component \cite{2010Sci...327..306M,2020AJ....159...12M}. New SPHERE data have shown the details of the connecting filament between the two main stars as well as a northern disk heavily affected by the dynamical interaction as can be seen in Figure~\ref{fig:interacting} \cite{2022MNRAS.tmp.3245W}. 

\subsection{Radio interferometric observations: ALMA and VLA}

The power of radio interferometry has revolutionized our understanding of young multiple systems thanks to high-resolution observations showing how multiplicity affects the properties of disks. The expansion of the Karl G. Jansky Very Large Array (VLA), but most notably the advent of the ALMA Observatory, has provided stunning images from parsec down to au scales of the processes of binary formation, dynamical interaction of disks, as well as circumbinary warps and disk misalignments. In this section, we give an overview of the main ideas related to the formation of multiple systems and review examples of observations of young multiple systems in this context. 





\begin{figure}
    \centering
    \includegraphics[width=\textwidth]{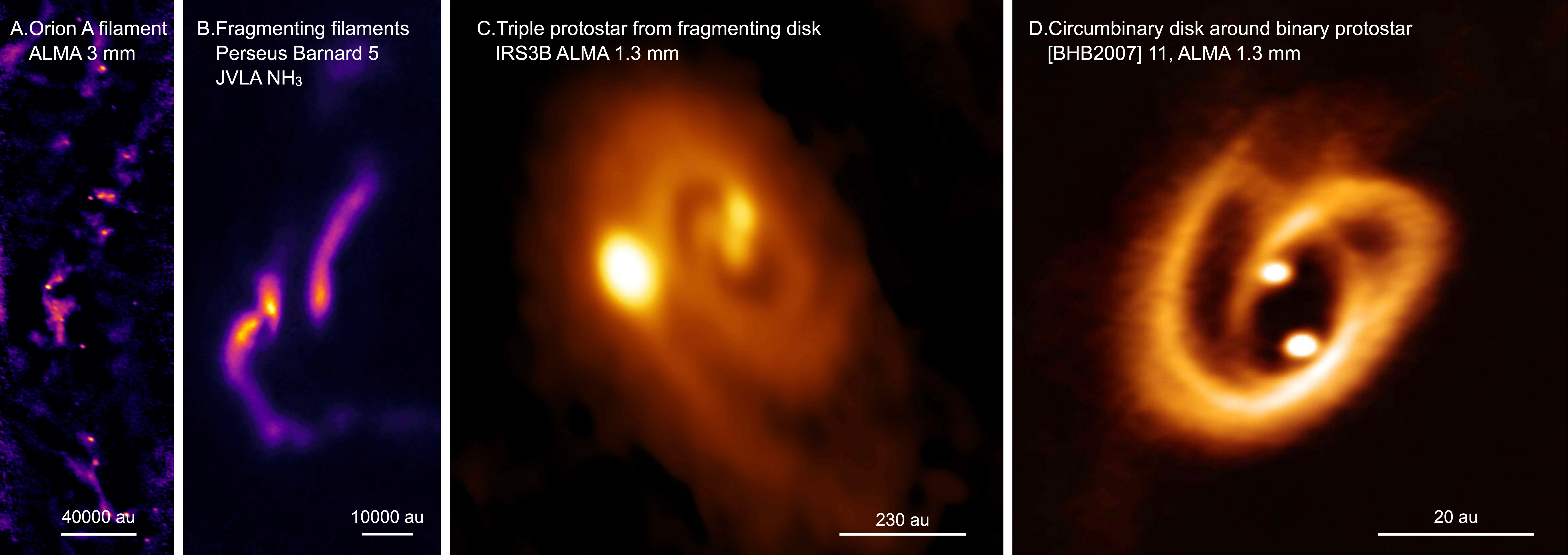}
    \caption{A gallery displaying a few examples of observations of mechanisms of multiple formation, from fragmenting filaments, through the striking morphology of a gravitationally unstable disk, to the formation of a binary protostar. From left to right, the ALMA view of the Orion A filament breaking into clumps \cite{2017A&A...600A.141K}, the JVLA image of the formation of a quadruple star system in the Perseus Barnard 5 cloud \cite{2015Natur.518..213P}, the ALMA high-resolution view of the L1448 IRS3B system with two young stars at the center and a third distant from them surrounded by spiral structure indicating the fragmentation of the disk \cite{2016Natur.538..483T}, and ALMA image of the dust filaments surrounding the protobinary system in [BHB2007] 11 \cite{2019Sci...366...90A}. }
    \label{fig:frag}
\end{figure}

The formation of multiple systems has been witnessed through various observations revealing the intricate processes at play at different spatial scales. A notable example of fragmentation is the Perseus Barnard 5 cloud. At 5000 au scales, the VLA images of Perseus B5 show that the core is resolved into several fragmenting filaments and will likely produce a quadruple system \cite{2015Natur.518..213P}. On the other hand, ALMA observations of filaments in the Orion molecular cloud show that, at $\sim$1000 au scales, the filaments appear fragmenting into several groups of dense clumps, half of which are associated with protostars \cite{2017A&A...600A.141K}. Going down to a hundred astronomical units in spatial scale, ALMA and VLA observations of the triple protostar system L1448 IRS3B show the striking morphology of a fragmenting disk \cite{2016Natur.538..483T}. At 8 mm, VLA allows seeing the locations of the three protostellar cores within the disk. At the same time, at 1.3 mm, an ALMA image recovers the emission from the circum-multiple disk showing the spiral structures expected from { gravitationally unstable massive disk} \cite{2012MNRAS.419.3115B, 2016ARA&A..54..271K}. See Figure~\ref{fig:frag} for a summary of the observations described in this paragraph. 

The complexity of the star formation process produces a wide variety of disks with different configurations, morphologies, and orientations \cite{2018MNRAS.475.5618B}. There are binaries with primary and secondary disks and circum-binaries with a single disk surrounding the multiple systems. High-resolution images of disks in the context of multiple systems have identified structures and substructures covering spatial scales over several orders of magnitudes down to the current resolution limits. Morphology diversity can be grouped into at least four categories: multiple misaligned disks, circumbinary disks with warped inner regions, multiple disks with signs of interaction, and disks without clear signs of interaction. { Here `misalignment’ refers to a situation where the plane of a disk is different from the orbital plane of the binary system it surrounds or belongs to, while `warp’ refers to a situation where a disk’s inner region is inclined with respect to the outer region of the same disk.} The last few years have seen an advent of observations of these categories. Resolving these systems is important because the relative orientation between the orbital plane and any disk in the multiple system can provide important information about the system's geometry and formation pathway.

Some of the most spectacular and complex morphologies and structures are observed in young interacting systems. It is thought that some of these structures, namely tidal arms, spirals, and bridges, can be evidence of sustained dynamical interaction or fly-by encounters (see review by \cite{2023EPJP..138...11C} and references therein). The RW Aurigae binary is a promising candidate for a fly-by as it is composed of two disks that show signatures of dynamical disruption \cite{2016A&A...596A..38F}. A kinematic analysis of the RW Aur circumstellar gas shows that the disks are strongly misaligned \cite{2018ApJ...859..150R}. Another example of a fly-by candidate is the quadrupole system UX Tau. Multiwavelength observations at near-infrared and radio wavelengths show that this system is composed of two disks whose velocity maps show misalignment, which, in turn, display evidence of perturbations that indicate a fly-by encounter between its stellar components $\sim$1000 years ago \cite{2020A&A...639L...1M}. Another intriguing multiple system is ISO-Oph-2 whose recent ALMA observations revealed two transition disks, one star surrounded by a disk with a cavity of 50 au in radius seen in millimeter emission, and a second much more compact disk with a cavity of only 2 au in radius, both disks connected by a bridge of CO gas \cite{2020ApJ...902L..33G}. 

Other notable examples of interacting disks are given by the AS 205 and SR24 systems. AS205 is a hierarchical triple whose overall molecular line emission shows extended structures thought to be due to a combination of disk interactions and winds \cite{2014ApJ...792...68S}. The high-resolution ALMA continuum images resolve this system into two separate disks, one of which shows prominent spiral arms while the other, circumbinary in nature, shows substructures in the form of a gap and an inner disk \cite{2018ApJ...869L..44K}. Similarly, recent ALMA observations of the SR 24 system have revealed that its southern disk is a transitional disk with an inner cavity \cite{2019ApJ...878...16P}. Earlier radio observations of SR 24 showed that the northern disk was deprived of dust emission but it was strong in CO emission \cite{Andrews2005}. Later, early ALMA data detected continuum signal from SR 24 N as well as a striking bridge of molecular emission connecting the stellar components \cite{2017ApJ...845...10F}. A re-analysis of recent and archival ALMA continuum data allowed us to detect and resolve the northern component into two point sources, suggesting each of the binary stars is surrounded by a compact disk \cite{2023MNRAS.518.5620W}. Figure~\ref{fig:interacting} compares images at near-IR and millimeter wavelengths, at similar resolutions, of the SR24 and AS205 systems. It is striking that the millimeter maps show structures that are typical of isolated disks (gaps, cavities, and a hint of spiral structure) but the reflected light reveals complex structures that evidence dynamical interactions.

A different class of objects that are frequently found in pairs and that also often exhibit signs of dynamical interaction are the so-called FUor objects. Their name comes from the prototype system FU Orionis whose light curve showed a dramatic increase in 5 magnitudes in 1936-37. These systems are thought to be in an active state of outburst due to episodes of enhanced accretion (see \cite{2014prpl.conf..387A} and references therein). FU Orionis itself is a binary \cite{2004ApJ...601L..83W} composed of two circumstellar disks \cite{2015ApJ...812..134H}. The disks appear compact ($\sim$10 au in radius), and although they do not show apparent substructures at the current resolution limits, their spectral energy distributions suggest several dust populations are coexisting within the ten au radii \cite{2017A&A...602A..19L, 2019ApJ...884...97L}. High-resolution ALMA images of the dust emission show that both disks share similar inclinations, while the gas emission traces a complex kinematic environment with signs of interaction and possibly mass transfer between the stellar components \cite{2020ApJ...889...59P}.  
    
The three main triggering mechanisms behind episodic accretion are the magnetorotational instability, the falling of a clump in a self-gravitating disk, and a dynamical encounter with an intruder star or flyby \cite{2021A&A...647A..44V}. Each process produces distinct kinematic signatures that are distinguishable in ALMA molecular gas observations \cite{2021A&A...647A..44V}. Although it is yet to be determined which of these mechanisms FU Orionis has undergone, there is growing interest in the flyby scenario \cite{2022MNRAS.517.4436B}, given the increased number of binaries within the group of episodically accreting objects. The Z Canis Majoris (Z CMa) system is a binary protostar where one of the components is outbursting and emanating a large streamer detected in scattered light \cite{2015A&A...578L...1C, 2016SciA....2E0875L}. Recent ALMA and VLA images of Z CMa discover the third protostar at the end of the streamer, suggesting it is a flyby around a binary young stellar object \cite{2022NatAs...6..331D}.   


\begin{figure}
    \centering
    \includegraphics[width=\textwidth]{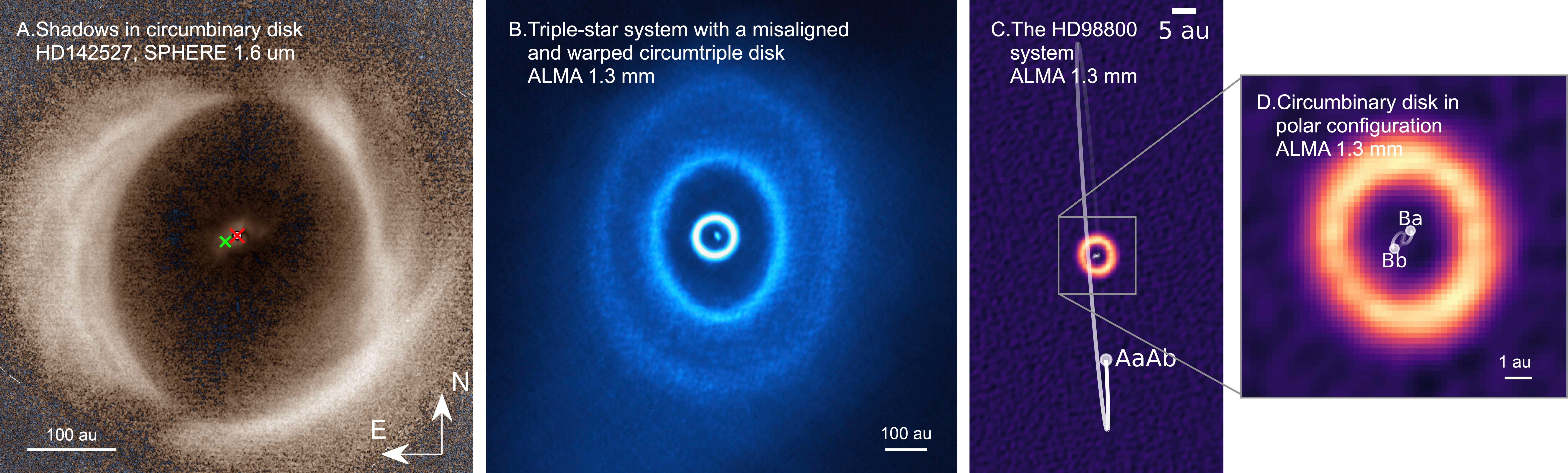}
    \caption{Observational examples of circum-multiple disks with misalignments. From left to right, the circumbinary disk around HD142527 as seen in polarized scattered light by SPHERE\cite{2017AJ....154...33A}, the circum-triple system GW Orionis displaying a set of nested misaligned dusty rings at 1.3 mm \cite{2020Sci...369.1233K}, the long-lived circumbinary disk in a polar configuration in the quadruple system HD98800 as revealed by ALMA \cite{2019NatAs...3..230K}. }
    \label{fig:circum}
\end{figure}

In the case of circumbinary disks, if the inner binary's orbit is inclined with respect to the disk, it can warp the inner regions and dynamically break the disk into a configuration where the inner and outer parts have different orientations. This is the case of the HD 142527 protoplanetary disk, where a low-mass stellar companion on a near-polar orbit around a Herbig star has cleared a deep cavity at the same time that it has produced a dramatic warp in the disk. This configuration was first identified to explain two oppositely-directed shadows cast onto the outer disk of HD 142527 \cite{2015ApJ...798L..44M} (see Figure~\ref{fig:circum}A) and the peculiar kinematics inside the cavity \cite{2015ApJ...812..126C}. Dynamical models have shown that such inner binary in a polar orientation is indeed a stable configuration able to explain the warp and the cavity, as well as several other observational features \cite{2018MNRAS.477.1270P}.

A rare example of disk misalignment is the GW Orionis system, a young star system hosting three stars and a disk. The inner region of this system has two stars GW Ori A and B that orbit each other and are separated by 1 au, and the third star, GW Ori C, orbits its siblings at a distance of just 8 au \cite{2020Sci...369.1233K, 2020ApJ...895L..18B}. The triple system is surrounded by a disk consisting of several rings with different orientations. The precise shapes of these nested rings were revealed by ALMA and can be appreciated in Figure~\ref{fig:circum}B. 

The well-known quadrupole young stellar system HD 98800 consists of two pairs of binaries that themselves orbit each other \cite{2019NatAs...3..230K}. One of the pairs is surrounded by a narrow circumbinary disk in a polar configuration, i.e., perpendicular to the inner binary's orbital plane (see Figure~\ref{fig:circum}C). Furthermore, the second binary pair orbits external to the polar disk in a wide orbit on a plane that, although less constrained, is likely misaligned with the rest of the system. This system { is a prime example} of a multiple system with a {\it polar} circumbinary disk. As the number of observational and theoretical studies of polar disks increases, it becomes more apparent that a polar misalignment may be a natural result of the intricate nature of multiple formation. 

T Tauri, the prototypical system of its class, is another clear case of misalignment in a multiple stellar system \cite{2020ApJ...902..132B}. In T Tauri, there is an interplay of stellar components being obscured by a circumbinary ring, with misaligned disks and interacting outflows, making it an exciting laboratory to understand early dynamical processes in multiple star formation \cite{2022arXiv220310066O}.

\section{Relative inclination for circumbinary and multiple disks}

The relative inclination $\theta$\ of the binary orbital planes and circumbinary disks and between multiple disks may provide important details on how disks evolve with time and star forms. In general, we expect that the spatial orientation of disks is due to the direction of the angular momentum vector of the material accreting from the interstellar matter organized in filaments \cite{Low1984, Miville2010, Ward-Thompson2010, Andre2010} onto the disk. The orientation of this vector may evolve with time as a consequence of the relative motion between the center of gravity of the system and the filament due e.g. to turbulence \cite{Offner2016} or to the effect of gravity by nearby objects. On the other hand, large-scale interstellar magnetic fields may help to keep constant the direction of the filaments for a long interval of time \cite{Galli2006, Hennebelle2013, Wang2011}. Magnetic fields likely play an important role in star formation \cite{Shu1987}, though it is unclear that the effect of magnetic fields is still relevant at scales $<1000$~au \cite{Zhang2014} and then if they may keep constant the angular momentum vector of the material accreting from the interstellar matter onto protostellar disks. In addition, the role of magnetic fields during star formation may be complex and vary with the age of the system \cite{Targon2011, Chapman2013}. If during the star formation phase companions are generated by turbulent fragmentation within disks \cite{Adams1989, Kratter2010}, generating a multiple system, the further accretion on the various components depends on the relative orientation of the orbital plane of these companions and of the plane of the disk. If these two planes are aligned { and orbits are not highly eccentric \cite{Dunhill2015}}, we expect that accretion mainly occurs on the secondary that in its orbital motion is closer to the disk and then intercepts the material that is slowly flowing from the outer to the inner parts of the system \cite{Clarke2012}; this process tends to generate equal mass binaries. On the other hand, if the two planes are misaligned, the secondary spends most of the time far from the disk plane and accretion mainly occurs on the primary (see Rigliaco et al. private communication), maintaining the mass ratio $q=M_B/M_A$ smaller than 1. In addition, we might expect that the torque exerted by the disk on the secondary \cite{Ward1997} is larger if the disk is coplanar to the orbital plane - causing a more efficient migration of the secondary toward the primary \cite{Moe2018, Arzamasskiy2018}. We may then expect that coplanarity may favor the generation of close binaries, that could not form in situ \cite{Larson1969, Bate2002, Bate2019}. On the other hand, the disk-secondary interactions should dampen the orbital inclination  at least in the linear regime ($\theta<(h/r)$, $h$ and $r$ being the disk height and radius) \cite{Tanaka2004}. { In addition, a high eccentricity may cause disk precession and a more even accretion on the two components \cite{Dunhill2015}. This may lead to keeping a mass ratio quite different from unity. We should then expect a correlation between eccentricity and mass ratio. Finally, the case of triple hierarchical systems has been studied by Ceppi and coworkers \cite{Ceppi2022}. They found that the smaller binary of a hierarchical triple accretes more mass than a single star of the same mass as the smaller binary. If the small binary is heavier than the third body, the standard differential accretion scenario (whereby the secondary accretes more of the mass) is hampered. Reciprocally, if the small binary is lighter than the third body, the standard differential accretion scenario is enhanced.}


From the model point of view, the relative frequency of aligned and misaligned systems is quite unconstrained at present, given our ignorance about the timescale of accretion from the ISM to the disks and of the relative role of turbulence and magnetic fields in star formation (for a discussion, see \cite{Tokovinin2020}). It is then very important to compare these expectations with actual observations. The alignment of disks and filaments in various star-forming regions was studied by \cite{Aizawa2020}, who found that disks orientation is in most cases consistent with a random orientation except for the Lupus III cloud, which indicates the departure from the random distribution at the 2~$\sigma$-level. Another possibility is to consider the relative orientation between jets (that are assumed to be perpendicular to disks) and filaments \cite{Bally2016}. Results are quite contradictory: while outflows look roughly perpendicular to filaments in the Coronet cluster \cite{Kumar2011}, in IRDC G28.34+0.06 \cite{Wang2011}, and in a number of protostellar cores explored by \cite{Chapman2013}, an almost random proportion of jet orientations have been found in the Perseus molecular cloud \cite{Lee2016}. \cite{Targon2011} found no alignment over a sample of 28 different regions around protostars, but some alignment for the less evolved among them (class 0 and 1 objects). Random alignment is also found for denser protocluster regions \cite{Baug2020}. This might indicate that both magnetic fields and turbulence play a role, but the relative importance is unclear. \cite{Tokovinin2017} rather considered the statistics of hierarchical triples, determining the relative inclination of the outer companion with respect to the inner binary. They found that typical relative inclination is of the order of 20 degrees for systems where the tertiary is at less than 50 au from the inner binary, while the relative inclination is essentially random for companions further than 1000 au. This might be related to the different origin of systems with separation below (disk fragmentation, \cite{Adams1989, Kratter2010}) and above (cloud fragmentation, \cite{Fisher2004, Goodwin2004, Offner2010}) these limits.

\begin{figure}
    
\centering
    
\includegraphics[width=0.9\textwidth]{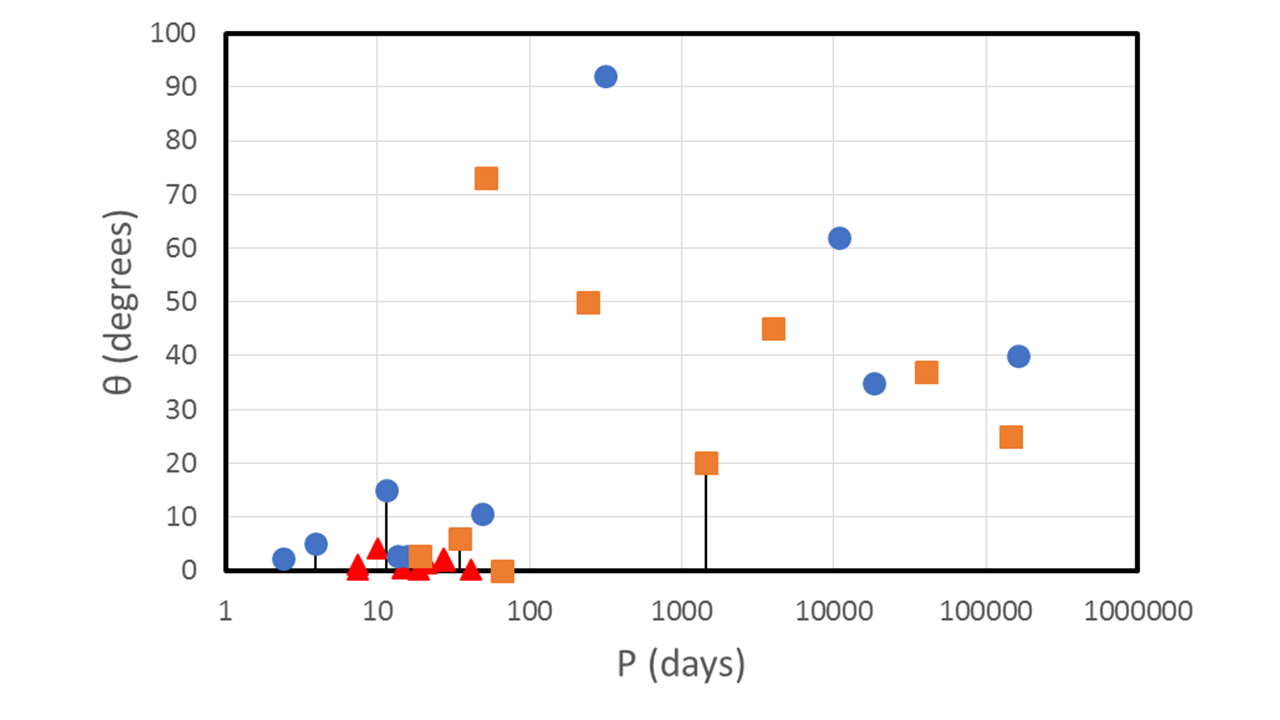}
    
\caption{Run of the relative inclination $\theta$ between circumbinary disks and orbital plane with orbital period. Orange squares are systems of higher multiplicity; blue circles are systems with only two known components. Points with lines are upper limits to $\theta$. Red triangles are circumbinary transiting planets from Kepler and TESS.}
    
\label{fig:rel_inc}

\end{figure}


Observations of circumbinary protoplanetary disks may provide more direct insight into this phenomenon. A similar analysis has been recently done by \cite{Czekala2019} who found that circumbinary disks are typically well aligned with the inner binary when the binary period is below a few tens of days, while the alignment is much poorer for wider binaries. Table~\ref{tab:rel_inc} presents the results for the stars with good and moderately good orbital parameters considered by \cite{Czekala2019} as well as a few others gathered from the literature. For each star, we give the total mass of the binary, $\theta$, the binary period $P$, semi-major axis $a$, the mass ratio $q$, the eccentricity $e$, and the inner radius of the circumbinary disk $r_{\rm in}$. The systems are sorted in order of increasing period. The median value of $\theta$ among these 20 targets is 25 degrees, with a standard deviation of 28 degrees and an uncertainty on the mean value of 6 degrees. This is fully consistent with the value obtained by \cite{Tokovinin2017} for triples with separations $<50$~au. A similar value of $\theta=28\pm 3$ degrees (r.m.s. of 13 degrees) has been obtained by \cite{Bohn2022} when considering the misalignment between inner and outer segments in transition disks.

\begin{table}[]

\label{tab:rel_inc}

\footnotesize
    
\caption[width=0.9\textwidth]{Relative disk-orbit inclination in circumbinary disks.}
\centering
    
\begin{tabular}{lcccccccl}

\hline

Star	&	Tot. mass	& $\theta$	&	$P$	&	$a$	&	$q$	&	$e$	&	$r_{\rm in}$ & Ref.	\\

	&	M$_\odot$	&	degrees	&	d	&	au	&		&		&	au	\\

\hline
\hline

V4046 Sgr	&	1.75	&	$<$2.3	&	2.4	&	0.04	&	0.94	&	0.00	&	 5	&\cite{D'Orazi2019, Martinez2022}\\

Corot 2239	&	1.17	&	$<$5	&	3.9	&	0.05	&	0.74	&	0.00	&	--	&\cite{Gillen2014, Terquem2015, Gillen2017}\\

HD131511	&	1.24	&	$<$15	&	11.5	&	0.11	&	0.57	&	0.51	&	--	&\cite{Kennedy2015}\\

AK Sco	&	2.50	&	$<$2.7	&	13.6	&	0.15	&	1.00	&	0.47	&	13	&\cite{Alencar2003, Anthonioz2015, Czekala2019, Janson2016}\\

DQ Tau	&	1.22	&	$<$2.7	&	15.8	&	0.13	&	0.94	&	0.57	&	33.6	&\cite{Czekala2016, Guilloteau2011}\\

UZ Tau E	&	1.31	&	$<$2.7	&	19.1	&	0.15	&	0.30	&	0.33	&	26	&\cite{Czekala2019, Long2018}\\

TWA 3A	&	0.53	&	$<$6	&	34.9	&	0.17	&	0.84	&	0.63	&	0.63	&\cite{Andrews2010, Kellogg2017, Czekala2021}\\

KH15D	&	1.30	&	5 - 16	&	48.4	&	0.28	&	0.85	&	0.74	&	1	&\cite{Herbst2002, Capelo2012, Johnson2004, Poon2021}\\

V773 Tau	&	2.91	&		73	&	51.1	&	0.38	&	0.86	&	0.27	&	--	&\cite{Welty1995, Boden2007, Kenworthy2022}\\

R CrA	&	5.34	&		0	&	65.8	&	0.56	&	0.77	&	0.00	&	--	&\cite{Sissa2019, Mesa2019}\\

GW Ori A-B	&	4.48	&		50	&	241.5	&	1.25	&	0.60	&	0.13	&	--	&\cite{Czekala2017}\\

HD98800	&	1.30	&		92	&	315.0	&	0.99	&	0.86	&	0.77	&	3.5	&\cite{Kennedy2012, Andrews2010}\\

HD200775	&	9.77	&	$<$20	&	1431.8	&	5.3	&	0.82	&	0.30	&	100	&\cite{Monnier2006, Okamoto2009, Benisty2013}\\

GW Ori AB-C	&	5.63	&		45	&	4017.7	&	8.8	&	0.26	&	0.13	&	--	&\cite{Czekala2017}\\

GG TauAb1-Ab2	&	0.68	&	--		&	5113.4	&	5.1	&	0.80	&	0.44	&	60	&\cite{Aly2018, Guilloteau1999}\\

T CrA	&	2.60	&		62	&	10811	&	13.2	&	0.53	&	0.00	&	25	& { \cite{Rigliaco2023}}\\

HD142527	&	2.21	&		35	&	18300	&	17.7	&	0.05	&	0.50	&	140	&\cite{Claudi2019, Hunziker2021}\\

SR 24N	&	1.27	&		37	&	40500.0	&	25	&	0.50	&	0.64	&	--	&\cite{Andrews2005, Fernandez2017, Schaefer2018}\\

GG Tau Aa-Ab	&	2.13	&		25	&	146000	&	70	&	0.88	&	0.50	&	230	&\cite{Cazzoletti2017}
\\
IRS 43	&	2.00	&	$>$40	&	164000	&	74	&	1.00	&	--	&	132	&\cite{Brinch2016}\\

\hline
    
\end{tabular}

\end{table}


Data in Table~\ref{tab:rel_inc} were used to construct the plot shown in Fig~\ref{fig:rel_inc}. With a slightly larger sample, we confirm the result obtained by \cite{Czekala2019} that circumbinary disks around short-period binaries are coplanar while those around longer-period systems may have a wide range of relative inclinations. Indeed, the distribution of points in this diagram may well be represented by two different families of systems: short-period-coplanar and long-period-high inclination, with a cut-off at about 50-100 d. The distribution of systems with period in our sample - with half of the systems having a $P<100$~d - is strongly affected by selection effects, because orbits may be better determined for close binaries. To put this value in context, we notice that $\sim 9$\% of the solar-like binaries (itself some 50\% of the stars) has period shorter than 100~d using the period distribution by \cite{Raghavan2010}; on the other hand, about 40\% of the solar-like binaries have periods between $10^2$ and $10^5$~d. The coplanarity of orbit and disk in short-period binaries can be understood in a context where migration of the secondary is very efficient in co-planar disks, while it is much less efficient in case of misalignment \cite{Arzamasskiy2018}. In addition, both families include similar proportions of purely binary systems and systems of higher multiplicity. These last cases are in principle more complex from a dynamical point of view because the perturbations by additional objects can excite the inclination of the circumstellar disk \cite{Lidov1962}, but this distinction is not clear in this figure perhaps because of the small number of systems considered. Finally, as noticed by \cite{Czekala2019}, the near co-planarity of circumbinary disks around short-period binaries implies near co-planarity of circumbinary planets discovered by Kepler \cite{Winn2015, Orosz2019} and TESS \cite{Kostov2020}, which in turn implies that the occurrence rate of close-in circumbinary planets is similar to that around single stars. The location of circumbinary planets matches very well the family of short-period-coplanar circumbinary disks, which are then very likely the parent of the transiting circumbinary planet population. On the other hand, the agreement between the (high) frequency of circumbinary planets and of planets around single stars suggests that most if not all close-period binaries have coplanar circumbinary disks at some phase of their evolution.

\begin{figure}
    
\centering
    
\includegraphics[width=0.9\textwidth]{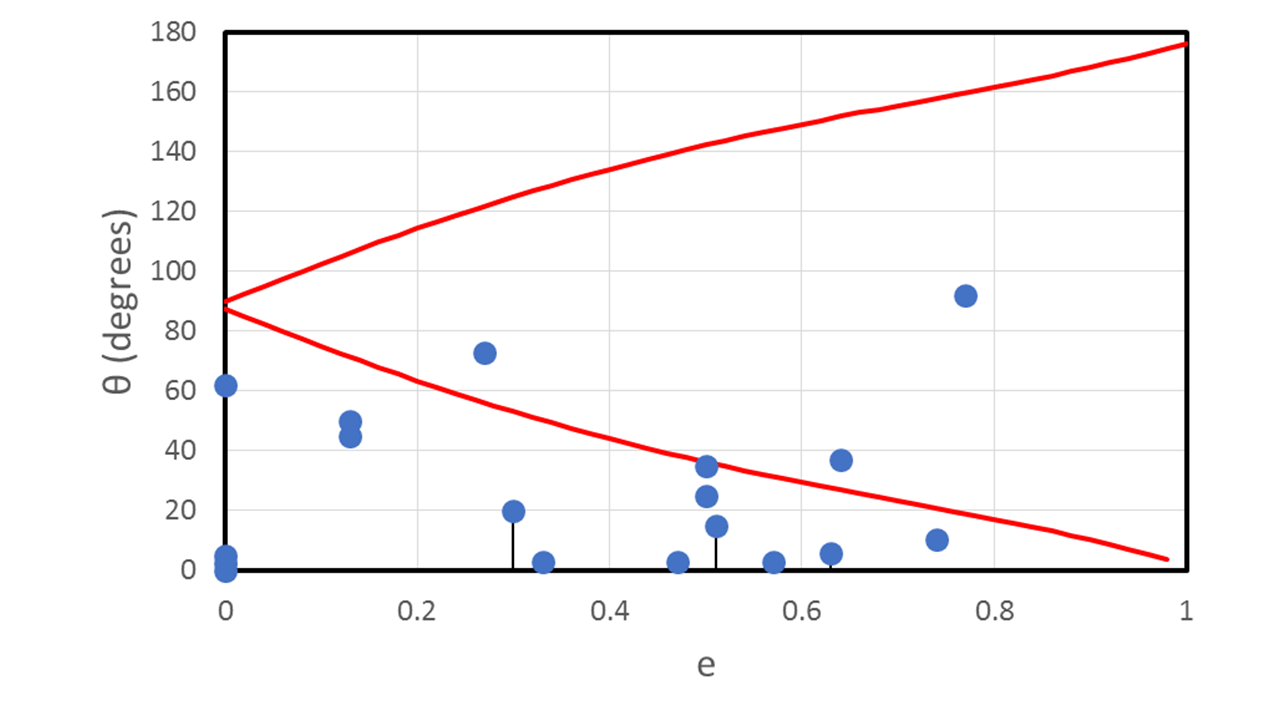}
    
\caption{Run of the relative inclination $\theta$ between circumbinary disks and orbital plane with the eccentricity of the binary orbit. The solid lines represent the lower and upper inclination limits for the nodal libration mechanism  with the inclination oscillating around $\theta =90$~degrees according to \cite{Doolin2011}. }
    
\label{fig:disk_eccentricity}
\end{figure}


Various authors (see e.g. \cite{Verrier2009}) studied the pumping of inclination of circumbinary disks that is due to a highly eccentric orbit of the binary; this may cause a nodal libration, with the inclination oscillating around $\theta =90$~degrees. \cite{Doolin2011} found that this may occur if the initial inclination of the disk is higher than a value that depends on the eccentricity of the binary orbit. Due to the damping of oscillations in a viscous disk, these systems may end up in a polar configuration \cite{Martin2018}. Figure~\ref{fig:disk_eccentricity} shows the correlation existing between relative disk inclination and eccentricity. This figure shows that only a small fraction of the systems listed in Table~\ref{tab:rel_inc} are in this regime: HD98800, V773 Tau, and SR24N. { Some of these systems have a higher multiplicity; recently, Ceppi and coworkers\cite{Ceppi2023} considered the case of hierarchical systems and found that this can have an important effect on the stability of polar disks, possibly preventing it. In particular, they found that discs orbiting the innermost level of an HS can go polar but radially extended discs orbiting the outer levels of an HS cannot polarly align and evolve as orbiting around a circular binary. }

\begin{figure}
    
\centering
    
\includegraphics[width=0.9\textwidth]{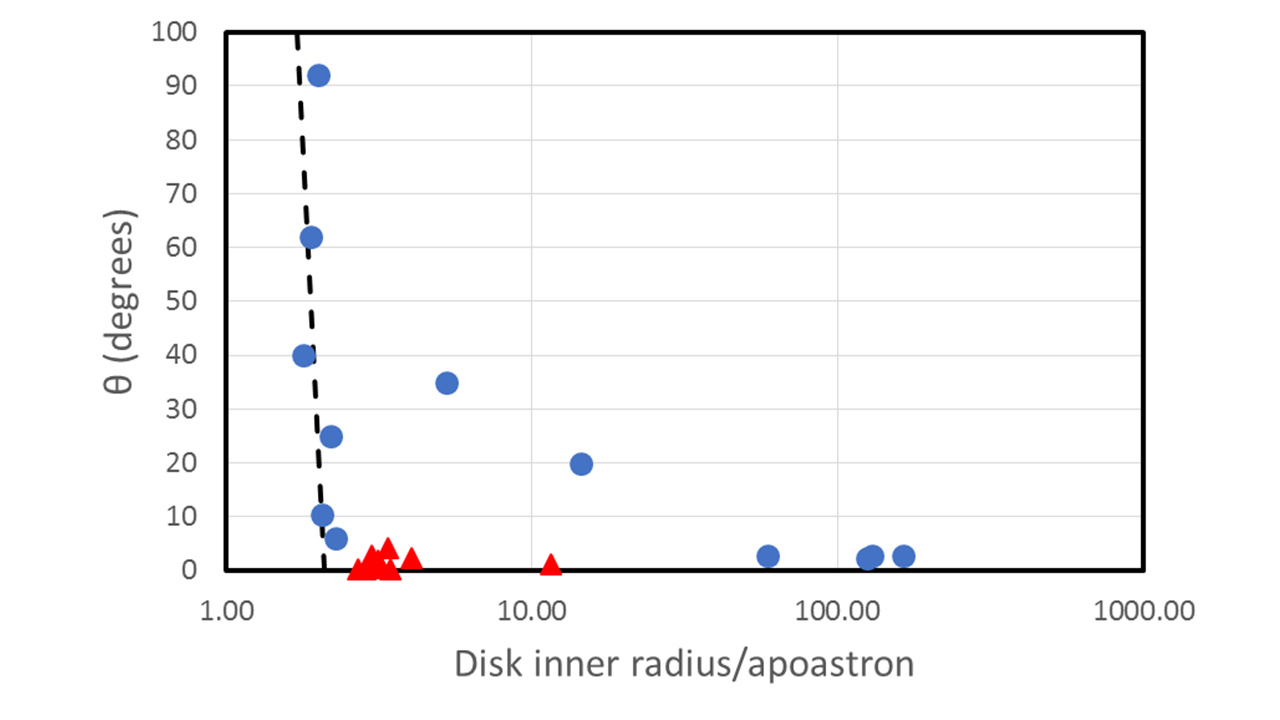}
    
\caption{Run of the relative inclination $\theta$ between circumbinary disks and orbital plane with the ratio between the disk inner radius and the orbit apoastron. Red triangles are circumbinary transiting planets from Kepler and TESS. The dashed line represents the minimum stable orbital radius around a binary with mass ratio $q=0.5$ according to \cite{Chen2020}. }
    
\label{fig:disk_radius}

\end{figure}


Another interesting plot shows the run of the relative inclination $\theta$ between circumbinary disks and the  orbital plane with the ratio between the disk inner radius and the orbit apoastron (see Fig~\ref{fig:disk_radius}). This plot shows that there is an anti-correlation between these two quantities. An anti-correlation was indeed expected because the distance from the binary where orbits become unstable is larger for the coplanar case { \cite{Doolin2011, Martin2018, CuelloGiuppone2019, Chen2020}, though some of these results refer to planetary orbits rather than gaseous disks}. However, the values of the ratio between the disk inner radius and the orbit apoastron for the coplanar systems in most cases largely exceed the minimum expected value. This can be due to a lack of recognition of the existence of disk structures closer to the binary, so this result should be taken with care. If true, it would be very intriguing, suggesting that some mechanism was able to clean up a substantial fraction of the inner disk regions. In this respect, an analysis of the mechanisms forming short-period binaries by \cite{Moe2018} concluded that most of these systems should form through substantial extra energy dissipation due to interactions with primordial gas (leading to compact equal mass objects), with a minority possibly formed by the Kozai-Lidov mechanism in triple system { \cite{1962AJ.....67..591K}}; in addition, a few others of the systems with slightly longer periods $P\sim 10--100$ days and rather a high eccentricity might be explained by dynamical ejections and disruptions of unstable coplanar triples in the disk, producing solitary binaries. We notice here that only one of the four systems with a substantial fraction of the disk cleaned up (that is a ratio between the disk inner radius and the orbit apoastron larger than 30), UZ Tau E, has additional companions (the visual binary UZ Tau W), but at a large separation of $\sim 500$~au, so the Kozai-Lidov mechanism is not likely important even for this system. UZ Tau E has the longest period among these four and it is the only one with $q<0.9$.

\begin{figure}
    
\centering
    
\includegraphics[width=0.9\textwidth]{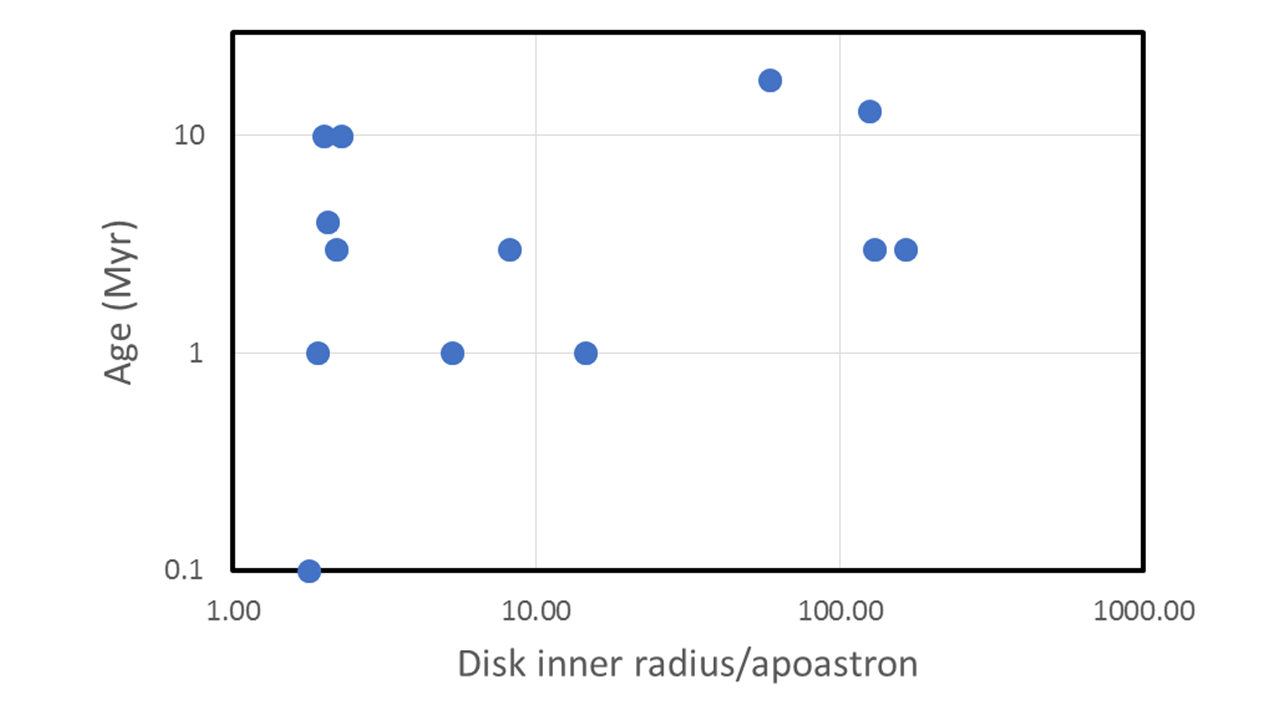}
    
\caption{Run of the age of systems with the ratio between the disk inner radius and the orbit apoastron. Ages are taken from the compilation by \cite{Czekala2019}, but for HD200775 for which we adopted an age of 1 Myr following the analysis of nearby objects by \cite{Saha2020}.}
    
\label{fig:age_radius}

\end{figure}


We also notice that the systems with a large ratio between the disk inner radius and the orbit apoastron have ages that are older than 3~Myr (ages for very young objects are quite uncertain and may then be affected by rather large errors), as shown by Fig.~\ref{fig:age_radius}; they are then relatively old systems, where the disk have undergone a significant evolution. However, there are stars with comparable ages (HD98800 and TWA3) where the disk extends much closer to the central star, in units of the orbit apoastron.


Finally, { while the search is obviously biased towards coplanar systems, the unexpected very high frequency of circumbinary planets discovered by Kepler and TESS \cite{Winn2015, Czekala2019} actually fit well in this scenario, again suggesting that they may be the descendants of these circumbinary disks (see Fig~\ref{fig:disk_radius}).}


\section{Summary and Future directions} 

Given their small physical sizes ($\lesssim$ 100 au) and large distances ($\gtrsim$ 140 pc) involved,  the observations of protoplanetary disks in multiple systems are challenging. However,  thanks to advanced instrumentation  8-m class telescopes and (sub)mm/radio interferometers,  these observations have significantly improved in the past decade.   Here we have reviewed recent demography studies of stellar and disk multiplicity as well as  detailed studies of noteworthy individual objects. The main results can be summarized as follows:  \\

\noindent 1) The combination of AO near-IR multiplicity survey and dust continuum ALMA studies  show that the distributions of disk masses and sizes are similar for apparently single (within the available limits)  stars and multiple systems for M$_{Dust}$ $\lesssim$  100 M$_\oplus$ and R$_{dust}$ $\lesssim$ 100 au. Such disks represent 80$\%$ to 90$\%$ of the samples in nearby star-forming regions. Larger and more massive disks  are almost exclusively found around single stars.  A handful of large and massive  circumbinary disks can still be found around tight separation binaries (a $\lesssim$ 1 au).  \\

\noindent 2) High-resolution optical/near-IR and (sub)mm/radio observations of large and massive disks reveal complex disk morphologies and structures in interacting disks in both  bound and unbound (flybys) systems. These features include tidal arms, spirals, and bridges and depend on a wide range of variables, such as orbital parameters,  stellar mass ratios, and disk sizes and orientations. \\ 

\noindent 3) The fragmentation of self-gravitating  structures is believed to play an important role in the formation of multiple systems and is central to most formation models. Thanks to ALMA and  VLA observations, fragmentation has been seen at the scales of filaments, cores, and disks.  The multi-scale fragmentation process produces binary and circumbinary disks with a variety of relative orientations.  \\ 

\noindent 4) The relative disk-orbit inclination in circumbinary disks can provide important information on the formation and evolution of disks in multiple systems.  For instance, observations show that circumbinary disks around short-period ($\lesssim$ 100 d) systems lie in the orbital plane, while longer periods have a wide range of inclination.  This could be due to the fast migration expected in co-planar systems. Observations also  suggest that highly eccentric orbits can pump up  the inclination of circumbinary disks.     
\\

Most of the above conclusions are based on incomplete samples and are still limited by the spatial resolution of current surveys.  While ALMA can technically reach a resolution of 10 mas (1.4 au at 140 pc), most disk surveys have been conducted at $\sim$200 mas ($\sim$ 30 au at 140 pc), a scale at which most disks in nearby star-forming regions remain unresolved.  Carrying out ALMA surveys at a higher spatial resolution to  measure the sizes of the smallest disks in  molecular clouds is desirable to better quantify the tidal truncation effects on binary systems.  Increasing the resolution of optical and near-IR binary surveys is also critical to identify tight visual binaries.  The next generation of  extremely large telescopes will have a factor of 4-5 better resolution than current instruments,  allowing the identification of close binaries down to $\sim$2 au.  The use of aperture masking interferometry in such telescopes could push the limits to sub-au separations.  Similarly, scattered, and polarized light imaging with  extremely large telescopes will allow the study of  faint and compact disks that are currently beyond current capabilities.  In terms of studying the formation of binaries,  the next-generation VLA  will provide a factor of 10 improvements in resolution with respect to ALMA at 3 mm, allowing the study at a high resolution of large samples of binary systems in very young regions such as the Orion molecular cloud at a distance of 415 pc. \\

\backmatter

\bmhead{Acknowledgments}

We thank Philipp Weber for his helpful comments and the adaptation of figure~\ref{fig:interacting}. A.Z. acknowledges support from the FONDECYT Iniciaci\'on en investigaci\'on project number 11190837. S.P. acknowledges support from FONDECYT grant 1191934. L.C. acknowledges support from the FONDECYT project number 1211656. This work was funded by ANID -- Millennium Science Initiative Program -- Center Code NCN2021\_080.

Data Availability Statement: No Data associated with the manuscript.

\bibliography{diskbin.bib}


\begin{thebibliography}{171}
\ifx \bisbn   \undefined \def \bisbn  #1{ISBN #1}\fi
\ifx \binits  \undefined \def \binits#1{#1}\fi
\ifx \bauthor  \undefined \def \bauthor#1{#1}\fi
\ifx \batitle  \undefined \def \batitle#1{#1}\fi
\ifx \bjtitle  \undefined \def \bjtitle#1{#1}\fi
\ifx \bvolume  \undefined \def \bvolume#1{\textbf{#1}}\fi
\ifx \byear  \undefined \def \byear#1{#1}\fi
\ifx \bissue  \undefined \def \bissue#1{#1}\fi
\ifx \bfpage  \undefined \def \bfpage#1{#1}\fi
\ifx \blpage  \undefined \def \blpage #1{#1}\fi
\ifx \burl  \undefined \def \burl#1{\textsf{#1}}\fi
\ifx \doiurl  \undefined \def \doiurl#1{\url{https://doi.org/#1}}\fi
\ifx \betal  \undefined \def \betal{\textit{et al.}}\fi
\ifx \binstitute  \undefined \def \binstitute#1{#1}\fi
\ifx \binstitutionaled  \undefined \def \binstitutionaled#1{#1}\fi
\ifx \bctitle  \undefined \def \bctitle#1{#1}\fi
\ifx \beditor  \undefined \def \beditor#1{#1}\fi
\ifx \bpublisher  \undefined \def \bpublisher#1{#1}\fi
\ifx \bbtitle  \undefined \def \bbtitle#1{#1}\fi
\ifx \bedition  \undefined \def \bedition#1{#1}\fi
\ifx \bseriesno  \undefined \def \bseriesno#1{#1}\fi
\ifx \blocation  \undefined \def \blocation#1{#1}\fi
\ifx \bsertitle  \undefined \def \bsertitle#1{#1}\fi
\ifx \bsnm \undefined \def \bsnm#1{#1}\fi
\ifx \bsuffix \undefined \def \bsuffix#1{#1}\fi
\ifx \bparticle \undefined \def \bparticle#1{#1}\fi
\ifx \barticle \undefined \def \barticle#1{#1}\fi
\bibcommenthead
\ifx \bconfdate \undefined \def \bconfdate #1{#1}\fi
\ifx \botherref \undefined \def \botherref #1{#1}\fi
\ifx \url \undefined \def \url#1{\textsf{#1}}\fi
\ifx \bchapter \undefined \def \bchapter#1{#1}\fi
\ifx \bbook \undefined \def \bbook#1{#1}\fi
\ifx \bcomment \undefined \def \bcomment#1{#1}\fi
\ifx \oauthor \undefined \def \oauthor#1{#1}\fi
\ifx \citeauthoryear \undefined \def \citeauthoryear#1{#1}\fi
\ifx \endbibitem  \undefined \def \endbibitem {}\fi
\ifx \bconflocation  \undefined \def \bconflocation#1{#1}\fi
\ifx \arxivurl  \undefined \def \arxivurl#1{\textsf{#1}}\fi
\csname PreBibitemsHook\endcsname

\bibitem[\protect\citeauthoryear{{Duch{\^e}ne} and
  {Kraus}}{2013}]{2013ARA&A..51..269D}
\begin{barticle}
\bauthor{\bsnm{{Duch{\^e}ne}}, \binits{G.}},
\bauthor{\bsnm{{Kraus}}, \binits{A.}}:
\batitle{{Stellar Multiplicity}}.
\bjtitle{\araa}
\bvolume{51}(\bissue{1}),
\bfpage{269}--\blpage{310}
(\byear{2013})
\doiurl{10.1146/annurev-astro-081710-102602}
{\href{https://arxiv.org/abs/1303.3028}{{arXiv:1303.3028}}}
{[astro-ph.SR]}
\end{barticle}
\endbibitem

\bibitem[\protect\citeauthoryear{{C{\'a}novas}
  et~al.}{2019}]{2019A&A...626A..80C}
\begin{barticle}
\bauthor{\bsnm{{C{\'a}novas}}, \binits{H.}},
\bauthor{\bsnm{{Cantero}}, \binits{C.}},
\bauthor{\bsnm{{Cieza}}, \binits{L.}},
\bauthor{\bsnm{{Bombrun}}, \binits{A.}},
\bauthor{\bsnm{{Lammers}}, \binits{U.}},
\bauthor{\bsnm{{Mer{\'\i}n}}, \binits{B.}},
\bauthor{\bsnm{{Mora}}, \binits{A.}},
\bauthor{\bsnm{{Ribas}}, \binits{{\'A}.}},
\bauthor{\bsnm{{Ru{\'\i}z-Rodr{\'\i}guez}}, \binits{D.}}:
\batitle{{Census of {\ensuremath{\rho}} Ophiuchi candidate members from Gaia
  Data Release 2}}.
\bjtitle{\aap}
\bvolume{626},
\bfpage{80}
(\byear{2019})
\doiurl{10.1051/0004-6361/201935321}
{\href{https://arxiv.org/abs/1902.07600}{{arXiv:1902.07600}}}
{[astro-ph.EP]}
\end{barticle}
\endbibitem

\bibitem[\protect\citeauthoryear{{Lovell} et~al.}{2021}]{2021MNRAS.500.4878L}
\begin{barticle}
\bauthor{\bsnm{{Lovell}}, \binits{J.B.}},
\bauthor{\bsnm{{Wyatt}}, \binits{M.C.}},
\bauthor{\bsnm{{Ansdell}}, \binits{M.}},
\bauthor{\bsnm{{Kama}}, \binits{M.}},
\bauthor{\bsnm{{Kennedy}}, \binits{G.M.}},
\bauthor{\bsnm{{Manara}}, \binits{C.F.}},
\bauthor{\bsnm{{Marino}}, \binits{S.}},
\bauthor{\bsnm{{Matr{\`a}}}, \binits{L.}},
\bauthor{\bsnm{{Rosotti}}, \binits{G.}},
\bauthor{\bsnm{{Tazzari}}, \binits{M.}},
\bauthor{\bsnm{{Testi}}, \binits{L.}},
\bauthor{\bsnm{{Williams}}, \binits{J.P.}}:
\batitle{{ALMA survey of Lupus class III stars: Early planetesimal belt
  formation and rapid disc dispersal}}.
\bjtitle{\mnras}
\bvolume{500}(\bissue{4}),
\bfpage{4878}--\blpage{4900}
(\byear{2021})
\doiurl{10.1093/mnras/staa3335}
{\href{https://arxiv.org/abs/2010.12657}{{arXiv:2010.12657}}}
{[astro-ph.EP]}
\end{barticle}
\endbibitem

\bibitem[\protect\citeauthoryear{{Cieza} et~al.}{2009}]{2009ApJ...696L..84C}
\begin{barticle}
\bauthor{\bsnm{{Cieza}}, \binits{L.A.}},
\bauthor{\bsnm{{Padgett}}, \binits{D.L.}},
\bauthor{\bsnm{{Allen}}, \binits{L.E.}},
\bauthor{\bsnm{{McCabe}}, \binits{C.E.}},
\bauthor{\bsnm{{Brooke}}, \binits{T.Y.}},
\bauthor{\bsnm{{Carey}}, \binits{S.J.}},
\bauthor{\bsnm{{Chapman}}, \binits{N.L.}},
\bauthor{\bsnm{{Fukagawa}}, \binits{M.}},
\bauthor{\bsnm{{Huard}}, \binits{T.L.}},
\bauthor{\bsnm{{Noriga-Crespo}}, \binits{A.}},
\bauthor{\bsnm{{Peterson}}, \binits{D.E.}},
\bauthor{\bsnm{{Rebull}}, \binits{L.M.}}:
\batitle{{Primordial Circumstellar Disks in Binary Systems: Evidence for
  Reduced Lifetimes}}.
\bjtitle{\apjl}
\bvolume{696}(\bissue{1}),
\bfpage{84}--\blpage{88}
(\byear{2009})
\doiurl{10.1088/0004-637X/696/1/L84}
{\href{https://arxiv.org/abs/0903.3057}{{arXiv:0903.3057}}}
{[astro-ph.SR]}
\end{barticle}
\endbibitem

\bibitem[\protect\citeauthoryear{{Kraus} and
  {Hillenbrand}}{2012}]{2012ApJ...757..141K}
\begin{barticle}
\bauthor{\bsnm{{Kraus}}, \binits{A.L.}},
\bauthor{\bsnm{{Hillenbrand}}, \binits{L.A.}}:
\batitle{{Multiple Star Formation to the Bottom of the Initial Mass Function}}.
\bjtitle{\apj}
\bvolume{757}(\bissue{2}),
\bfpage{141}
(\byear{2012})
\doiurl{10.1088/0004-637X/757/2/141}
{\href{https://arxiv.org/abs/1206.4995}{{arXiv:1206.4995}}}
{[astro-ph.SR]}
\end{barticle}
\endbibitem

\bibitem[\protect\citeauthoryear{{Andrews} and
  {Williams}}{2005}]{2005ApJ...631.1134A}
\begin{barticle}
\bauthor{\bsnm{{Andrews}}, \binits{S.M.}},
\bauthor{\bsnm{{Williams}}, \binits{J.P.}}:
\batitle{{Circumstellar Dust Disks in Taurus-Auriga: The Submillimeter
  Perspective}}.
\bjtitle{\apj}
\bvolume{631}(\bissue{2}),
\bfpage{1134}--\blpage{1160}
(\byear{2005})
\doiurl{10.1086/432712}
{\href{https://arxiv.org/abs/astro-ph/0506187}{{arXiv:astro-ph/0506187}}}
{[astro-ph]}
\end{barticle}
\endbibitem

\bibitem[\protect\citeauthoryear{{Cox} et~al.}{2017}]{2017ApJ...851...83C}
\begin{barticle}
\bauthor{\bsnm{{Cox}}, \binits{E.G.}},
\bauthor{\bsnm{{Harris}}, \binits{R.J.}},
\bauthor{\bsnm{{Looney}}, \binits{L.W.}},
\bauthor{\bsnm{{Chiang}}, \binits{H.-F.}},
\bauthor{\bsnm{{Chandler}}, \binits{C.}},
\bauthor{\bsnm{{Kratter}}, \binits{K.}},
\bauthor{\bsnm{{Li}}, \binits{Z.-Y.}},
\bauthor{\bsnm{{Perez}}, \binits{L.}},
\bauthor{\bsnm{{Tobin}}, \binits{J.J.}}:
\batitle{{Protoplanetary Disks in {\ensuremath{\rho}} Ophiuchus as Seen from
  ALMA}}.
\bjtitle{\apj}
\bvolume{851}(\bissue{2}),
\bfpage{83}
(\byear{2017})
\doiurl{10.3847/1538-4357/aa97e2}
{\href{https://arxiv.org/abs/1711.03974}{{arXiv:1711.03974}}}
{[astro-ph.SR]}
\end{barticle}
\endbibitem

\bibitem[\protect\citeauthoryear{{Harris} et~al.}{2012}]{2012ApJ...751..115H}
\begin{barticle}
\bauthor{\bsnm{{Harris}}, \binits{R.J.}},
\bauthor{\bsnm{{Andrews}}, \binits{S.M.}},
\bauthor{\bsnm{{Wilner}}, \binits{D.J.}},
\bauthor{\bsnm{{Kraus}}, \binits{A.L.}}:
\batitle{{A Resolved Census of Millimeter Emission from Taurus Multiple Star
  Systems}}.
\bjtitle{\apj}
\bvolume{751}(\bissue{2}),
\bfpage{115}
(\byear{2012})
\doiurl{10.1088/0004-637X/751/2/115}
{\href{https://arxiv.org/abs/1203.6353}{{arXiv:1203.6353}}}
{[astro-ph.SR]}
\end{barticle}
\endbibitem

\bibitem[\protect\citeauthoryear{{Papaloizou} and
  {Pringle}}{1977}]{1977MNRAS.181..441P}
\begin{barticle}
\bauthor{\bsnm{{Papaloizou}}, \binits{J.}},
\bauthor{\bsnm{{Pringle}}, \binits{J.E.}}:
\batitle{{Tidal torques on accretion discs in close binary systems.}}
\bjtitle{\mnras}
\bvolume{181},
\bfpage{441}--\blpage{454}
(\byear{1977})
\doiurl{10.1093/mnras/181.3.441}
\end{barticle}
\endbibitem

\bibitem[\protect\citeauthoryear{{Artymowicz} and
  {Lubow}}{1994}]{1994ApJ...421..651A}
\begin{barticle}
\bauthor{\bsnm{{Artymowicz}}, \binits{P.}},
\bauthor{\bsnm{{Lubow}}, \binits{S.H.}}:
\batitle{{Dynamics of Binary-Disk Interaction. I. Resonances and Disk Gap
  Sizes}}.
\bjtitle{\apj}
\bvolume{421},
\bfpage{651}
(\byear{1994})
\doiurl{10.1086/173679}
\end{barticle}
\endbibitem

\bibitem[\protect\citeauthoryear{{Lubow} et~al.}{2015}]{2015ApJ...800...96L}
\begin{barticle}
\bauthor{\bsnm{{Lubow}}, \binits{S.H.}},
\bauthor{\bsnm{{Martin}}, \binits{R.G.}},
\bauthor{\bsnm{{Nixon}}, \binits{C.}}:
\batitle{{Tidal Torques on Misaligned Disks in Binary Systems}}.
\bjtitle{\apj}
\bvolume{800}(\bissue{2}),
\bfpage{96}
(\byear{2015})
\doiurl{10.1088/0004-637X/800/2/96}
{\href{https://arxiv.org/abs/1412.7741}{{arXiv:1412.7741}}}
{[astro-ph.SR]}
\end{barticle}
\endbibitem

\bibitem[\protect\citeauthoryear{{Miranda} and
  {Lai}}{2015}]{2015MNRAS.452.2396M}
\begin{barticle}
\bauthor{\bsnm{{Miranda}}, \binits{R.}},
\bauthor{\bsnm{{Lai}}, \binits{D.}}:
\batitle{{Tidal truncation of inclined circumstellar and circumbinary discs in
  young stellar binaries}}.
\bjtitle{\mnras}
\bvolume{452}(\bissue{3}),
\bfpage{2396}--\blpage{2409}
(\byear{2015})
\doiurl{10.1093/mnras/stv1450}
{\href{https://arxiv.org/abs/1504.02917}{{arXiv:1504.02917}}}
{[astro-ph.EP]}
\end{barticle}
\endbibitem

\bibitem[\protect\citeauthoryear{{Zagaria} et~al.}{2023}]{2023EPJP..138...25Z}
\begin{barticle}
\bauthor{\bsnm{{Zagaria}}, \binits{F.}},
\bauthor{\bsnm{{Rosotti}}, \binits{G.P.}},
\bauthor{\bsnm{{Alexander}}, \binits{R.D.}},
\bauthor{\bsnm{{Clarke}}, \binits{C.J.}}:
\batitle{{Dust dynamics in planet-forming discs in binary systems}}.
\bjtitle{European Physical Journal Plus}
\bvolume{138}(\bissue{1}),
\bfpage{25}
(\byear{2023})
\doiurl{10.1140/epjp/s13360-022-03616-4}
{\href{https://arxiv.org/abs/2212.07711}{{arXiv:2212.07711}}}
{[astro-ph.EP]}
\end{barticle}
\endbibitem

\bibitem[\protect\citeauthoryear{{Wilking} et~al.}{2008}]{2008hsf2.book..351W}
\begin{bchapter}
\bauthor{\bsnm{{Wilking}}, \binits{B.A.}},
\bauthor{\bsnm{{Gagn{\'e}}}, \binits{M.}},
\bauthor{\bsnm{{Allen}}, \binits{L.E.}}:
\bctitle{{Star Formation in the {\ensuremath{\rho}} Ophiuchi Molecular Cloud}}.
In: \beditor{\bsnm{{Reipurth}}, \binits{B.}} (ed.)
\bbtitle{Handbook of Star Forming Regions, Volume II}
vol. \bseriesno{5},
p. \bfpage{351}
(\byear{2008})
\end{bchapter}
\endbibitem

\bibitem[\protect\citeauthoryear{{Cieza} et~al.}{2019}]{2019MNRAS.482..698C}
\begin{barticle}
\bauthor{\bsnm{{Cieza}}, \binits{L.A.}},
\bauthor{\bsnm{{Ru{\'\i}z-Rodr{\'\i}guez}}, \binits{D.}},
\bauthor{\bsnm{{Hales}}, \binits{A.}},
\bauthor{\bsnm{{Casassus}}, \binits{S.}},
\bauthor{\bsnm{{P{\'e}rez}}, \binits{S.}},
\bauthor{\bsnm{{Gonzalez-Ruilova}}, \binits{C.}},
\bauthor{\bsnm{{C{\'a}novas}}, \binits{H.}},
\bauthor{\bsnm{{Williams}}, \binits{J.P.}},
\bauthor{\bsnm{{Zurlo}}, \binits{A.}},
\bauthor{\bsnm{{Ansdell}}, \binits{M.}},
\bauthor{\bsnm{{Avenhaus}}, \binits{H.}},
\bauthor{\bsnm{{Bayo}}, \binits{A.}},
\bauthor{\bsnm{{Bertrang}}, \binits{G.H.-M.}},
\bauthor{\bsnm{{Christiaens}}, \binits{V.}},
\bauthor{\bsnm{{Dent}}, \binits{W.}},
\bauthor{\bsnm{{Ferrero}}, \binits{G.}},
\bauthor{\bsnm{{Gamen}}, \binits{R.}},
\bauthor{\bsnm{{Olofsson}}, \binits{J.}},
\bauthor{\bsnm{{Orcajo}}, \binits{S.}},
\bauthor{\bsnm{{Pe{\~n}a Ram{\'\i}rez}}, \binits{K.}},
\bauthor{\bsnm{{Principe}}, \binits{D.}},
\bauthor{\bsnm{{Schreiber}}, \binits{M.R.}},
\bauthor{\bsnm{{van der Plas}}, \binits{G.}}:
\batitle{{The Ophiuchus DIsc Survey Employing ALMA (ODISEA) - I: project
  description and continuum images at 28 au resolution}}.
\bjtitle{\mnras}
\bvolume{482}(\bissue{1}),
\bfpage{698}--\blpage{714}
(\byear{2019})
\doiurl{10.1093/mnras/sty2653}
{\href{https://arxiv.org/abs/1809.08844}{{arXiv:1809.08844}}}
{[astro-ph.EP]}
\end{barticle}
\endbibitem

\bibitem[\protect\citeauthoryear{{Zurlo} et~al.}{2020}]{2020MNRAS.496.5089Z}
\begin{barticle}
\bauthor{\bsnm{{Zurlo}}, \binits{A.}},
\bauthor{\bsnm{{Cieza}}, \binits{L.A.}},
\bauthor{\bsnm{{P{\'e}rez}}, \binits{S.}},
\bauthor{\bsnm{{Christiaens}}, \binits{V.}},
\bauthor{\bsnm{{Williams}}, \binits{J.P.}},
\bauthor{\bsnm{{Guidi}}, \binits{G.}},
\bauthor{\bsnm{{C{\'a}novas}}, \binits{H.}},
\bauthor{\bsnm{{Casassus}}, \binits{S.}},
\bauthor{\bsnm{{Hales}}, \binits{A.}},
\bauthor{\bsnm{{Principe}}, \binits{D.A.}},
\bauthor{\bsnm{{Ru{\'\i}z-Rodr{\'\i}guez}}, \binits{D.}},
\bauthor{\bsnm{{Fernandez-Figueroa}}, \binits{A.}}:
\batitle{{The Ophiuchus DIsc Survey Employing ALMA (ODISEA) - II. The effect of
  stellar multiplicity on disc properties}}.
\bjtitle{\mnras}
\bvolume{496}(\bissue{4}),
\bfpage{5089}--\blpage{5100}
(\byear{2020})
\doiurl{10.1093/mnras/staa1886}
{\href{https://arxiv.org/abs/2006.16259}{{arXiv:2006.16259}}}
{[astro-ph.SR]}
\end{barticle}
\endbibitem

\bibitem[\protect\citeauthoryear{{K{\"o}hler}
  et~al.}{2008}]{2008A&A...488..997K}
\begin{barticle}
\bauthor{\bsnm{{K{\"o}hler}}, \binits{R.}},
\bauthor{\bsnm{{Neuh{\"a}user}}, \binits{R.}},
\bauthor{\bsnm{{Kr{\"a}mer}}, \binits{S.}},
\bauthor{\bsnm{{Leinert}}, \binits{C.}},
\bauthor{\bsnm{{Ott}}, \binits{T.}},
\bauthor{\bsnm{{Eckart}}, \binits{A.}}:
\batitle{{Multiplicity of young stars in and around R Coronae Australis}}.
\bjtitle{\aap}
\bvolume{488}(\bissue{3}),
\bfpage{997}--\blpage{1006}
(\byear{2008})
\doiurl{10.1051/0004-6361:200809897}
{\href{https://arxiv.org/abs/0807.4393}{{arXiv:0807.4393}}}
{[astro-ph]}
\end{barticle}
\endbibitem

\bibitem[\protect\citeauthoryear{{Leinert} et~al.}{1993}]{1993A&A...278..129L}
\begin{barticle}
\bauthor{\bsnm{{Leinert}}, \binits{C.}},
\bauthor{\bsnm{{Zinnecker}}, \binits{H.}},
\bauthor{\bsnm{{Weitzel}}, \binits{N.}},
\bauthor{\bsnm{{Christou}}, \binits{J.}},
\bauthor{\bsnm{{Ridgway}}, \binits{S.T.}},
\bauthor{\bsnm{{Jameson}}, \binits{R.}},
\bauthor{\bsnm{{Haas}}, \binits{M.}},
\bauthor{\bsnm{{Lenzen}}, \binits{R.}}:
\batitle{{A systematic search for young binaries in Taurus.}}
\bjtitle{\aap}
\bvolume{278},
\bfpage{129}--\blpage{149}
(\byear{1993})
\end{barticle}
\endbibitem

\bibitem[\protect\citeauthoryear{{K{\"o}hler} and
  {Leinert}}{1998}]{1998A&A...331..977K}
\begin{barticle}
\bauthor{\bsnm{{K{\"o}hler}}, \binits{R.}},
\bauthor{\bsnm{{Leinert}}, \binits{C.}}:
\batitle{{Multiplicity of T Tauri stars in Taurus after ROSAT}}.
\bjtitle{\aap}
\bvolume{331},
\bfpage{977}--\blpage{988}
(\byear{1998})
\end{barticle}
\endbibitem

\bibitem[\protect\citeauthoryear{{Gaia Collaboration}
  et~al.}{2018}]{2018A&A...616A...1G}
\begin{barticle}
\bauthor{\bsnm{{Gaia Collaboration}}},
\bauthor{\bsnm{{Brown}}, \binits{A.G.A.}},
\bauthor{\bsnm{{Vallenari}}, \binits{A.}},
\bauthor{\bsnm{{Prusti}}, \binits{T.}},
\bauthor{\bsnm{{de Bruijne}}, \binits{J.H.J.}},
\bauthor{\bsnm{{Babusiaux}}, \binits{C.}},
\bauthor{\bsnm{{Bailer-Jones}}, \binits{C.A.L.}},
\bauthor{\bsnm{{Biermann}}, \binits{M.}},
\bauthor{\bsnm{{Evans}}, \binits{D.W.}},
\bauthor{\bsnm{{Eyer}}, \binits{L.}},
\bauthor{\bsnm{{Jansen}}, \binits{F.}},
\bauthor{\bsnm{{Jordi}}, \binits{C.}},
\bauthor{\bsnm{{Klioner}}, \binits{S.A.}},
\bauthor{\bsnm{{Lammers}}, \binits{U.}},
\bauthor{\bsnm{{Lindegren}}, \binits{L.}},
\bauthor{\bsnm{{Luri}}, \binits{X.}},
\bauthor{\bsnm{{Mignard}}, \binits{F.}},
\bauthor{\bsnm{{Panem}}, \binits{C.}},
\bauthor{\bsnm{{Pourbaix}}, \binits{D.}},
\bauthor{\bsnm{{Randich}}, \binits{S.}},
\bauthor{\bsnm{{Sartoretti}}, \binits{P.}},
\bauthor{\bsnm{{Siddiqui}}, \binits{H.I.}},
\bauthor{\bsnm{{Soubiran}}, \binits{C.}},
\bauthor{\bsnm{{van Leeuwen}}, \binits{F.}},
\bauthor{\bsnm{{Walton}}, \binits{N.A.}},
\bauthor{\bsnm{{Arenou}}, \binits{F.}},
\bauthor{\bsnm{{Bastian}}, \binits{U.}},
\bauthor{\bsnm{{Cropper}}, \binits{M.}},
\bauthor{\bsnm{{Drimmel}}, \binits{R.}},
\bauthor{\bsnm{{Katz}}, \binits{D.}},
\bauthor{\bsnm{{Lattanzi}}, \binits{M.G.}},
\bauthor{\bsnm{{Bakker}}, \binits{J.}},
\bauthor{\bsnm{{Cacciari}}, \binits{C.}},
\bauthor{\bsnm{{Casta{\~n}eda}}, \binits{J.}},
\bauthor{\bsnm{{Chaoul}}, \binits{L.}},
\bauthor{\bsnm{{Cheek}}, \binits{N.}},
\bauthor{\bsnm{{De Angeli}}, \binits{F.}},
\bauthor{\bsnm{{Fabricius}}, \binits{C.}},
\bauthor{\bsnm{{Guerra}}, \binits{R.}},
\bauthor{\bsnm{{Holl}}, \binits{B.}},
\bauthor{\bsnm{{Masana}}, \binits{E.}},
\bauthor{\bsnm{{Messineo}}, \binits{R.}},
\bauthor{\bsnm{{Mowlavi}}, \binits{N.}},
\bauthor{\bsnm{{Nienartowicz}}, \binits{K.}},
\bauthor{\bsnm{{Panuzzo}}, \binits{P.}},
\bauthor{\bsnm{{Portell}}, \binits{J.}},
\bauthor{\bsnm{{Riello}}, \binits{M.}},
\bauthor{\bsnm{{Seabroke}}, \binits{G.M.}},
\bauthor{\bsnm{{Tanga}}, \binits{P.}},
\bauthor{\bsnm{{Th{\'e}venin}}, \binits{F.}},
\bauthor{\bsnm{{Gracia-Abril}}, \binits{G.}},
\bauthor{\bsnm{{Comoretto}}, \binits{G.}},
\bauthor{\bsnm{{Garcia-Reinaldos}}, \binits{M.}},
\bauthor{\bsnm{{Teyssier}}, \binits{D.}},
\bauthor{\bsnm{{Altmann}}, \binits{M.}},
\bauthor{\bsnm{{Andrae}}, \binits{R.}},
\bauthor{\bsnm{{Audard}}, \binits{M.}},
\bauthor{\bsnm{{Bellas-Velidis}}, \binits{I.}},
\bauthor{\bsnm{{Benson}}, \binits{K.}},
\bauthor{\bsnm{{Berthier}}, \binits{J.}},
\bauthor{\bsnm{{Blomme}}, \binits{R.}},
\bauthor{\bsnm{{Burgess}}, \binits{P.}},
\bauthor{\bsnm{{Busso}}, \binits{G.}},
\bauthor{\bsnm{{Carry}}, \binits{B.}},
\bauthor{\bsnm{{Cellino}}, \binits{A.}},
\bauthor{\bsnm{{Clementini}}, \binits{G.}},
\bauthor{\bsnm{{Clotet}}, \binits{M.}},
\bauthor{\bsnm{{Creevey}}, \binits{O.}},
\bauthor{\bsnm{{Davidson}}, \binits{M.}},
\bauthor{\bsnm{{De Ridder}}, \binits{J.}},
\bauthor{\bsnm{{Delchambre}}, \binits{L.}},
\bauthor{\bsnm{{Dell'Oro}}, \binits{A.}},
\bauthor{\bsnm{{Ducourant}}, \binits{C.}},
\bauthor{\bsnm{{Fern{\'a}ndez-Hern{\'a}ndez}}, \binits{J.}},
\bauthor{\bsnm{{Fouesneau}}, \binits{M.}},
\bauthor{\bsnm{{Fr{\'e}mat}}, \binits{Y.}},
\bauthor{\bsnm{{Galluccio}}, \binits{L.}},
\bauthor{\bsnm{{Garc{\'\i}a-Torres}}, \binits{M.}},
\bauthor{\bsnm{{Gonz{\'a}lez-N{\'u}{\~n}ez}}, \binits{J.}},
\bauthor{\bsnm{{Gonz{\'a}lez-Vidal}}, \binits{J.J.}},
\bauthor{\bsnm{{Gosset}}, \binits{E.}},
\bauthor{\bsnm{{Guy}}, \binits{L.P.}},
\bauthor{\bsnm{{Halbwachs}}, \binits{J.-L.}},
\bauthor{\bsnm{{Hambly}}, \binits{N.C.}},
\bauthor{\bsnm{{Harrison}}, \binits{D.L.}},
\bauthor{\bsnm{{Hern{\'a}ndez}}, \binits{J.}},
\bauthor{\bsnm{{Hestroffer}}, \binits{D.}},
\bauthor{\bsnm{{Hodgkin}}, \binits{S.T.}},
\bauthor{\bsnm{{Hutton}}, \binits{A.}},
\bauthor{\bsnm{{Jasniewicz}}, \binits{G.}},
\bauthor{\bsnm{{Jean-Antoine-Piccolo}}, \binits{A.}},
\bauthor{\bsnm{{Jordan}}, \binits{S.}},
\bauthor{\bsnm{{Korn}}, \binits{A.J.}},
\bauthor{\bsnm{{Krone-Martins}}, \binits{A.}},
\bauthor{\bsnm{{Lanzafame}}, \binits{A.C.}},
\bauthor{\bsnm{{Lebzelter}}, \binits{T.}},
\bauthor{\bsnm{{L{\"o}ffler}}, \binits{W.}},
\bauthor{\bsnm{{Manteiga}}, \binits{M.}},
\bauthor{\bsnm{{Marrese}}, \binits{P.M.}},
\bauthor{\bsnm{{Mart{\'\i}n-Fleitas}}, \binits{J.M.}},
\bauthor{\bsnm{{Moitinho}}, \binits{A.}},
\bauthor{\bsnm{{Mora}}, \binits{A.}},
\bauthor{\bsnm{{Muinonen}}, \binits{K.}},
\bauthor{\bsnm{{Osinde}}, \binits{J.}},
\bauthor{\bsnm{{Pancino}}, \binits{E.}},
\bauthor{\bsnm{{Pauwels}}, \binits{T.}},
\bauthor{\bsnm{{Petit}}, \binits{J.-M.}},
\bauthor{\bsnm{{Recio-Blanco}}, \binits{A.}},
\bauthor{\bsnm{{Richards}}, \binits{P.J.}},
\bauthor{\bsnm{{Rimoldini}}, \binits{L.}},
\bauthor{\bsnm{{Robin}}, \binits{A.C.}},
\bauthor{\bsnm{{Sarro}}, \binits{L.M.}},
\bauthor{\bsnm{{Siopis}}, \binits{C.}},
\bauthor{\bsnm{{Smith}}, \binits{M.}},
\bauthor{\bsnm{{Sozzetti}}, \binits{A.}},
\bauthor{\bsnm{{S{\"u}veges}}, \binits{M.}},
\bauthor{\bsnm{{Torra}}, \binits{J.}},
\bauthor{\bsnm{{van Reeven}}, \binits{W.}},
\bauthor{\bsnm{{Abbas}}, \binits{U.}},
\bauthor{\bsnm{{Abreu Aramburu}}, \binits{A.}},
\bauthor{\bsnm{{Accart}}, \binits{S.}},
\bauthor{\bsnm{{Aerts}}, \binits{C.}},
\bauthor{\bsnm{{Altavilla}}, \binits{G.}},
\bauthor{\bsnm{{{\'A}lvarez}}, \binits{M.A.}},
\bauthor{\bsnm{{Alvarez}}, \binits{R.}},
\bauthor{\bsnm{{Alves}}, \binits{J.}},
\bauthor{\bsnm{{Anderson}}, \binits{R.I.}},
\bauthor{\bsnm{{Andrei}}, \binits{A.H.}},
\bauthor{\bsnm{{Anglada Varela}}, \binits{E.}},
\bauthor{\bsnm{{Antiche}}, \binits{E.}},
\bauthor{\bsnm{{Antoja}}, \binits{T.}},
\bauthor{\bsnm{{Arcay}}, \binits{B.}},
\bauthor{\bsnm{{Astraatmadja}}, \binits{T.L.}},
\bauthor{\bsnm{{Bach}}, \binits{N.}},
\bauthor{\bsnm{{Baker}}, \binits{S.G.}},
\bauthor{\bsnm{{Balaguer-N{\'u}{\~n}ez}}, \binits{L.}},
\bauthor{\bsnm{{Balm}}, \binits{P.}},
\bauthor{\bsnm{{Barache}}, \binits{C.}},
\bauthor{\bsnm{{Barata}}, \binits{C.}},
\bauthor{\bsnm{{Barbato}}, \binits{D.}},
\bauthor{\bsnm{{Barblan}}, \binits{F.}},
\bauthor{\bsnm{{Barklem}}, \binits{P.S.}},
\bauthor{\bsnm{{Barrado}}, \binits{D.}},
\bauthor{\bsnm{{Barros}}, \binits{M.}},
\bauthor{\bsnm{{Barstow}}, \binits{M.A.}},
\bauthor{\bsnm{{Bartholom{\'e} Mu{\~n}oz}}, \binits{S.}},
\bauthor{\bsnm{{Bassilana}}, \binits{J.-L.}},
\bauthor{\bsnm{{Becciani}}, \binits{U.}},
\bauthor{\bsnm{{Bellazzini}}, \binits{M.}},
\bauthor{\bsnm{{Berihuete}}, \binits{A.}},
\bauthor{\bsnm{{Bertone}}, \binits{S.}},
\bauthor{\bsnm{{Bianchi}}, \binits{L.}},
\bauthor{\bsnm{{Bienaym{\'e}}}, \binits{O.}},
\bauthor{\bsnm{{Blanco-Cuaresma}}, \binits{S.}},
\bauthor{\bsnm{{Boch}}, \binits{T.}},
\bauthor{\bsnm{{Boeche}}, \binits{C.}},
\bauthor{\bsnm{{Bombrun}}, \binits{A.}},
\bauthor{\bsnm{{Borrachero}}, \binits{R.}},
\bauthor{\bsnm{{Bossini}}, \binits{D.}},
\bauthor{\bsnm{{Bouquillon}}, \binits{S.}},
\bauthor{\bsnm{{Bourda}}, \binits{G.}},
\bauthor{\bsnm{{Bragaglia}}, \binits{A.}},
\bauthor{\bsnm{{Bramante}}, \binits{L.}},
\bauthor{\bsnm{{Breddels}}, \binits{M.A.}},
\bauthor{\bsnm{{Bressan}}, \binits{A.}},
\bauthor{\bsnm{{Brouillet}}, \binits{N.}},
\bauthor{\bsnm{{Br{\"u}semeister}}, \binits{T.}},
\bauthor{\bsnm{{Brugaletta}}, \binits{E.}},
\bauthor{\bsnm{{Bucciarelli}}, \binits{B.}},
\bauthor{\bsnm{{Burlacu}}, \binits{A.}},
\bauthor{\bsnm{{Busonero}}, \binits{D.}},
\bauthor{\bsnm{{Butkevich}}, \binits{A.G.}},
\bauthor{\bsnm{{Buzzi}}, \binits{R.}},
\bauthor{\bsnm{{Caffau}}, \binits{E.}},
\bauthor{\bsnm{{Cancelliere}}, \binits{R.}},
\bauthor{\bsnm{{Cannizzaro}}, \binits{G.}},
\bauthor{\bsnm{{Cantat-Gaudin}}, \binits{T.}},
\bauthor{\bsnm{{Carballo}}, \binits{R.}},
\bauthor{\bsnm{{Carlucci}}, \binits{T.}},
\bauthor{\bsnm{{Carrasco}}, \binits{J.M.}},
\bauthor{\bsnm{{Casamiquela}}, \binits{L.}},
\bauthor{\bsnm{{Castellani}}, \binits{M.}},
\bauthor{\bsnm{{Castro-Ginard}}, \binits{A.}},
\bauthor{\bsnm{{Charlot}}, \binits{P.}},
\bauthor{\bsnm{{Chemin}}, \binits{L.}},
\bauthor{\bsnm{{Chiavassa}}, \binits{A.}},
\bauthor{\bsnm{{Cocozza}}, \binits{G.}},
\bauthor{\bsnm{{Costigan}}, \binits{G.}},
\bauthor{\bsnm{{Cowell}}, \binits{S.}},
\bauthor{\bsnm{{Crifo}}, \binits{F.}},
\bauthor{\bsnm{{Crosta}}, \binits{M.}},
\bauthor{\bsnm{{Crowley}}, \binits{C.}},
\bauthor{\bsnm{{Cuypers}}, \binits{J.}},
\bauthor{\bsnm{{Dafonte}}, \binits{C.}},
\bauthor{\bsnm{{Damerdji}}, \binits{Y.}},
\bauthor{\bsnm{{Dapergolas}}, \binits{A.}},
\bauthor{\bsnm{{David}}, \binits{P.}},
\bauthor{\bsnm{{David}}, \binits{M.}},
\bauthor{\bsnm{{de Laverny}}, \binits{P.}},
\bauthor{\bsnm{{De Luise}}, \binits{F.}},
\bauthor{\bsnm{{De March}}, \binits{R.}},
\bauthor{\bsnm{{de Martino}}, \binits{D.}},
\bauthor{\bsnm{{de Souza}}, \binits{R.}},
\bauthor{\bsnm{{de Torres}}, \binits{A.}},
\bauthor{\bsnm{{Debosscher}}, \binits{J.}},
\bauthor{\bsnm{{del Pozo}}, \binits{E.}},
\bauthor{\bsnm{{Delbo}}, \binits{M.}},
\bauthor{\bsnm{{Delgado}}, \binits{A.}},
\bauthor{\bsnm{{Delgado}}, \binits{H.E.}},
\bauthor{\bsnm{{Di Matteo}}, \binits{P.}},
\bauthor{\bsnm{{Diakite}}, \binits{S.}},
\bauthor{\bsnm{{Diener}}, \binits{C.}},
\bauthor{\bsnm{{Distefano}}, \binits{E.}},
\bauthor{\bsnm{{Dolding}}, \binits{C.}},
\bauthor{\bsnm{{Drazinos}}, \binits{P.}},
\bauthor{\bsnm{{Dur{\'a}n}}, \binits{J.}},
\bauthor{\bsnm{{Edvardsson}}, \binits{B.}},
\bauthor{\bsnm{{Enke}}, \binits{H.}},
\bauthor{\bsnm{{Eriksson}}, \binits{K.}},
\bauthor{\bsnm{{Esquej}}, \binits{P.}},
\bauthor{\bsnm{{Eynard Bontemps}}, \binits{G.}},
\bauthor{\bsnm{{Fabre}}, \binits{C.}},
\bauthor{\bsnm{{Fabrizio}}, \binits{M.}},
\bauthor{\bsnm{{Faigler}}, \binits{S.}},
\bauthor{\bsnm{{Falc{\~a}o}}, \binits{A.J.}},
\bauthor{\bsnm{{Farr{\`a}s Casas}}, \binits{M.}},
\bauthor{\bsnm{{Federici}}, \binits{L.}},
\bauthor{\bsnm{{Fedorets}}, \binits{G.}},
\bauthor{\bsnm{{Fernique}}, \binits{P.}},
\bauthor{\bsnm{{Figueras}}, \binits{F.}},
\bauthor{\bsnm{{Filippi}}, \binits{F.}},
\bauthor{\bsnm{{Findeisen}}, \binits{K.}},
\bauthor{\bsnm{{Fonti}}, \binits{A.}},
\bauthor{\bsnm{{Fraile}}, \binits{E.}},
\bauthor{\bsnm{{Fraser}}, \binits{M.}},
\bauthor{\bsnm{{Fr{\'e}zouls}}, \binits{B.}},
\bauthor{\bsnm{{Gai}}, \binits{M.}},
\bauthor{\bsnm{{Galleti}}, \binits{S.}},
\bauthor{\bsnm{{Garabato}}, \binits{D.}},
\bauthor{\bsnm{{Garc{\'\i}a-Sedano}}, \binits{F.}},
\bauthor{\bsnm{{Garofalo}}, \binits{A.}},
\bauthor{\bsnm{{Garralda}}, \binits{N.}},
\bauthor{\bsnm{{Gavel}}, \binits{A.}},
\bauthor{\bsnm{{Gavras}}, \binits{P.}},
\bauthor{\bsnm{{Gerssen}}, \binits{J.}},
\bauthor{\bsnm{{Geyer}}, \binits{R.}},
\bauthor{\bsnm{{Giacobbe}}, \binits{P.}},
\bauthor{\bsnm{{Gilmore}}, \binits{G.}},
\bauthor{\bsnm{{Girona}}, \binits{S.}},
\bauthor{\bsnm{{Giuffrida}}, \binits{G.}},
\bauthor{\bsnm{{Glass}}, \binits{F.}},
\bauthor{\bsnm{{Gomes}}, \binits{M.}},
\bauthor{\bsnm{{Granvik}}, \binits{M.}},
\bauthor{\bsnm{{Gueguen}}, \binits{A.}},
\bauthor{\bsnm{{Guerrier}}, \binits{A.}},
\bauthor{\bsnm{{Guiraud}}, \binits{J.}},
\bauthor{\bsnm{{Guti{\'e}rrez-S{\'a}nchez}}, \binits{R.}},
\bauthor{\bsnm{{Haigron}}, \binits{R.}},
\bauthor{\bsnm{{Hatzidimitriou}}, \binits{D.}},
\bauthor{\bsnm{{Hauser}}, \binits{M.}},
\bauthor{\bsnm{{Haywood}}, \binits{M.}},
\bauthor{\bsnm{{Heiter}}, \binits{U.}},
\bauthor{\bsnm{{Helmi}}, \binits{A.}},
\bauthor{\bsnm{{Heu}}, \binits{J.}},
\bauthor{\bsnm{{Hilger}}, \binits{T.}},
\bauthor{\bsnm{{Hobbs}}, \binits{D.}},
\bauthor{\bsnm{{Hofmann}}, \binits{W.}},
\bauthor{\bsnm{{Holland}}, \binits{G.}},
\bauthor{\bsnm{{Huckle}}, \binits{H.E.}},
\bauthor{\bsnm{{Hypki}}, \binits{A.}},
\bauthor{\bsnm{{Icardi}}, \binits{V.}},
\bauthor{\bsnm{{Jan{\ss}en}}, \binits{K.}},
\bauthor{\bsnm{{Jevardat de Fombelle}}, \binits{G.}},
\bauthor{\bsnm{{Jonker}}, \binits{P.G.}},
\bauthor{\bsnm{{Juh{\'a}sz}}, \binits{{\'A}.L.}},
\bauthor{\bsnm{{Julbe}}, \binits{F.}},
\bauthor{\bsnm{{Karampelas}}, \binits{A.}},
\bauthor{\bsnm{{Kewley}}, \binits{A.}},
\bauthor{\bsnm{{Klar}}, \binits{J.}},
\bauthor{\bsnm{{Kochoska}}, \binits{A.}},
\bauthor{\bsnm{{Kohley}}, \binits{R.}},
\bauthor{\bsnm{{Kolenberg}}, \binits{K.}},
\bauthor{\bsnm{{Kontizas}}, \binits{M.}},
\bauthor{\bsnm{{Kontizas}}, \binits{E.}},
\bauthor{\bsnm{{Koposov}}, \binits{S.E.}},
\bauthor{\bsnm{{Kordopatis}}, \binits{G.}},
\bauthor{\bsnm{{Kostrzewa-Rutkowska}}, \binits{Z.}},
\bauthor{\bsnm{{Koubsky}}, \binits{P.}},
\bauthor{\bsnm{{Lambert}}, \binits{S.}},
\bauthor{\bsnm{{Lanza}}, \binits{A.F.}},
\bauthor{\bsnm{{Lasne}}, \binits{Y.}},
\bauthor{\bsnm{{Lavigne}}, \binits{J.-B.}},
\bauthor{\bsnm{{Le Fustec}}, \binits{Y.}},
\bauthor{\bsnm{{Le Poncin-Lafitte}}, \binits{C.}},
\bauthor{\bsnm{{Lebreton}}, \binits{Y.}},
\bauthor{\bsnm{{Leccia}}, \binits{S.}},
\bauthor{\bsnm{{Leclerc}}, \binits{N.}},
\bauthor{\bsnm{{Lecoeur-Taibi}}, \binits{I.}},
\bauthor{\bsnm{{Lenhardt}}, \binits{H.}},
\bauthor{\bsnm{{Leroux}}, \binits{F.}},
\bauthor{\bsnm{{Liao}}, \binits{S.}},
\bauthor{\bsnm{{Licata}}, \binits{E.}},
\bauthor{\bsnm{{Lindstr{\o}m}}, \binits{H.E.P.}},
\bauthor{\bsnm{{Lister}}, \binits{T.A.}},
\bauthor{\bsnm{{Livanou}}, \binits{E.}},
\bauthor{\bsnm{{Lobel}}, \binits{A.}},
\bauthor{\bsnm{{L{\'o}pez}}, \binits{M.}},
\bauthor{\bsnm{{Managau}}, \binits{S.}},
\bauthor{\bsnm{{Mann}}, \binits{R.G.}},
\bauthor{\bsnm{{Mantelet}}, \binits{G.}},
\bauthor{\bsnm{{Marchal}}, \binits{O.}},
\bauthor{\bsnm{{Marchant}}, \binits{J.M.}},
\bauthor{\bsnm{{Marconi}}, \binits{M.}},
\bauthor{\bsnm{{Marinoni}}, \binits{S.}},
\bauthor{\bsnm{{Marschalk{\'o}}}, \binits{G.}},
\bauthor{\bsnm{{Marshall}}, \binits{D.J.}},
\bauthor{\bsnm{{Martino}}, \binits{M.}},
\bauthor{\bsnm{{Marton}}, \binits{G.}},
\bauthor{\bsnm{{Mary}}, \binits{N.}},
\bauthor{\bsnm{{Massari}}, \binits{D.}},
\bauthor{\bsnm{{Matijevi{\v{c}}}}, \binits{G.}},
\bauthor{\bsnm{{Mazeh}}, \binits{T.}},
\bauthor{\bsnm{{McMillan}}, \binits{P.J.}},
\bauthor{\bsnm{{Messina}}, \binits{S.}},
\bauthor{\bsnm{{Michalik}}, \binits{D.}},
\bauthor{\bsnm{{Millar}}, \binits{N.R.}},
\bauthor{\bsnm{{Molina}}, \binits{D.}},
\bauthor{\bsnm{{Molinaro}}, \binits{R.}},
\bauthor{\bsnm{{Moln{\'a}r}}, \binits{L.}},
\bauthor{\bsnm{{Montegriffo}}, \binits{P.}},
\bauthor{\bsnm{{Mor}}, \binits{R.}},
\bauthor{\bsnm{{Morbidelli}}, \binits{R.}},
\bauthor{\bsnm{{Morel}}, \binits{T.}},
\bauthor{\bsnm{{Morris}}, \binits{D.}},
\bauthor{\bsnm{{Mulone}}, \binits{A.F.}},
\bauthor{\bsnm{{Muraveva}}, \binits{T.}},
\bauthor{\bsnm{{Musella}}, \binits{I.}},
\bauthor{\bsnm{{Nelemans}}, \binits{G.}},
\bauthor{\bsnm{{Nicastro}}, \binits{L.}},
\bauthor{\bsnm{{Noval}}, \binits{L.}},
\bauthor{\bsnm{{O'Mullane}}, \binits{W.}},
\bauthor{\bsnm{{Ord{\'e}novic}}, \binits{C.}},
\bauthor{\bsnm{{Ord{\'o}{\~n}ez-Blanco}}, \binits{D.}},
\bauthor{\bsnm{{Osborne}}, \binits{P.}},
\bauthor{\bsnm{{Pagani}}, \binits{C.}},
\bauthor{\bsnm{{Pagano}}, \binits{I.}},
\bauthor{\bsnm{{Pailler}}, \binits{F.}},
\bauthor{\bsnm{{Palacin}}, \binits{H.}},
\bauthor{\bsnm{{Palaversa}}, \binits{L.}},
\bauthor{\bsnm{{Panahi}}, \binits{A.}},
\bauthor{\bsnm{{Pawlak}}, \binits{M.}},
\bauthor{\bsnm{{Piersimoni}}, \binits{A.M.}},
\bauthor{\bsnm{{Pineau}}, \binits{F.-X.}},
\bauthor{\bsnm{{Plachy}}, \binits{E.}},
\bauthor{\bsnm{{Plum}}, \binits{G.}},
\bauthor{\bsnm{{Poggio}}, \binits{E.}},
\bauthor{\bsnm{{Poujoulet}}, \binits{E.}},
\bauthor{\bsnm{{Pr{\v{s}}a}}, \binits{A.}},
\bauthor{\bsnm{{Pulone}}, \binits{L.}},
\bauthor{\bsnm{{Racero}}, \binits{E.}},
\bauthor{\bsnm{{Ragaini}}, \binits{S.}},
\bauthor{\bsnm{{Rambaux}}, \binits{N.}},
\bauthor{\bsnm{{Ramos-Lerate}}, \binits{M.}},
\bauthor{\bsnm{{Regibo}}, \binits{S.}},
\bauthor{\bsnm{{Reyl{\'e}}}, \binits{C.}},
\bauthor{\bsnm{{Riclet}}, \binits{F.}},
\bauthor{\bsnm{{Ripepi}}, \binits{V.}},
\bauthor{\bsnm{{Riva}}, \binits{A.}},
\bauthor{\bsnm{{Rivard}}, \binits{A.}},
\bauthor{\bsnm{{Rixon}}, \binits{G.}},
\bauthor{\bsnm{{Roegiers}}, \binits{T.}},
\bauthor{\bsnm{{Roelens}}, \binits{M.}},
\bauthor{\bsnm{{Romero-G{\'o}mez}}, \binits{M.}},
\bauthor{\bsnm{{Rowell}}, \binits{N.}},
\bauthor{\bsnm{{Royer}}, \binits{F.}},
\bauthor{\bsnm{{Ruiz-Dern}}, \binits{L.}},
\bauthor{\bsnm{{Sadowski}}, \binits{G.}},
\bauthor{\bsnm{{Sagrist{\`a} Sell{\'e}s}}, \binits{T.}},
\bauthor{\bsnm{{Sahlmann}}, \binits{J.}},
\bauthor{\bsnm{{Salgado}}, \binits{J.}},
\bauthor{\bsnm{{Salguero}}, \binits{E.}},
\bauthor{\bsnm{{Sanna}}, \binits{N.}},
\bauthor{\bsnm{{Santana-Ros}}, \binits{T.}},
\bauthor{\bsnm{{Sarasso}}, \binits{M.}},
\bauthor{\bsnm{{Savietto}}, \binits{H.}},
\bauthor{\bsnm{{Schultheis}}, \binits{M.}},
\bauthor{\bsnm{{Sciacca}}, \binits{E.}},
\bauthor{\bsnm{{Segol}}, \binits{M.}},
\bauthor{\bsnm{{Segovia}}, \binits{J.C.}},
\bauthor{\bsnm{{S{\'e}gransan}}, \binits{D.}},
\bauthor{\bsnm{{Shih}}, \binits{I.-C.}},
\bauthor{\bsnm{{Siltala}}, \binits{L.}},
\bauthor{\bsnm{{Silva}}, \binits{A.F.}},
\bauthor{\bsnm{{Smart}}, \binits{R.L.}},
\bauthor{\bsnm{{Smith}}, \binits{K.W.}},
\bauthor{\bsnm{{Solano}}, \binits{E.}},
\bauthor{\bsnm{{Solitro}}, \binits{F.}},
\bauthor{\bsnm{{Sordo}}, \binits{R.}},
\bauthor{\bsnm{{Soria Nieto}}, \binits{S.}},
\bauthor{\bsnm{{Souchay}}, \binits{J.}},
\bauthor{\bsnm{{Spagna}}, \binits{A.}},
\bauthor{\bsnm{{Spoto}}, \binits{F.}},
\bauthor{\bsnm{{Stampa}}, \binits{U.}},
\bauthor{\bsnm{{Steele}}, \binits{I.A.}},
\bauthor{\bsnm{{Steidelm{\"u}ller}}, \binits{H.}},
\bauthor{\bsnm{{Stephenson}}, \binits{C.A.}},
\bauthor{\bsnm{{Stoev}}, \binits{H.}},
\bauthor{\bsnm{{Suess}}, \binits{F.F.}},
\bauthor{\bsnm{{Surdej}}, \binits{J.}},
\bauthor{\bsnm{{Szabados}}, \binits{L.}},
\bauthor{\bsnm{{Szegedi-Elek}}, \binits{E.}},
\bauthor{\bsnm{{Tapiador}}, \binits{D.}},
\bauthor{\bsnm{{Taris}}, \binits{F.}},
\bauthor{\bsnm{{Tauran}}, \binits{G.}},
\bauthor{\bsnm{{Taylor}}, \binits{M.B.}},
\bauthor{\bsnm{{Teixeira}}, \binits{R.}},
\bauthor{\bsnm{{Terrett}}, \binits{D.}},
\bauthor{\bsnm{{Teyssandier}}, \binits{P.}},
\bauthor{\bsnm{{Thuillot}}, \binits{W.}},
\bauthor{\bsnm{{Titarenko}}, \binits{A.}},
\bauthor{\bsnm{{Torra Clotet}}, \binits{F.}},
\bauthor{\bsnm{{Turon}}, \binits{C.}},
\bauthor{\bsnm{{Ulla}}, \binits{A.}},
\bauthor{\bsnm{{Utrilla}}, \binits{E.}},
\bauthor{\bsnm{{Uzzi}}, \binits{S.}},
\bauthor{\bsnm{{Vaillant}}, \binits{M.}},
\bauthor{\bsnm{{Valentini}}, \binits{G.}},
\bauthor{\bsnm{{Valette}}, \binits{V.}},
\bauthor{\bsnm{{van Elteren}}, \binits{A.}},
\bauthor{\bsnm{{Van Hemelryck}}, \binits{E.}},
\bauthor{\bsnm{{van Leeuwen}}, \binits{M.}},
\bauthor{\bsnm{{Vaschetto}}, \binits{M.}},
\bauthor{\bsnm{{Vecchiato}}, \binits{A.}},
\bauthor{\bsnm{{Veljanoski}}, \binits{J.}},
\bauthor{\bsnm{{Viala}}, \binits{Y.}},
\bauthor{\bsnm{{Vicente}}, \binits{D.}},
\bauthor{\bsnm{{Vogt}}, \binits{S.}},
\bauthor{\bsnm{{von Essen}}, \binits{C.}},
\bauthor{\bsnm{{Voss}}, \binits{H.}},
\bauthor{\bsnm{{Votruba}}, \binits{V.}},
\bauthor{\bsnm{{Voutsinas}}, \binits{S.}},
\bauthor{\bsnm{{Walmsley}}, \binits{G.}},
\bauthor{\bsnm{{Weiler}}, \binits{M.}},
\bauthor{\bsnm{{Wertz}}, \binits{O.}},
\bauthor{\bsnm{{Wevers}}, \binits{T.}},
\bauthor{\bsnm{{Wyrzykowski}}, \binits{{\L}.}},
\bauthor{\bsnm{{Yoldas}}, \binits{A.}},
\bauthor{\bsnm{{{\v{Z}}erjal}}, \binits{M.}},
\bauthor{\bsnm{{Ziaeepour}}, \binits{H.}},
\bauthor{\bsnm{{Zorec}}, \binits{J.}},
\bauthor{\bsnm{{Zschocke}}, \binits{S.}},
\bauthor{\bsnm{{Zucker}}, \binits{S.}},
\bauthor{\bsnm{{Zurbach}}, \binits{C.}},
\bauthor{\bsnm{{Zwitter}}, \binits{T.}}:
\batitle{{Gaia Data Release 2. Summary of the contents and survey properties}}.
\bjtitle{\aap}
\bvolume{616},
\bfpage{1}
(\byear{2018})
\doiurl{10.1051/0004-6361/201833051}
{\href{https://arxiv.org/abs/1804.09365}{{arXiv:1804.09365}}}
{[astro-ph.GA]}
\end{barticle}
\endbibitem

\bibitem[\protect\citeauthoryear{{Galli} et~al.}{2020}]{2020A&A...643A.148G}
\begin{barticle}
\bauthor{\bsnm{{Galli}}, \binits{P.A.B.}},
\bauthor{\bsnm{{Bouy}}, \binits{H.}},
\bauthor{\bsnm{{Olivares}}, \binits{J.}},
\bauthor{\bsnm{{Miret-Roig}}, \binits{N.}},
\bauthor{\bsnm{{Vieira}}, \binits{R.G.}},
\bauthor{\bsnm{{Sarro}}, \binits{L.M.}},
\bauthor{\bsnm{{Barrado}}, \binits{D.}},
\bauthor{\bsnm{{Berihuete}}, \binits{A.}},
\bauthor{\bsnm{{Bertout}}, \binits{C.}},
\bauthor{\bsnm{{Bertin}}, \binits{E.}},
\bauthor{\bsnm{{Cuillandre}}, \binits{J.-C.}}:
\batitle{{Lupus DANCe. Census of stars and 6D structure with Gaia-DR2 data}}.
\bjtitle{\aap}
\bvolume{643},
\bfpage{148}
(\byear{2020})
\doiurl{10.1051/0004-6361/202038717}
{\href{https://arxiv.org/abs/2010.00233}{{arXiv:2010.00233}}}
{[astro-ph.SR]}
\end{barticle}
\endbibitem

\bibitem[\protect\citeauthoryear{{Ansdell} et~al.}{2016}]{2016ApJ...828...46A}
\begin{barticle}
\bauthor{\bsnm{{Ansdell}}, \binits{M.}},
\bauthor{\bsnm{{Williams}}, \binits{J.P.}},
\bauthor{\bsnm{{van der Marel}}, \binits{N.}},
\bauthor{\bsnm{{Carpenter}}, \binits{J.M.}},
\bauthor{\bsnm{{Guidi}}, \binits{G.}},
\bauthor{\bsnm{{Hogerheijde}}, \binits{M.}},
\bauthor{\bsnm{{Mathews}}, \binits{G.S.}},
\bauthor{\bsnm{{Manara}}, \binits{C.F.}},
\bauthor{\bsnm{{Miotello}}, \binits{A.}},
\bauthor{\bsnm{{Natta}}, \binits{A.}},
\bauthor{\bsnm{{Oliveira}}, \binits{I.}},
\bauthor{\bsnm{{Tazzari}}, \binits{M.}},
\bauthor{\bsnm{{Testi}}, \binits{L.}},
\bauthor{\bsnm{{van Dishoeck}}, \binits{E.F.}},
\bauthor{\bsnm{{van Terwisga}}, \binits{S.E.}}:
\batitle{{ALMA Survey of Lupus Protoplanetary Disks. I. Dust and Gas Masses}}.
\bjtitle{\apj}
\bvolume{828}(\bissue{1}),
\bfpage{46}
(\byear{2016})
\doiurl{10.3847/0004-637X/828/1/46}
{\href{https://arxiv.org/abs/1604.05719}{{arXiv:1604.05719}}}
{[astro-ph.EP]}
\end{barticle}
\endbibitem

\bibitem[\protect\citeauthoryear{{Ansdell} et~al.}{2018}]{2018ApJ...859...21A}
\begin{barticle}
\bauthor{\bsnm{{Ansdell}}, \binits{M.}},
\bauthor{\bsnm{{Williams}}, \binits{J.P.}},
\bauthor{\bsnm{{Trapman}}, \binits{L.}},
\bauthor{\bsnm{{van Terwisga}}, \binits{S.E.}},
\bauthor{\bsnm{{Facchini}}, \binits{S.}},
\bauthor{\bsnm{{Manara}}, \binits{C.F.}},
\bauthor{\bsnm{{van der Marel}}, \binits{N.}},
\bauthor{\bsnm{{Miotello}}, \binits{A.}},
\bauthor{\bsnm{{Tazzari}}, \binits{M.}},
\bauthor{\bsnm{{Hogerheijde}}, \binits{M.}},
\bauthor{\bsnm{{Guidi}}, \binits{G.}},
\bauthor{\bsnm{{Testi}}, \binits{L.}},
\bauthor{\bsnm{{van Dishoeck}}, \binits{E.F.}}:
\batitle{{ALMA Survey of Lupus Protoplanetary Disks. II. Gas Disk Radii}}.
\bjtitle{\apj}
\bvolume{859}(\bissue{1}),
\bfpage{21}
(\byear{2018})
\doiurl{10.3847/1538-4357/aab890}
{\href{https://arxiv.org/abs/1803.05923}{{arXiv:1803.05923}}}
{[astro-ph.EP]}
\end{barticle}
\endbibitem

\bibitem[\protect\citeauthoryear{{Zurlo} et~al.}{2021}]{2021MNRAS.501.2305Z}
\begin{barticle}
\bauthor{\bsnm{{Zurlo}}, \binits{A.}},
\bauthor{\bsnm{{Cieza}}, \binits{L.A.}},
\bauthor{\bsnm{{Ansdell}}, \binits{M.}},
\bauthor{\bsnm{{Christiaens}}, \binits{V.}},
\bauthor{\bsnm{{P{\'e}rez}}, \binits{S.}},
\bauthor{\bsnm{{Lovell}}, \binits{J.}},
\bauthor{\bsnm{{Mesa}}, \binits{D.}},
\bauthor{\bsnm{{Williams}}, \binits{J.P.}},
\bauthor{\bsnm{{Gonzalez-Ruilova}}, \binits{C.}},
\bauthor{\bsnm{{Carraro}}, \binits{R.}},
\bauthor{\bsnm{{Ru{\'\i}z-Rodr{\'\i}guez}}, \binits{D.}},
\bauthor{\bsnm{{Wyatt}}, \binits{M.}}:
\batitle{{The effect of stellar multiplicity on protoplanetary discs: a
  near-infrared survey of the Lupus star-forming region}}.
\bjtitle{\mnras}
\bvolume{501}(\bissue{2}),
\bfpage{2305}--\blpage{2315}
(\byear{2021})
\doiurl{10.1093/mnras/staa3674}
{\href{https://arxiv.org/abs/2011.12297}{{arXiv:2011.12297}}}
{[astro-ph.EP]}
\end{barticle}
\endbibitem

\bibitem[\protect\citeauthoryear{{Galli} et~al.}{2019}]{2019A&A...630A.137G}
\begin{barticle}
\bauthor{\bsnm{{Galli}}, \binits{P.A.B.}},
\bauthor{\bsnm{{Loinard}}, \binits{L.}},
\bauthor{\bsnm{{Bouy}}, \binits{H.}},
\bauthor{\bsnm{{Sarro}}, \binits{L.M.}},
\bauthor{\bsnm{{Ortiz-Le{\'o}n}}, \binits{G.N.}},
\bauthor{\bsnm{{Dzib}}, \binits{S.A.}},
\bauthor{\bsnm{{Olivares}}, \binits{J.}},
\bauthor{\bsnm{{Heyer}}, \binits{M.}},
\bauthor{\bsnm{{Hernandez}}, \binits{J.}},
\bauthor{\bsnm{{Rom{\'a}n-Z{\'u}{\~n}iga}}, \binits{C.}},
\bauthor{\bsnm{{Kounkel}}, \binits{M.}},
\bauthor{\bsnm{{Covey}}, \binits{K.}}:
\batitle{{Structure and kinematics of the Taurus star-forming region from
  Gaia-DR2 and VLBI astrometry}}.
\bjtitle{\aap}
\bvolume{630},
\bfpage{137}
(\byear{2019})
\doiurl{10.1051/0004-6361/201935928}
{\href{https://arxiv.org/abs/1909.01118}{{arXiv:1909.01118}}}
{[astro-ph.SR]}
\end{barticle}
\endbibitem

\bibitem[\protect\citeauthoryear{{Luhman}}{2023}]{2023AJ....165...37L}
\begin{barticle}
\bauthor{\bsnm{{Luhman}}, \binits{K.L.}}:
\batitle{{A Census of the Taurus Star-forming Region and Neighboring
  Associations with Gaia}}.
\bjtitle{\aj}
\bvolume{165}(\bissue{2}),
\bfpage{37}
(\byear{2023})
\doiurl{10.3847/1538-3881/ac9da3}
{\href{https://arxiv.org/abs/2211.09785}{{arXiv:2211.09785}}}
{[astro-ph.GA]}
\end{barticle}
\endbibitem

\bibitem[\protect\citeauthoryear{{Long} et~al.}{2019}]{2019ApJ...882...49L}
\begin{barticle}
\bauthor{\bsnm{{Long}}, \binits{F.}},
\bauthor{\bsnm{{Herczeg}}, \binits{G.J.}},
\bauthor{\bsnm{{Harsono}}, \binits{D.}},
\bauthor{\bsnm{{Pinilla}}, \binits{P.}},
\bauthor{\bsnm{{Tazzari}}, \binits{M.}},
\bauthor{\bsnm{{Manara}}, \binits{C.F.}},
\bauthor{\bsnm{{Pascucci}}, \binits{I.}},
\bauthor{\bsnm{{Cabrit}}, \binits{S.}},
\bauthor{\bsnm{{Nisini}}, \binits{B.}},
\bauthor{\bsnm{{Johnstone}}, \binits{D.}},
\bauthor{\bsnm{{Edwards}}, \binits{S.}},
\bauthor{\bsnm{{Salyk}}, \binits{C.}},
\bauthor{\bsnm{{Menard}}, \binits{F.}},
\bauthor{\bsnm{{Lodato}}, \binits{G.}},
\bauthor{\bsnm{{Boehler}}, \binits{Y.}},
\bauthor{\bsnm{{Mace}}, \binits{G.N.}},
\bauthor{\bsnm{{Liu}}, \binits{Y.}},
\bauthor{\bsnm{{Mulders}}, \binits{G.D.}},
\bauthor{\bsnm{{Hendler}}, \binits{N.}},
\bauthor{\bsnm{{Ragusa}}, \binits{E.}},
\bauthor{\bsnm{{Fischer}}, \binits{W.J.}},
\bauthor{\bsnm{{Banzatti}}, \binits{A.}},
\bauthor{\bsnm{{Rigliaco}}, \binits{E.}},
\bauthor{\bsnm{{van de Plas}}, \binits{G.}},
\bauthor{\bsnm{{Dipierro}}, \binits{G.}},
\bauthor{\bsnm{{Gully-Santiago}}, \binits{M.}},
\bauthor{\bsnm{{Lopez-Valdivia}}, \binits{R.}}:
\batitle{{Compact Disks in a High-resolution ALMA Survey of Dust Structures in
  the Taurus Molecular Cloud}}.
\bjtitle{\apj}
\bvolume{882}(\bissue{1}),
\bfpage{49}
(\byear{2019})
\doiurl{10.3847/1538-4357/ab2d2d}
{\href{https://arxiv.org/abs/1906.10809}{{arXiv:1906.10809}}}
{[astro-ph.SR]}
\end{barticle}
\endbibitem

\bibitem[\protect\citeauthoryear{{Manara} et~al.}{2019}]{2019A&A...628A..95M}
\begin{barticle}
\bauthor{\bsnm{{Manara}}, \binits{C.F.}},
\bauthor{\bsnm{{Tazzari}}, \binits{M.}},
\bauthor{\bsnm{{Long}}, \binits{F.}},
\bauthor{\bsnm{{Herczeg}}, \binits{G.J.}},
\bauthor{\bsnm{{Lodato}}, \binits{G.}},
\bauthor{\bsnm{{Rota}}, \binits{A.A.}},
\bauthor{\bsnm{{Cazzoletti}}, \binits{P.}},
\bauthor{\bsnm{{van der Plas}}, \binits{G.}},
\bauthor{\bsnm{{Pinilla}}, \binits{P.}},
\bauthor{\bsnm{{Dipierro}}, \binits{G.}},
\bauthor{\bsnm{{Edwards}}, \binits{S.}},
\bauthor{\bsnm{{Harsono}}, \binits{D.}},
\bauthor{\bsnm{{Johnstone}}, \binits{D.}},
\bauthor{\bsnm{{Liu}}, \binits{Y.}},
\bauthor{\bsnm{{Menard}}, \binits{F.}},
\bauthor{\bsnm{{Nisini}}, \binits{B.}},
\bauthor{\bsnm{{Ragusa}}, \binits{E.}},
\bauthor{\bsnm{{Boehler}}, \binits{Y.}},
\bauthor{\bsnm{{Cabrit}}, \binits{S.}}:
\batitle{{Observational constraints on dust disk sizes in tidally truncated
  protoplanetary disks in multiple systems in the Taurus region}}.
\bjtitle{\aap}
\bvolume{628},
\bfpage{95}
(\byear{2019})
\doiurl{10.1051/0004-6361/201935964}
{\href{https://arxiv.org/abs/1907.03846}{{arXiv:1907.03846}}}
{[astro-ph.EP]}
\end{barticle}
\endbibitem

\bibitem[\protect\citeauthoryear{{Rota} et~al.}{2022}]{2022A&A...662A.121R}
\begin{barticle}
\bauthor{\bsnm{{Rota}}, \binits{A.A.}},
\bauthor{\bsnm{{Manara}}, \binits{C.F.}},
\bauthor{\bsnm{{Miotello}}, \binits{A.}},
\bauthor{\bsnm{{Lodato}}, \binits{G.}},
\bauthor{\bsnm{{Facchini}}, \binits{S.}},
\bauthor{\bsnm{{Koutoulaki}}, \binits{M.}},
\bauthor{\bsnm{{Herczeg}}, \binits{G.}},
\bauthor{\bsnm{{Long}}, \binits{F.}},
\bauthor{\bsnm{{Tazzari}}, \binits{M.}},
\bauthor{\bsnm{{Cabrit}}, \binits{S.}},
\bauthor{\bsnm{{Harsono}}, \binits{D.}},
\bauthor{\bsnm{{M{\'e}nard}}, \binits{F.}},
\bauthor{\bsnm{{Pinilla}}, \binits{P.}},
\bauthor{\bsnm{{van der Plas}}, \binits{G.}},
\bauthor{\bsnm{{Ragusa}}, \binits{E.}},
\bauthor{\bsnm{{Yen}}, \binits{H.-W.}}:
\batitle{{Observational constraints on gas disc sizes in the protoplanetary
  discs of multiple systems in the Taurus region}}.
\bjtitle{\aap}
\bvolume{662},
\bfpage{121}
(\byear{2022})
\doiurl{10.1051/0004-6361/202141035}
{\href{https://arxiv.org/abs/2201.03588}{{arXiv:2201.03588}}}
{[astro-ph.EP]}
\end{barticle}
\endbibitem

\bibitem[\protect\citeauthoryear{{Offner} et~al.}{2022}]{2022arXiv220310066O}
\begin{botherref}
\oauthor{\bsnm{{Offner}}, \binits{S.S.R.}},
\oauthor{\bsnm{{Moe}}, \binits{M.}},
\oauthor{\bsnm{{Kratter}}, \binits{K.M.}},
\oauthor{\bsnm{{Sadavoy}}, \binits{S.I.}},
\oauthor{\bsnm{{Jensen}}, \binits{E.L.N.}},
\oauthor{\bsnm{{Tobin}}, \binits{J.J.}}:
{The Origin and Evolution of Multiple Star Systems}.
arXiv e-prints,
2203--10066
(2022)
{\href{https://arxiv.org/abs/2203.10066}{{arXiv:2203.10066}}}
{[astro-ph.SR]}
\end{botherref}
\endbibitem

\bibitem[\protect\citeauthoryear{{Beuzit} et~al.}{2019}]{2019A&A...631A.155B}
\begin{barticle}
\bauthor{\bsnm{{Beuzit}}, \binits{J.-L.}},
\bauthor{\bsnm{{Vigan}}, \binits{A.}},
\bauthor{\bsnm{{Mouillet}}, \binits{D.}},
\bauthor{\bsnm{{Dohlen}}, \binits{K.}},
\bauthor{\bsnm{{Gratton}}, \binits{R.}},
\bauthor{\bsnm{{Boccaletti}}, \binits{A.}},
\bauthor{\bsnm{{Sauvage}}, \binits{J.-F.}},
\bauthor{\bsnm{{Schmid}}, \binits{H.M.}},
\bauthor{\bsnm{{Langlois}}, \binits{M.}},
\bauthor{\bsnm{{Petit}}, \binits{C.}},
\bauthor{\bsnm{{Baruffolo}}, \binits{A.}},
\bauthor{\bsnm{{Feldt}}, \binits{M.}},
\bauthor{\bsnm{{Milli}}, \binits{J.}},
\bauthor{\bsnm{{Wahhaj}}, \binits{Z.}},
\bauthor{\bsnm{{Abe}}, \binits{L.}},
\bauthor{\bsnm{{Anselmi}}, \binits{U.}},
\bauthor{\bsnm{{Antichi}}, \binits{J.}},
\bauthor{\bsnm{{Barette}}, \binits{R.}},
\bauthor{\bsnm{{Baudrand}}, \binits{J.}},
\bauthor{\bsnm{{Baudoz}}, \binits{P.}},
\bauthor{\bsnm{{Bazzon}}, \binits{A.}},
\bauthor{\bsnm{{Bernardi}}, \binits{P.}},
\bauthor{\bsnm{{Blanchard}}, \binits{P.}},
\bauthor{\bsnm{{Brast}}, \binits{R.}},
\bauthor{\bsnm{{Bruno}}, \binits{P.}},
\bauthor{\bsnm{{Buey}}, \binits{T.}},
\bauthor{\bsnm{{Carbillet}}, \binits{M.}},
\bauthor{\bsnm{{Carle}}, \binits{M.}},
\bauthor{\bsnm{{Cascone}}, \binits{E.}},
\bauthor{\bsnm{{Chapron}}, \binits{F.}},
\bauthor{\bsnm{{Charton}}, \binits{J.}},
\bauthor{\bsnm{{Chauvin}}, \binits{G.}},
\bauthor{\bsnm{{Claudi}}, \binits{R.}},
\bauthor{\bsnm{{Costille}}, \binits{A.}},
\bauthor{\bsnm{{De Caprio}}, \binits{V.}},
\bauthor{\bsnm{{de Boer}}, \binits{J.}},
\bauthor{\bsnm{{Delboulb{\'e}}}, \binits{A.}},
\bauthor{\bsnm{{Desidera}}, \binits{S.}},
\bauthor{\bsnm{{Dominik}}, \binits{C.}},
\bauthor{\bsnm{{Downing}}, \binits{M.}},
\bauthor{\bsnm{{Dupuis}}, \binits{O.}},
\bauthor{\bsnm{{Fabron}}, \binits{C.}},
\bauthor{\bsnm{{Fantinel}}, \binits{D.}},
\bauthor{\bsnm{{Farisato}}, \binits{G.}},
\bauthor{\bsnm{{Feautrier}}, \binits{P.}},
\bauthor{\bsnm{{Fedrigo}}, \binits{E.}},
\bauthor{\bsnm{{Fusco}}, \binits{T.}},
\bauthor{\bsnm{{Gigan}}, \binits{P.}},
\bauthor{\bsnm{{Ginski}}, \binits{C.}},
\bauthor{\bsnm{{Girard}}, \binits{J.}},
\bauthor{\bsnm{{Giro}}, \binits{E.}},
\bauthor{\bsnm{{Gisler}}, \binits{D.}},
\bauthor{\bsnm{{Gluck}}, \binits{L.}},
\bauthor{\bsnm{{Gry}}, \binits{C.}},
\bauthor{\bsnm{{Henning}}, \binits{T.}},
\bauthor{\bsnm{{Hubin}}, \binits{N.}},
\bauthor{\bsnm{{Hugot}}, \binits{E.}},
\bauthor{\bsnm{{Incorvaia}}, \binits{S.}},
\bauthor{\bsnm{{Jaquet}}, \binits{M.}},
\bauthor{\bsnm{{Kasper}}, \binits{M.}},
\bauthor{\bsnm{{Lagadec}}, \binits{E.}},
\bauthor{\bsnm{{Lagrange}}, \binits{A.-M.}},
\bauthor{\bsnm{{Le Coroller}}, \binits{H.}},
\bauthor{\bsnm{{Le Mignant}}, \binits{D.}},
\bauthor{\bsnm{{Le Ruyet}}, \binits{B.}},
\bauthor{\bsnm{{Lessio}}, \binits{G.}},
\bauthor{\bsnm{{Lizon}}, \binits{J.-L.}},
\bauthor{\bsnm{{Llored}}, \binits{M.}},
\bauthor{\bsnm{{Lundin}}, \binits{L.}},
\bauthor{\bsnm{{Madec}}, \binits{F.}},
\bauthor{\bsnm{{Magnard}}, \binits{Y.}},
\bauthor{\bsnm{{Marteaud}}, \binits{M.}},
\bauthor{\bsnm{{Martinez}}, \binits{P.}},
\bauthor{\bsnm{{Maurel}}, \binits{D.}},
\bauthor{\bsnm{{M{\'e}nard}}, \binits{F.}},
\bauthor{\bsnm{{Mesa}}, \binits{D.}},
\bauthor{\bsnm{{M{\"o}ller-Nilsson}}, \binits{O.}},
\bauthor{\bsnm{{Moulin}}, \binits{T.}},
\bauthor{\bsnm{{Moutou}}, \binits{C.}},
\bauthor{\bsnm{{Orign{\'e}}}, \binits{A.}},
\bauthor{\bsnm{{Parisot}}, \binits{J.}},
\bauthor{\bsnm{{Pavlov}}, \binits{A.}},
\bauthor{\bsnm{{Perret}}, \binits{D.}},
\bauthor{\bsnm{{Pragt}}, \binits{J.}},
\bauthor{\bsnm{{Puget}}, \binits{P.}},
\bauthor{\bsnm{{Rabou}}, \binits{P.}},
\bauthor{\bsnm{{Ramos}}, \binits{J.}},
\bauthor{\bsnm{{Reess}}, \binits{J.-M.}},
\bauthor{\bsnm{{Rigal}}, \binits{F.}},
\bauthor{\bsnm{{Rochat}}, \binits{S.}},
\bauthor{\bsnm{{Roelfsema}}, \binits{R.}},
\bauthor{\bsnm{{Rousset}}, \binits{G.}},
\bauthor{\bsnm{{Roux}}, \binits{A.}},
\bauthor{\bsnm{{Saisse}}, \binits{M.}},
\bauthor{\bsnm{{Salasnich}}, \binits{B.}},
\bauthor{\bsnm{{Santambrogio}}, \binits{E.}},
\bauthor{\bsnm{{Scuderi}}, \binits{S.}},
\bauthor{\bsnm{{Segransan}}, \binits{D.}},
\bauthor{\bsnm{{Sevin}}, \binits{A.}},
\bauthor{\bsnm{{Siebenmorgen}}, \binits{R.}},
\bauthor{\bsnm{{Soenke}}, \binits{C.}},
\bauthor{\bsnm{{Stadler}}, \binits{E.}},
\bauthor{\bsnm{{Suarez}}, \binits{M.}},
\bauthor{\bsnm{{Tiph{\`e}ne}}, \binits{D.}},
\bauthor{\bsnm{{Turatto}}, \binits{M.}},
\bauthor{\bsnm{{Udry}}, \binits{S.}},
\bauthor{\bsnm{{Vakili}}, \binits{F.}},
\bauthor{\bsnm{{Waters}}, \binits{L.B.F.M.}},
\bauthor{\bsnm{{Weber}}, \binits{L.}},
\bauthor{\bsnm{{Wildi}}, \binits{F.}},
\bauthor{\bsnm{{Zins}}, \binits{G.}},
\bauthor{\bsnm{{Zurlo}}, \binits{A.}}:
\batitle{{SPHERE: the exoplanet imager for the Very Large Telescope}}.
\bjtitle{\aap}
\bvolume{631},
\bfpage{155}
(\byear{2019})
\doiurl{10.1051/0004-6361/201935251}
{\href{https://arxiv.org/abs/1902.04080}{{arXiv:1902.04080}}}
{[astro-ph.IM]}
\end{barticle}
\endbibitem

\bibitem[\protect\citeauthoryear{{Macintosh}
  et~al.}{2014}]{2014PNAS..11112661M}
\begin{barticle}
\bauthor{\bsnm{{Macintosh}}, \binits{B.}},
\bauthor{\bsnm{{Graham}}, \binits{J.R.}},
\bauthor{\bsnm{{Ingraham}}, \binits{P.}},
\bauthor{\bsnm{{Konopacky}}, \binits{Q.}},
\bauthor{\bsnm{{Marois}}, \binits{C.}},
\bauthor{\bsnm{{Perrin}}, \binits{M.}},
\bauthor{\bsnm{{Poyneer}}, \binits{L.}},
\bauthor{\bsnm{{Bauman}}, \binits{B.}},
\bauthor{\bsnm{{Barman}}, \binits{T.}},
\bauthor{\bsnm{{Burrows}}, \binits{A.S.}},
\bauthor{\bsnm{{Cardwell}}, \binits{A.}},
\bauthor{\bsnm{{Chilcote}}, \binits{J.}},
\bauthor{\bsnm{{De Rosa}}, \binits{R.J.}},
\bauthor{\bsnm{{Dillon}}, \binits{D.}},
\bauthor{\bsnm{{Doyon}}, \binits{R.}},
\bauthor{\bsnm{{Dunn}}, \binits{J.}},
\bauthor{\bsnm{{Erikson}}, \binits{D.}},
\bauthor{\bsnm{{Fitzgerald}}, \binits{M.P.}},
\bauthor{\bsnm{{Gavel}}, \binits{D.}},
\bauthor{\bsnm{{Goodsell}}, \binits{S.}},
\bauthor{\bsnm{{Hartung}}, \binits{M.}},
\bauthor{\bsnm{{Hibon}}, \binits{P.}},
\bauthor{\bsnm{{Kalas}}, \binits{P.}},
\bauthor{\bsnm{{Larkin}}, \binits{J.}},
\bauthor{\bsnm{{Maire}}, \binits{J.}},
\bauthor{\bsnm{{Marchis}}, \binits{F.}},
\bauthor{\bsnm{{Marley}}, \binits{M.S.}},
\bauthor{\bsnm{{McBride}}, \binits{J.}},
\bauthor{\bsnm{{Millar-Blanchaer}}, \binits{M.}},
\bauthor{\bsnm{{Morzinski}}, \binits{K.}},
\bauthor{\bsnm{{Norton}}, \binits{A.}},
\bauthor{\bsnm{{Oppenheimer}}, \binits{B.R.}},
\bauthor{\bsnm{{Palmer}}, \binits{D.}},
\bauthor{\bsnm{{Patience}}, \binits{J.}},
\bauthor{\bsnm{{Pueyo}}, \binits{L.}},
\bauthor{\bsnm{{Rantakyro}}, \binits{F.}},
\bauthor{\bsnm{{Sadakuni}}, \binits{N.}},
\bauthor{\bsnm{{Saddlemyer}}, \binits{L.}},
\bauthor{\bsnm{{Savransky}}, \binits{D.}},
\bauthor{\bsnm{{Serio}}, \binits{A.}},
\bauthor{\bsnm{{Soummer}}, \binits{R.}},
\bauthor{\bsnm{{Sivaramakrishnan}}, \binits{A.}},
\bauthor{\bsnm{{Song}}, \binits{I.}},
\bauthor{\bsnm{{Thomas}}, \binits{S.}},
\bauthor{\bsnm{{Wallace}}, \binits{J.K.}},
\bauthor{\bsnm{{Wiktorowicz}}, \binits{S.}},
\bauthor{\bsnm{{Wolff}}, \binits{S.}}:
\batitle{{First light of the Gemini Planet Imager}}.
\bjtitle{Proceedings of the National Academy of Science}
\bvolume{111}(\bissue{35}),
\bfpage{12661}--\blpage{12666}
(\byear{2014})
\doiurl{10.1073/pnas.1304215111}
{\href{https://arxiv.org/abs/1403.7520}{{arXiv:1403.7520}}}
{[astro-ph.EP]}
\end{barticle}
\endbibitem

\bibitem[\protect\citeauthoryear{{Suzuki} et~al.}{2010}]{2010SPIE.7735E..30S}
\begin{bchapter}
\bauthor{\bsnm{{Suzuki}}, \binits{R.}},
\bauthor{\bsnm{{Kudo}}, \binits{T.}},
\bauthor{\bsnm{{Hashimoto}}, \binits{J.}},
\bauthor{\bsnm{{Carson}}, \binits{J.}},
\bauthor{\bsnm{{Egner}}, \binits{S.}},
\bauthor{\bsnm{{Goto}}, \binits{M.}},
\bauthor{\bsnm{{Hattori}}, \binits{M.}},
\bauthor{\bsnm{{Hayano}}, \binits{Y.}},
\bauthor{\bsnm{{Hodapp}}, \binits{K.}},
\bauthor{\bsnm{{Ito}}, \binits{M.}},
\bauthor{\bsnm{{Iye}}, \binits{M.}},
\bauthor{\bsnm{{Jacobson}}, \binits{S.}},
\bauthor{\bsnm{{Kandori}}, \binits{R.}},
\bauthor{\bsnm{{Kusakabe}}, \binits{N.}},
\bauthor{\bsnm{{Kuzuhara}}, \binits{M.}},
\bauthor{\bsnm{{Matsuo}}, \binits{T.}},
\bauthor{\bsnm{{Mcelwain}}, \binits{M.}},
\bauthor{\bsnm{{Morino}}, \binits{J.-I.}},
\bauthor{\bsnm{{Oya}}, \binits{S.}},
\bauthor{\bsnm{{Saito}}, \binits{Y.}},
\bauthor{\bsnm{{Shelton}}, \binits{R.}},
\bauthor{\bsnm{{Stahlberger}}, \binits{V.}},
\bauthor{\bsnm{{Suto}}, \binits{H.}},
\bauthor{\bsnm{{Takami}}, \binits{H.}},
\bauthor{\bsnm{{Thalmann}}, \binits{C.}},
\bauthor{\bsnm{{Watanabe}}, \binits{M.}},
\bauthor{\bsnm{{Yamada}}, \binits{H.}},
\bauthor{\bsnm{{Tamura}}, \binits{M.}}:
\bctitle{{Performance characterization of the HiCIAO instrument for the Subaru
  Telescope}}.
In: \beditor{\bsnm{{McLean}}, \binits{I.S.}},
\beditor{\bsnm{{Ramsay}}, \binits{S.K.}},
\beditor{\bsnm{{Takami}}, \binits{H.}} (eds.)
\bbtitle{Ground-based and Airborne Instrumentation for Astronomy III}.
\bsertitle{Society of Photo-Optical Instrumentation Engineers (SPIE) Conference
  Series},
vol. \bseriesno{7735},
p. \bfpage{773530}
(\byear{2010}).
\doiurl{10.1117/12.857361}
\end{bchapter}
\endbibitem

\bibitem[\protect\citeauthoryear{{Ginski} et~al.}{2020}]{2020A&A...642A.119G}
\begin{barticle}
\bauthor{\bsnm{{Ginski}}, \binits{C.}},
\bauthor{\bsnm{{M{\'e}nard}}, \binits{F.}},
\bauthor{\bsnm{{Rab}}, \binits{C.}},
\bauthor{\bsnm{{Mamajek}}, \binits{E.E.}},
\bauthor{\bsnm{{van Holstein}}, \binits{R.G.}},
\bauthor{\bsnm{{Benisty}}, \binits{M.}},
\bauthor{\bsnm{{Manara}}, \binits{C.F.}},
\bauthor{\bsnm{{Asensio Torres}}, \binits{R.}},
\bauthor{\bsnm{{Bohn}}, \binits{A.}},
\bauthor{\bsnm{{Birnstiel}}, \binits{T.}},
\bauthor{\bsnm{{Delorme}}, \binits{P.}},
\bauthor{\bsnm{{Facchini}}, \binits{S.}},
\bauthor{\bsnm{{Garufi}}, \binits{A.}},
\bauthor{\bsnm{{Gratton}}, \binits{R.}},
\bauthor{\bsnm{{Hogerheijde}}, \binits{M.}},
\bauthor{\bsnm{{Huang}}, \binits{J.}},
\bauthor{\bsnm{{Kenworthy}}, \binits{M.}},
\bauthor{\bsnm{{Langlois}}, \binits{M.}},
\bauthor{\bsnm{{Pinilla}}, \binits{P.}},
\bauthor{\bsnm{{Pinte}}, \binits{C.}},
\bauthor{\bsnm{{Ribas}}, \binits{{\'A}.}},
\bauthor{\bsnm{{Rosotti}}, \binits{G.}},
\bauthor{\bsnm{{Schmidt}}, \binits{T.O.B.}},
\bauthor{\bsnm{{van den Ancker}}, \binits{M.}},
\bauthor{\bsnm{{Wahhaj}}, \binits{Z.}},
\bauthor{\bsnm{{Waters}}, \binits{L.B.F.M.}},
\bauthor{\bsnm{{Williams}}, \binits{J.}},
\bauthor{\bsnm{{Zurlo}}, \binits{A.}}:
\batitle{{Disk Evolution Study Through Imaging of Nearby Young Stars
  (DESTINYS): A close low-mass companion to ET Cha}}.
\bjtitle{\aap}
\bvolume{642},
\bfpage{119}
(\byear{2020})
\doiurl{10.1051/0004-6361/202038383}
{\href{https://arxiv.org/abs/2007.05274}{{arXiv:2007.05274}}}
{[astro-ph.SR]}
\end{barticle}
\endbibitem

\bibitem[\protect\citeauthoryear{{Ginski} et~al.}{2021}]{2021ApJ...908L..25G}
\begin{barticle}
\bauthor{\bsnm{{Ginski}}, \binits{C.}},
\bauthor{\bsnm{{Facchini}}, \binits{S.}},
\bauthor{\bsnm{{Huang}}, \binits{J.}},
\bauthor{\bsnm{{Benisty}}, \binits{M.}},
\bauthor{\bsnm{{Vaendel}}, \binits{D.}},
\bauthor{\bsnm{{Stapper}}, \binits{L.}},
\bauthor{\bsnm{{Dominik}}, \binits{C.}},
\bauthor{\bsnm{{Bae}}, \binits{J.}},
\bauthor{\bsnm{{M{\'e}nard}}, \binits{F.}},
\bauthor{\bsnm{{Muro-Arena}}, \binits{G.}},
\bauthor{\bsnm{{Hogerheijde}}, \binits{M.R.}},
\bauthor{\bsnm{{McClure}}, \binits{M.}},
\bauthor{\bsnm{{van Holstein}}, \binits{R.G.}},
\bauthor{\bsnm{{Birnstiel}}, \binits{T.}},
\bauthor{\bsnm{{Boehler}}, \binits{Y.}},
\bauthor{\bsnm{{Bohn}}, \binits{A.}},
\bauthor{\bsnm{{Flock}}, \binits{M.}},
\bauthor{\bsnm{{Mamajek}}, \binits{E.E.}},
\bauthor{\bsnm{{Manara}}, \binits{C.F.}},
\bauthor{\bsnm{{Pinilla}}, \binits{P.}},
\bauthor{\bsnm{{Pinte}}, \binits{C.}},
\bauthor{\bsnm{{Ribas}}, \binits{{\'A}.}}:
\batitle{{Disk Evolution Study Through Imaging of Nearby Young Stars
  (DESTINYS): Late Infall Causing Disk Misalignment and Dynamic Structures in
  SU Aur}}.
\bjtitle{\apjl}
\bvolume{908}(\bissue{2}),
\bfpage{25}
(\byear{2021})
\doiurl{10.3847/2041-8213/abdf57}
{\href{https://arxiv.org/abs/2102.08781}{{arXiv:2102.08781}}}
{[astro-ph.EP]}
\end{barticle}
\endbibitem

\bibitem[\protect\citeauthoryear{{Zhang} et~al.}{2023}]{2023A&A...672A.145Z}
\begin{barticle}
\bauthor{\bsnm{{Zhang}}, \binits{Y.}},
\bauthor{\bsnm{{Ginski}}, \binits{C.}},
\bauthor{\bsnm{{Huang}}, \binits{J.}},
\bauthor{\bsnm{{Zurlo}}, \binits{A.}},
\bauthor{\bsnm{{Beust}}, \binits{H.}},
\bauthor{\bsnm{{Bae}}, \binits{J.}},
\bauthor{\bsnm{{Benisty}}, \binits{M.}},
\bauthor{\bsnm{{Garufi}}, \binits{A.}},
\bauthor{\bsnm{{Hogerheijde}}, \binits{M.R.}},
\bauthor{\bsnm{{van Holstein}}, \binits{R.G.}},
\bauthor{\bsnm{{Kenworthy}}, \binits{M.}},
\bauthor{\bsnm{{Langlois}}, \binits{M.}},
\bauthor{\bsnm{{Manara}}, \binits{C.F.}},
\bauthor{\bsnm{{Pinilla}}, \binits{P.}},
\bauthor{\bsnm{{Rab}}, \binits{C.}},
\bauthor{\bsnm{{Ribas}}, \binits{{\'A}.}},
\bauthor{\bsnm{{Rosotti}}, \binits{G.P.}},
\bauthor{\bsnm{{Williams}}, \binits{J.}}:
\batitle{{Disk Evolution Study Through Imaging of Nearby Young Stars
  (DESTINYS): Diverse outcomes of binary-disk interactions}}.
\bjtitle{\aap}
\bvolume{672},
\bfpage{145}
(\byear{2023})
\doiurl{10.1051/0004-6361/202245577}
{\href{https://arxiv.org/abs/2302.12824}{{arXiv:2302.12824}}}
{[astro-ph.SR]}
\end{barticle}
\endbibitem

\bibitem[\protect\citeauthoryear{{Daemgen} et~al.}{2012}]{2012A&A...540A..46D}
\begin{barticle}
\bauthor{\bsnm{{Daemgen}}, \binits{S.}},
\bauthor{\bsnm{{Correia}}, \binits{S.}},
\bauthor{\bsnm{{Petr-Gotzens}}, \binits{M.G.}}:
\batitle{{Protoplanetary disks of T Tauri binary systems in the Orion nebula
  cluster}}.
\bjtitle{\aap}
\bvolume{540},
\bfpage{46}
(\byear{2012})
\doiurl{10.1051/0004-6361/201118314}
{\href{https://arxiv.org/abs/1201.2421}{{arXiv:1201.2421}}}
{[astro-ph.SR]}
\end{barticle}
\endbibitem

\bibitem[\protect\citeauthoryear{{Gahm} et~al.}{2018}]{2018A&A...614A.117G}
\begin{barticle}
\bauthor{\bsnm{{Gahm}}, \binits{G.F.}},
\bauthor{\bsnm{{Petrov}}, \binits{P.P.}},
\bauthor{\bsnm{{Tambovsteva}}, \binits{L.V.}},
\bauthor{\bsnm{{Grinin}}, \binits{V.P.}},
\bauthor{\bsnm{{Stempels}}, \binits{H.C.}},
\bauthor{\bsnm{{Walter}}, \binits{F.M.}}:
\batitle{{S Coronae Australis: a T Tauri twin}}.
\bjtitle{\aap}
\bvolume{614},
\bfpage{117}
(\byear{2018})
\doiurl{10.1051/0004-6361/201832891}
{\href{https://arxiv.org/abs/1804.04971}{{arXiv:1804.04971}}}
{[astro-ph.SR]}
\end{barticle}
\endbibitem

\bibitem[\protect\citeauthoryear{{Chelli} et~al.}{1988}]{1988A&A...207...46C}
\begin{barticle}
\bauthor{\bsnm{{Chelli}}, \binits{A.}},
\bauthor{\bsnm{{Zinnecker}}, \binits{H.}},
\bauthor{\bsnm{{Carrasco}}, \binits{L.}},
\bauthor{\bsnm{{Cruz-Gonzalez}}, \binits{I.}},
\bauthor{\bsnm{{Perrier}}, \binits{C.}}:
\batitle{{Infrared companions to T Tauri stars.}}
\bjtitle{\aap}
\bvolume{207},
\bfpage{46}--\blpage{54}
(\byear{1988})
\end{barticle}
\endbibitem

\bibitem[\protect\citeauthoryear{{Daemgen} et~al.}{2013}]{2013A&A...554A..43D}
\begin{barticle}
\bauthor{\bsnm{{Daemgen}}, \binits{S.}},
\bauthor{\bsnm{{Petr-Gotzens}}, \binits{M.G.}},
\bauthor{\bsnm{{Correia}}, \binits{S.}},
\bauthor{\bsnm{{Teixeira}}, \binits{P.S.}},
\bauthor{\bsnm{{Brandner}}, \binits{W.}},
\bauthor{\bsnm{{Kley}}, \binits{W.}},
\bauthor{\bsnm{{Zinnecker}}, \binits{H.}}:
\batitle{{Protoplanetary disk evolution and stellar parameters of T Tauri
  binaries in Chamaeleon I}}.
\bjtitle{\aap}
\bvolume{554},
\bfpage{43}
(\byear{2013})
\doiurl{10.1051/0004-6361/201321220}
{\href{https://arxiv.org/abs/1304.1150}{{arXiv:1304.1150}}}
{[astro-ph.SR]}
\end{barticle}
\endbibitem

\bibitem[\protect\citeauthoryear{{Schmidt} et~al.}{2013}]{2013A&A...557A..80S}
\begin{barticle}
\bauthor{\bsnm{{Schmidt}}, \binits{T.O.B.}},
\bauthor{\bsnm{{Vogt}}, \binits{N.}},
\bauthor{\bsnm{{Neuh{\"a}user}}, \binits{R.}},
\bauthor{\bsnm{{Bedalov}}, \binits{A.}},
\bauthor{\bsnm{{Roell}}, \binits{T.}}:
\batitle{{New companions in the stellar systems of DI Cha, Sz 22, CHXR 32, and
  Cha H{\ensuremath{\alpha}} 5 in the Chamaeleon I star-forming region}}.
\bjtitle{\aap}
\bvolume{557},
\bfpage{80}
(\byear{2013})
\doiurl{10.1051/0004-6361/201220660}
{\href{https://arxiv.org/abs/1307.1082}{{arXiv:1307.1082}}}
{[astro-ph.SR]}
\end{barticle}
\endbibitem

\bibitem[\protect\citeauthoryear{{de Boer} et~al.}{2020}]{2020A&A...633A..63D}
\begin{barticle}
\bauthor{\bsnm{{de Boer}}, \binits{J.}},
\bauthor{\bsnm{{Langlois}}, \binits{M.}},
\bauthor{\bsnm{{van Holstein}}, \binits{R.G.}},
\bauthor{\bsnm{{Girard}}, \binits{J.H.}},
\bauthor{\bsnm{{Mouillet}}, \binits{D.}},
\bauthor{\bsnm{{Vigan}}, \binits{A.}},
\bauthor{\bsnm{{Dohlen}}, \binits{K.}},
\bauthor{\bsnm{{Snik}}, \binits{F.}},
\bauthor{\bsnm{{Keller}}, \binits{C.U.}},
\bauthor{\bsnm{{Ginski}}, \binits{C.}},
\bauthor{\bsnm{{Stam}}, \binits{D.M.}},
\bauthor{\bsnm{{Milli}}, \binits{J.}},
\bauthor{\bsnm{{Wahhaj}}, \binits{Z.}},
\bauthor{\bsnm{{Kasper}}, \binits{M.}},
\bauthor{\bsnm{{Schmid}}, \binits{H.M.}},
\bauthor{\bsnm{{Rabou}}, \binits{P.}},
\bauthor{\bsnm{{Gluck}}, \binits{L.}},
\bauthor{\bsnm{{Hugot}}, \binits{E.}},
\bauthor{\bsnm{{Perret}}, \binits{D.}},
\bauthor{\bsnm{{Martinez}}, \binits{P.}},
\bauthor{\bsnm{{Weber}}, \binits{L.}},
\bauthor{\bsnm{{Pragt}}, \binits{J.}},
\bauthor{\bsnm{{Sauvage}}, \binits{J.-F.}},
\bauthor{\bsnm{{Boccaletti}}, \binits{A.}},
\bauthor{\bsnm{{Le Coroller}}, \binits{H.}},
\bauthor{\bsnm{{Dominik}}, \binits{C.}},
\bauthor{\bsnm{{Henning}}, \binits{T.}},
\bauthor{\bsnm{{Lagadec}}, \binits{E.}},
\bauthor{\bsnm{{M{\'e}nard}}, \binits{F.}},
\bauthor{\bsnm{{Turatto}}, \binits{M.}},
\bauthor{\bsnm{{Udry}}, \binits{S.}},
\bauthor{\bsnm{{Chauvin}}, \binits{G.}},
\bauthor{\bsnm{{Feldt}}, \binits{M.}},
\bauthor{\bsnm{{Beuzit}}, \binits{J.-L.}}:
\batitle{{Polarimetric imaging mode of VLT/SPHERE/IRDIS. I. Description, data
  reduction, and observing strategy}}.
\bjtitle{\aap}
\bvolume{633},
\bfpage{63}
(\byear{2020})
\doiurl{10.1051/0004-6361/201834989}
{\href{https://arxiv.org/abs/1909.13107}{{arXiv:1909.13107}}}
{[astro-ph.IM]}
\end{barticle}
\endbibitem

\bibitem[\protect\citeauthoryear{{Dullemond}
  et~al.}{2019}]{2019A&A...628A..20D}
\begin{barticle}
\bauthor{\bsnm{{Dullemond}}, \binits{C.P.}},
\bauthor{\bsnm{{K{\"u}ffmeier}}, \binits{M.}},
\bauthor{\bsnm{{Goicovic}}, \binits{F.}},
\bauthor{\bsnm{{Fukagawa}}, \binits{M.}},
\bauthor{\bsnm{{Oehl}}, \binits{V.}},
\bauthor{\bsnm{{Kramer}}, \binits{M.}}:
\batitle{{Cloudlet capture by transitional disk and FU Orionis stars}}.
\bjtitle{\aap}
\bvolume{628},
\bfpage{20}
(\byear{2019})
\doiurl{10.1051/0004-6361/201832632}
{\href{https://arxiv.org/abs/1911.05158}{{arXiv:1911.05158}}}
{[astro-ph.EP]}
\end{barticle}
\endbibitem

\bibitem[\protect\citeauthoryear{{Francis} and {van der
  Marel}}{2020}]{2020ApJ...892..111F}
\begin{barticle}
\bauthor{\bsnm{{Francis}}, \binits{L.}},
\bauthor{\bsnm{{van der Marel}}, \binits{N.}}:
\batitle{{Dust-depleted Inner Disks in a Large Sample of Transition Disks
  through Long-baseline ALMA Observations}}.
\bjtitle{\apj}
\bvolume{892}(\bissue{2}),
\bfpage{111}
(\byear{2020})
\doiurl{10.3847/1538-4357/ab7b63}
{\href{https://arxiv.org/abs/2003.00079}{{arXiv:2003.00079}}}
{[astro-ph.EP]}
\end{barticle}
\endbibitem

\bibitem[\protect\citeauthoryear{{Mayama} et~al.}{2010}]{2010Sci...327..306M}
\begin{barticle}
\bauthor{\bsnm{{Mayama}}, \binits{S.}},
\bauthor{\bsnm{{Tamura}}, \binits{M.}},
\bauthor{\bsnm{{Hanawa}}, \binits{T.}},
\bauthor{\bsnm{{Matsumoto}}, \binits{T.}},
\bauthor{\bsnm{{Ishii}}, \binits{M.}},
\bauthor{\bsnm{{Pyo}}, \binits{T.-S.}},
\bauthor{\bsnm{{Suto}}, \binits{H.}},
\bauthor{\bsnm{{Naoi}}, \binits{T.}},
\bauthor{\bsnm{{Kudo}}, \binits{T.}},
\bauthor{\bsnm{{Hashimoto}}, \binits{J.}},
\bauthor{\bsnm{{Nishiyama}}, \binits{S.}},
\bauthor{\bsnm{{Kuzuhara}}, \binits{M.}},
\bauthor{\bsnm{{Hayashi}}, \binits{M.}}:
\batitle{{Direct Imaging of Bridged Twin Protoplanetary Disks in a Young
  Multiple Star}}.
\bjtitle{Science}
\bvolume{327}(\bissue{5963}),
\bfpage{306}
(\byear{2010})
\doiurl{10.1126/science.1179679}
\end{barticle}
\endbibitem

\bibitem[\protect\citeauthoryear{{Mayama} et~al.}{2020}]{2020AJ....159...12M}
\begin{barticle}
\bauthor{\bsnm{{Mayama}}, \binits{S.}},
\bauthor{\bsnm{{P{\'e}rez}}, \binits{S.}},
\bauthor{\bsnm{{Kusakabe}}, \binits{N.}},
\bauthor{\bsnm{{Muto}}, \binits{T.}},
\bauthor{\bsnm{{Tsukagoshi}}, \binits{T.}},
\bauthor{\bsnm{{Sitko}}, \binits{M.L.}},
\bauthor{\bsnm{{Takami}}, \binits{M.}},
\bauthor{\bsnm{{Hashimoto}}, \binits{J.}},
\bauthor{\bsnm{{Dong}}, \binits{R.}},
\bauthor{\bsnm{{Kwon}}, \binits{J.}},
\bauthor{\bsnm{{Hayashi}}, \binits{S.S.}},
\bauthor{\bsnm{{Kudo}}, \binits{T.}},
\bauthor{\bsnm{{Kuzuhara}}, \binits{M.}},
\bauthor{\bsnm{{Follette}}, \binits{K.}},
\bauthor{\bsnm{{Fukagawa}}, \binits{M.}},
\bauthor{\bsnm{{Momose}}, \binits{M.}},
\bauthor{\bsnm{{Oh}}, \binits{D.}},
\bauthor{\bsnm{{de Leon}}, \binits{J.}},
\bauthor{\bsnm{{Akiyama}}, \binits{E.}},
\bauthor{\bsnm{{Wisniewski}}, \binits{J.P.}},
\bauthor{\bsnm{{Yang}}, \binits{Y.}},
\bauthor{\bsnm{{Abe}}, \binits{L.}},
\bauthor{\bsnm{{Brandner}}, \binits{W.}},
\bauthor{\bsnm{{Brandt}}, \binits{T.D.}},
\bauthor{\bsnm{{Bonnefoy}}, \binits{M.}},
\bauthor{\bsnm{{Carson}}, \binits{J.C.}},
\bauthor{\bsnm{{Chilcote}}, \binits{J.}},
\bauthor{\bsnm{{Currie}}, \binits{T.}},
\bauthor{\bsnm{{Feldt}}, \binits{M.}},
\bauthor{\bsnm{{Goto}}, \binits{M.}},
\bauthor{\bsnm{{Grady}}, \binits{C.A.}},
\bauthor{\bsnm{{Groff}}, \binits{T.}},
\bauthor{\bsnm{{Guyon}}, \binits{O.}},
\bauthor{\bsnm{{Hayano}}, \binits{Y.}},
\bauthor{\bsnm{{Hayashi}}, \binits{M.}},
\bauthor{\bsnm{{Henning}}, \binits{T.}},
\bauthor{\bsnm{{Hodapp}}, \binits{K.W.}},
\bauthor{\bsnm{{Ishii}}, \binits{M.}},
\bauthor{\bsnm{{Iye}}, \binits{M.}},
\bauthor{\bsnm{{Janson}}, \binits{M.}},
\bauthor{\bsnm{{Jovanovic}}, \binits{N.}},
\bauthor{\bsnm{{Kandori}}, \binits{R.}},
\bauthor{\bsnm{{Kasdin}}, \binits{J.}},
\bauthor{\bsnm{{Knapp}}, \binits{G.R.}},
\bauthor{\bsnm{{Lozi}}, \binits{J.}},
\bauthor{\bsnm{{Martinache}}, \binits{F.}},
\bauthor{\bsnm{{Matsuo}}, \binits{T.}},
\bauthor{\bsnm{{McElwain}}, \binits{M.W.}},
\bauthor{\bsnm{{Miyama}}, \binits{S.}},
\bauthor{\bsnm{{Morino}}, \binits{J.-I.}},
\bauthor{\bsnm{{Moro-Martin}}, \binits{A.}},
\bauthor{\bsnm{{Nakagawa}}, \binits{T.}},
\bauthor{\bsnm{{Nishimura}}, \binits{T.}},
\bauthor{\bsnm{{Pyo}}, \binits{T.-S.}},
\bauthor{\bsnm{{Rich}}, \binits{E.A.}},
\bauthor{\bsnm{{Serabyn}}, \binits{E.}},
\bauthor{\bsnm{{Suto}}, \binits{H.}},
\bauthor{\bsnm{{Suzuki}}, \binits{R.}},
\bauthor{\bsnm{{Takato}}, \binits{N.}},
\bauthor{\bsnm{{Terada}}, \binits{H.}},
\bauthor{\bsnm{{Thalmann}}, \binits{C.}},
\bauthor{\bsnm{{Tomono}}, \binits{D.}},
\bauthor{\bsnm{{Turner}}, \binits{E.L.}},
\bauthor{\bsnm{{Watanabe}}, \binits{M.}},
\bauthor{\bsnm{{Yamada}}, \binits{T.}},
\bauthor{\bsnm{{Takami}}, \binits{H.}},
\bauthor{\bsnm{{Usuda}}, \binits{T.}},
\bauthor{\bsnm{{Uyama}}, \binits{T.}},
\bauthor{\bsnm{{Tamura}}, \binits{M.}}:
\batitle{{Subaru Near-infrared Imaging Polarimetry of Misaligned Disks around
  the SR 24 Hierarchical Triple System}}.
\bjtitle{\aj}
\bvolume{159}(\bissue{1}),
\bfpage{12}
(\byear{2020})
\doiurl{10.3847/1538-3881/ab5850}
{\href{https://arxiv.org/abs/1911.10941}{{arXiv:1911.10941}}}
{[astro-ph.SR]}
\end{barticle}
\endbibitem

\bibitem[\protect\citeauthoryear{{Weber} et~al.}{2022}]{2022MNRAS.tmp.3245W}
\begin{barticle}
\bauthor{\bsnm{{Weber}}, \binits{P.}},
\bauthor{\bsnm{{P{\'e}rez}}, \binits{S.}},
\bauthor{\bsnm{{Guidi}}, \binits{G.}},
\bauthor{\bsnm{{Kurtovic}}, \binits{N.T.}},
\bauthor{\bsnm{{Zurlo}}, \binits{A.}},
\bauthor{\bsnm{{Garufi}}, \binits{A.}},
\bauthor{\bsnm{{Pinilla}}, \binits{P.}},
\bauthor{\bsnm{{Mayama}}, \binits{S.}},
\bauthor{\bsnm{{van Holstein}}, \binits{R.G.}},
\bauthor{\bsnm{{Dullemond}}, \binits{C.P.}},
\bauthor{\bsnm{{Cuello}}, \binits{N.}},
\bauthor{\bsnm{{Principe}}, \binits{D.}},
\bauthor{\bsnm{{Cieza}}, \binits{L.}},
\bauthor{\bsnm{{Gonz{\'a}lez-Ruilova}}, \binits{C.}},
\bauthor{\bsnm{{Girard}}, \binits{J.}}:
\batitle{{The SPHERE view of three interacting twin disc systems in polarised
  light}}.
\bjtitle{\mnras}
(\byear{2022})
\doiurl{10.1093/mnras/stac3478}
{\href{https://arxiv.org/abs/2211.14322}{{arXiv:2211.14322}}}
{[astro-ph.SR]}
\end{barticle}
\endbibitem

\bibitem[\protect\citeauthoryear{{Weber} et~al.}{2023}]{2023MNRAS.518.5620W}
\begin{barticle}
\bauthor{\bsnm{{Weber}}, \binits{P.}},
\bauthor{\bsnm{{P{\'e}rez}}, \binits{S.}},
\bauthor{\bsnm{{Guidi}}, \binits{G.}},
\bauthor{\bsnm{{Kurtovic}}, \binits{N.T.}},
\bauthor{\bsnm{{Zurlo}}, \binits{A.}},
\bauthor{\bsnm{{Garufi}}, \binits{A.}},
\bauthor{\bsnm{{Pinilla}}, \binits{P.}},
\bauthor{\bsnm{{Mayama}}, \binits{S.}},
\bauthor{\bsnm{{van Holstein}}, \binits{R.G.}},
\bauthor{\bsnm{{Dullemond}}, \binits{C.P.}},
\bauthor{\bsnm{{Cuello}}, \binits{N.}},
\bauthor{\bsnm{{Principe}}, \binits{D.}},
\bauthor{\bsnm{{Cieza}}, \binits{L.}},
\bauthor{\bsnm{{Gonz{\'a}lez-Ruilova}}, \binits{C.}},
\bauthor{\bsnm{{Girard}}, \binits{J.}}:
\batitle{{The SPHERE view of three interacting twin disc systems in polarized
  light}}.
\bjtitle{\mnras}
\bvolume{518}(\bissue{4}),
\bfpage{5620}--\blpage{5642}
(\byear{2023})
\doiurl{10.1093/mnras/stac3478}
{\href{https://arxiv.org/abs/2211.14322}{{arXiv:2211.14322}}}
{[astro-ph.SR]}
\end{barticle}
\endbibitem

\bibitem[\protect\citeauthoryear{{Pineda} et~al.}{2015}]{2015Natur.518..213P}
\begin{barticle}
\bauthor{\bsnm{{Pineda}}, \binits{J.E.}},
\bauthor{\bsnm{{Offner}}, \binits{S.S.R.}},
\bauthor{\bsnm{{Parker}}, \binits{R.J.}},
\bauthor{\bsnm{{Arce}}, \binits{H.G.}},
\bauthor{\bsnm{{Goodman}}, \binits{A.A.}},
\bauthor{\bsnm{{Caselli}}, \binits{P.}},
\bauthor{\bsnm{{Fuller}}, \binits{G.A.}},
\bauthor{\bsnm{{Bourke}}, \binits{T.L.}},
\bauthor{\bsnm{{Corder}}, \binits{S.A.}}:
\batitle{{The formation of a quadruple star system with wide separation}}.
\bjtitle{\nat}
\bvolume{518}(\bissue{7538}),
\bfpage{213}--\blpage{215}
(\byear{2015})
\doiurl{10.1038/nature14166}
\end{barticle}
\endbibitem

\bibitem[\protect\citeauthoryear{{Kainulainen}
  et~al.}{2017}]{2017A&A...600A.141K}
\begin{barticle}
\bauthor{\bsnm{{Kainulainen}}, \binits{J.}},
\bauthor{\bsnm{{Stutz}}, \binits{A.M.}},
\bauthor{\bsnm{{Stanke}}, \binits{T.}},
\bauthor{\bsnm{{Abreu-Vicente}}, \binits{J.}},
\bauthor{\bsnm{{Beuther}}, \binits{H.}},
\bauthor{\bsnm{{Henning}}, \binits{T.}},
\bauthor{\bsnm{{Johnston}}, \binits{K.G.}},
\bauthor{\bsnm{{Megeath}}, \binits{S.T.}}:
\batitle{{Resolving the fragmentation of high line-mass filaments with ALMA:
  the integral shaped filament in Orion A}}.
\bjtitle{\aap}
\bvolume{600},
\bfpage{141}
(\byear{2017})
\doiurl{10.1051/0004-6361/201628481}
{\href{https://arxiv.org/abs/1603.05688}{{arXiv:1603.05688}}}
{[astro-ph.GA]}
\end{barticle}
\endbibitem

\bibitem[\protect\citeauthoryear{{Tobin} et~al.}{2016}]{2016Natur.538..483T}
\begin{barticle}
\bauthor{\bsnm{{Tobin}}, \binits{J.J.}},
\bauthor{\bsnm{{Kratter}}, \binits{K.M.}},
\bauthor{\bsnm{{Persson}}, \binits{M.V.}},
\bauthor{\bsnm{{Looney}}, \binits{L.W.}},
\bauthor{\bsnm{{Dunham}}, \binits{M.M.}},
\bauthor{\bsnm{{Segura-Cox}}, \binits{D.}},
\bauthor{\bsnm{{Li}}, \binits{Z.-Y.}},
\bauthor{\bsnm{{Chandler}}, \binits{C.J.}},
\bauthor{\bsnm{{Sadavoy}}, \binits{S.I.}},
\bauthor{\bsnm{{Harris}}, \binits{R.J.}},
\bauthor{\bsnm{{Melis}}, \binits{C.}},
\bauthor{\bsnm{{P{\'e}rez}}, \binits{L.M.}}:
\batitle{{A triple protostar system formed via fragmentation of a
  gravitationally unstable disk}}.
\bjtitle{\nat}
\bvolume{538}(\bissue{7626}),
\bfpage{483}--\blpage{486}
(\byear{2016})
\doiurl{10.1038/nature20094}
{\href{https://arxiv.org/abs/1610.08524}{{arXiv:1610.08524}}}
{[astro-ph.SR]}
\end{barticle}
\endbibitem

\bibitem[\protect\citeauthoryear{{Bate}}{2012}]{2012MNRAS.419.3115B}
\begin{barticle}
\bauthor{\bsnm{{Bate}}, \binits{M.R.}}:
\batitle{{Stellar, brown dwarf and multiple star properties from a radiation
  hydrodynamical simulation of star cluster formation}}.
\bjtitle{\mnras}
\bvolume{419}(\bissue{4}),
\bfpage{3115}--\blpage{3146}
(\byear{2012})
\doiurl{10.1111/j.1365-2966.2011.19955.x}
{\href{https://arxiv.org/abs/1110.1092}{{arXiv:1110.1092}}}
{[astro-ph.SR]}
\end{barticle}
\endbibitem

\bibitem[\protect\citeauthoryear{{Kratter} and
  {Lodato}}{2016}]{2016ARA&A..54..271K}
\begin{barticle}
\bauthor{\bsnm{{Kratter}}, \binits{K.}},
\bauthor{\bsnm{{Lodato}}, \binits{G.}}:
\batitle{{Gravitational Instabilities in Circumstellar Disks}}.
\bjtitle{\araa}
\bvolume{54},
\bfpage{271}--\blpage{311}
(\byear{2016})
\doiurl{10.1146/annurev-astro-081915-023307}
{\href{https://arxiv.org/abs/1603.01280}{{arXiv:1603.01280}}}
{[astro-ph.SR]}
\end{barticle}
\endbibitem

\bibitem[\protect\citeauthoryear{{Bate}}{2018}]{2018MNRAS.475.5618B}
\begin{barticle}
\bauthor{\bsnm{{Bate}}, \binits{M.R.}}:
\batitle{{On the diversity and statistical properties of protostellar discs}}.
\bjtitle{\mnras}
\bvolume{475}(\bissue{4}),
\bfpage{5618}--\blpage{5658}
(\byear{2018})
\doiurl{10.1093/mnras/sty169}
{\href{https://arxiv.org/abs/1801.07721}{{arXiv:1801.07721}}}
{[astro-ph.SR]}
\end{barticle}
\endbibitem

\bibitem[\protect\citeauthoryear{{Cuello} et~al.}{2023}]{2023EPJP..138...11C}
\begin{barticle}
\bauthor{\bsnm{{Cuello}}, \binits{N.}},
\bauthor{\bsnm{{M{\'e}nard}}, \binits{F.}},
\bauthor{\bsnm{{Price}}, \binits{D.J.}}:
\batitle{{Close encounters: How stellar flybys shape planet-forming discs}}.
\bjtitle{European Physical Journal Plus}
\bvolume{138}(\bissue{1}),
\bfpage{11}
(\byear{2023})
\doiurl{10.1140/epjp/s13360-022-03602-w}
{\href{https://arxiv.org/abs/2207.09752}{{arXiv:2207.09752}}}
{[astro-ph.EP]}
\end{barticle}
\endbibitem

\bibitem[\protect\citeauthoryear{{Facchini} et~al.}{2016}]{2016A&A...596A..38F}
\begin{barticle}
\bauthor{\bsnm{{Facchini}}, \binits{S.}},
\bauthor{\bsnm{{Manara}}, \binits{C.F.}},
\bauthor{\bsnm{{Schneider}}, \binits{P.C.}},
\bauthor{\bsnm{{Clarke}}, \binits{C.J.}},
\bauthor{\bsnm{{Bouvier}}, \binits{J.}},
\bauthor{\bsnm{{Rosotti}}, \binits{G.}},
\bauthor{\bsnm{{Booth}}, \binits{R.}},
\bauthor{\bsnm{{Haworth}}, \binits{T.J.}}:
\batitle{{Violent environment of the inner disk of RW Aurigae A probed by the
  2010 and 2015 dimming events}}.
\bjtitle{\aap}
\bvolume{596},
\bfpage{38}
(\byear{2016})
\doiurl{10.1051/0004-6361/201629607}
{\href{https://arxiv.org/abs/1610.08303}{{arXiv:1610.08303}}}
{[astro-ph.SR]}
\end{barticle}
\endbibitem

\bibitem[\protect\citeauthoryear{{Rodriguez}
  et~al.}{2018}]{2018ApJ...859..150R}
\begin{barticle}
\bauthor{\bsnm{{Rodriguez}}, \binits{J.E.}},
\bauthor{\bsnm{{Loomis}}, \binits{R.}},
\bauthor{\bsnm{{Cabrit}}, \binits{S.}},
\bauthor{\bsnm{{Haworth}}, \binits{T.J.}},
\bauthor{\bsnm{{Facchini}}, \binits{S.}},
\bauthor{\bsnm{{Dougados}}, \binits{C.}},
\bauthor{\bsnm{{Booth}}, \binits{R.A.}},
\bauthor{\bsnm{{Jensen}}, \binits{E.L.N.}},
\bauthor{\bsnm{{Clarke}}, \binits{C.J.}},
\bauthor{\bsnm{{Stassun}}, \binits{K.G.}},
\bauthor{\bsnm{{Dent}}, \binits{W.R.F.}},
\bauthor{\bsnm{{Pety}}, \binits{J.}}:
\batitle{{Multiple Stellar Flybys Sculpting the Circumstellar Architecture in
  RW Aurigae}}.
\bjtitle{\apj}
\bvolume{859}(\bissue{2}),
\bfpage{150}
(\byear{2018})
\doiurl{10.3847/1538-4357/aac08f}
{\href{https://arxiv.org/abs/1804.09190}{{arXiv:1804.09190}}}
{[astro-ph.SR]}
\end{barticle}
\endbibitem

\bibitem[\protect\citeauthoryear{{M{\'e}nard}
  et~al.}{2020}]{2020A&A...639L...1M}
\begin{barticle}
\bauthor{\bsnm{{M{\'e}nard}}, \binits{F.}},
\bauthor{\bsnm{{Cuello}}, \binits{N.}},
\bauthor{\bsnm{{Ginski}}, \binits{C.}},
\bauthor{\bsnm{{van der Plas}}, \binits{G.}},
\bauthor{\bsnm{{Villenave}}, \binits{M.}},
\bauthor{\bsnm{{Gonzalez}}, \binits{J.-F.}},
\bauthor{\bsnm{{Pinte}}, \binits{C.}},
\bauthor{\bsnm{{Benisty}}, \binits{M.}},
\bauthor{\bsnm{{Boccaletti}}, \binits{A.}},
\bauthor{\bsnm{{Price}}, \binits{D.J.}},
\bauthor{\bsnm{{Boehler}}, \binits{Y.}},
\bauthor{\bsnm{{Chripko}}, \binits{S.}},
\bauthor{\bsnm{{de Boer}}, \binits{J.}},
\bauthor{\bsnm{{Dominik}}, \binits{C.}},
\bauthor{\bsnm{{Garufi}}, \binits{A.}},
\bauthor{\bsnm{{Gratton}}, \binits{R.}},
\bauthor{\bsnm{{Hagelberg}}, \binits{J.}},
\bauthor{\bsnm{{Henning}}, \binits{T.}},
\bauthor{\bsnm{{Langlois}}, \binits{M.}},
\bauthor{\bsnm{{Maire}}, \binits{A.L.}},
\bauthor{\bsnm{{Pinilla}}, \binits{P.}},
\bauthor{\bsnm{{Ruane}}, \binits{G.J.}},
\bauthor{\bsnm{{Schmid}}, \binits{H.M.}},
\bauthor{\bsnm{{van Holstein}}, \binits{R.G.}},
\bauthor{\bsnm{{Vigan}}, \binits{A.}},
\bauthor{\bsnm{{Zurlo}}, \binits{A.}},
\bauthor{\bsnm{{Hubin}}, \binits{N.}},
\bauthor{\bsnm{{Pavlov}}, \binits{A.}},
\bauthor{\bsnm{{Rochat}}, \binits{S.}},
\bauthor{\bsnm{{Sauvage}}, \binits{J.-F.}},
\bauthor{\bsnm{{Stadler}}, \binits{E.}}:
\batitle{{Ongoing flyby in the young multiple system UX Tauri}}.
\bjtitle{\aap}
\bvolume{639},
\bfpage{1}
(\byear{2020})
\doiurl{10.1051/0004-6361/202038356}
{\href{https://arxiv.org/abs/2006.02439}{{arXiv:2006.02439}}}
{[astro-ph.SR]}
\end{barticle}
\endbibitem

\bibitem[\protect\citeauthoryear{{Gonz{\'a}lez-Ruilova}
  et~al.}{2020}]{2020ApJ...902L..33G}
\begin{barticle}
\bauthor{\bsnm{{Gonz{\'a}lez-Ruilova}}, \binits{C.}},
\bauthor{\bsnm{{Cieza}}, \binits{L.A.}},
\bauthor{\bsnm{{Hales}}, \binits{A.S.}},
\bauthor{\bsnm{{P{\'e}rez}}, \binits{S.}},
\bauthor{\bsnm{{Zurlo}}, \binits{A.}},
\bauthor{\bsnm{{Arce-Tord}}, \binits{C.}},
\bauthor{\bsnm{{Casassus}}, \binits{S.}},
\bauthor{\bsnm{{C{\'a}novas}}, \binits{H.}},
\bauthor{\bsnm{{Flock}}, \binits{M.}},
\bauthor{\bsnm{{Herczeg}}, \binits{G.J.}},
\bauthor{\bsnm{{Pinilla}}, \binits{P.}},
\bauthor{\bsnm{{Price}}, \binits{D.J.}},
\bauthor{\bsnm{{Principe}}, \binits{D.A.}},
\bauthor{\bsnm{{Ru{\'\i}z-Rodr{\'\i}guez}}, \binits{D.}},
\bauthor{\bsnm{{Williams}}, \binits{J.P.}}:
\batitle{{A Tale of Two Transition Disks: ALMA Long-baseline Observations of
  ISO-Oph 2 Reveal Two Closely Packed Nonaxisymmetric Rings and a
  {\ensuremath{\sim}}2 au Cavity}}.
\bjtitle{\apjl}
\bvolume{902}(\bissue{2}),
\bfpage{33}
(\byear{2020})
\doiurl{10.3847/2041-8213/abbcce}
{\href{https://arxiv.org/abs/2010.03650}{{arXiv:2010.03650}}}
{[astro-ph.EP]}
\end{barticle}
\endbibitem

\bibitem[\protect\citeauthoryear{{Salyk} et~al.}{2014}]{2014ApJ...792...68S}
\begin{barticle}
\bauthor{\bsnm{{Salyk}}, \binits{C.}},
\bauthor{\bsnm{{Pontoppidan}}, \binits{K.}},
\bauthor{\bsnm{{Corder}}, \binits{S.}},
\bauthor{\bsnm{{Mu{\~n}oz}}, \binits{D.}},
\bauthor{\bsnm{{Zhang}}, \binits{K.}},
\bauthor{\bsnm{{Blake}}, \binits{G.A.}}:
\batitle{{ALMA Observations of the T Tauri Binary System AS 205: Evidence for
  Molecular Winds and/or Binary Interactions}}.
\bjtitle{\apj}
\bvolume{792}(\bissue{1}),
\bfpage{68}
(\byear{2014})
\doiurl{10.1088/0004-637X/792/1/68}
{\href{https://arxiv.org/abs/1407.2652}{{arXiv:1407.2652}}}
{[astro-ph.SR]}
\end{barticle}
\endbibitem

\bibitem[\protect\citeauthoryear{{Kurtovic} et~al.}{2018}]{2018ApJ...869L..44K}
\begin{barticle}
\bauthor{\bsnm{{Kurtovic}}, \binits{N.T.}},
\bauthor{\bsnm{{P{\'e}rez}}, \binits{L.M.}},
\bauthor{\bsnm{{Benisty}}, \binits{M.}},
\bauthor{\bsnm{{Zhu}}, \binits{Z.}},
\bauthor{\bsnm{{Zhang}}, \binits{S.}},
\bauthor{\bsnm{{Huang}}, \binits{J.}},
\bauthor{\bsnm{{Andrews}}, \binits{S.M.}},
\bauthor{\bsnm{{Dullemond}}, \binits{C.P.}},
\bauthor{\bsnm{{Isella}}, \binits{A.}},
\bauthor{\bsnm{{Bai}}, \binits{X.-N.}},
\bauthor{\bsnm{{Carpenter}}, \binits{J.M.}},
\bauthor{\bsnm{{Guzm{\'a}n}}, \binits{V.V.}},
\bauthor{\bsnm{{Ricci}}, \binits{L.}},
\bauthor{\bsnm{{Wilner}}, \binits{D.J.}}:
\batitle{{The Disk Substructures at High Angular Resolution Project (DSHARP).
  IV. Characterizing Substructures and Interactions in Disks around Multiple
  Star Systems}}.
\bjtitle{\apjl}
\bvolume{869}(\bissue{2}),
\bfpage{44}
(\byear{2018})
\doiurl{10.3847/2041-8213/aaf746}
{\href{https://arxiv.org/abs/1812.04536}{{arXiv:1812.04536}}}
{[astro-ph.SR]}
\end{barticle}
\endbibitem

\bibitem[\protect\citeauthoryear{{Pinilla} et~al.}{2019}]{2019ApJ...878...16P}
\begin{barticle}
\bauthor{\bsnm{{Pinilla}}, \binits{P.}},
\bauthor{\bsnm{{Benisty}}, \binits{M.}},
\bauthor{\bsnm{{Cazzoletti}}, \binits{P.}},
\bauthor{\bsnm{{Harsono}}, \binits{D.}},
\bauthor{\bsnm{{P{\'e}rez}}, \binits{L.M.}},
\bauthor{\bsnm{{Tazzari}}, \binits{M.}}:
\batitle{{An Inner Disk in the Large Gap of the Transition Disk SR 24S}}.
\bjtitle{\apj}
\bvolume{878}(\bissue{1}),
\bfpage{16}
(\byear{2019})
\doiurl{10.3847/1538-4357/ab1cb8}
{\href{https://arxiv.org/abs/1904.11517}{{arXiv:1904.11517}}}
{[astro-ph.EP]}
\end{barticle}
\endbibitem

\bibitem[\protect\citeauthoryear{{Andrews} and {Williams}}{2005}]{Andrews2005}
\begin{barticle}
\bauthor{\bsnm{{Andrews}}, \binits{S.M.}},
\bauthor{\bsnm{{Williams}}, \binits{J.P.}}:
\batitle{{Submillimeter Array Observations of Disks in the SR 24 Multiple Star
  System}}.
\bjtitle{\apjl}
\bvolume{619}(\bissue{2}),
\bfpage{175}--\blpage{178}
(\byear{2005})
\doiurl{10.1086/427325}
{\href{https://arxiv.org/abs/astro-ph/0411131}{{arXiv:astro-ph/0411131}}}
{[astro-ph]}
\end{barticle}
\endbibitem

\bibitem[\protect\citeauthoryear{{Fern{\'a}ndez-L{\'o}pez}
  et~al.}{2017}]{2017ApJ...845...10F}
\begin{barticle}
\bauthor{\bsnm{{Fern{\'a}ndez-L{\'o}pez}}, \binits{M.}},
\bauthor{\bsnm{{Zapata}}, \binits{L.A.}},
\bauthor{\bsnm{{Gabbasov}}, \binits{R.}}:
\batitle{{Strongly Misaligned Triple System in SR 24 Revealed by ALMA}}.
\bjtitle{\apj}
\bvolume{845}(\bissue{1}),
\bfpage{10}
(\byear{2017})
\doiurl{10.3847/1538-4357/aa7d51}
{\href{https://arxiv.org/abs/1707.01128}{{arXiv:1707.01128}}}
{[astro-ph.SR]}
\end{barticle}
\endbibitem

\bibitem[\protect\citeauthoryear{{Alves} et~al.}{2019}]{2019Sci...366...90A}
\begin{barticle}
\bauthor{\bsnm{{Alves}}, \binits{F.O.}},
\bauthor{\bsnm{{Caselli}}, \binits{P.}},
\bauthor{\bsnm{{Girart}}, \binits{J.M.}},
\bauthor{\bsnm{{Segura-Cox}}, \binits{D.}},
\bauthor{\bsnm{{Franco}}, \binits{G.A.P.}},
\bauthor{\bsnm{{Schmiedeke}}, \binits{A.}},
\bauthor{\bsnm{{Zhao}}, \binits{B.}}:
\batitle{{Gas flow and accretion via spiral streamers and circumstellar disks
  in a young binary protostar}}.
\bjtitle{Science}
\bvolume{366}(\bissue{6461}),
\bfpage{90}--\blpage{93}
(\byear{2019})
\doiurl{10.1126/science.aaw3491}
{\href{https://arxiv.org/abs/1910.01141}{{arXiv:1910.01141}}}
{[astro-ph.SR]}
\end{barticle}
\endbibitem

\bibitem[\protect\citeauthoryear{{Audard} et~al.}{2014}]{2014prpl.conf..387A}
\begin{bchapter}
\bauthor{\bsnm{{Audard}}, \binits{M.}},
\bauthor{\bsnm{{{\'A}brah{\'a}m}}, \binits{P.}},
\bauthor{\bsnm{{Dunham}}, \binits{M.M.}},
\bauthor{\bsnm{{Green}}, \binits{J.D.}},
\bauthor{\bsnm{{Grosso}}, \binits{N.}},
\bauthor{\bsnm{{Hamaguchi}}, \binits{K.}},
\bauthor{\bsnm{{Kastner}}, \binits{J.H.}},
\bauthor{\bsnm{{K{\'o}sp{\'a}l}}, \binits{{\'A}.}},
\bauthor{\bsnm{{Lodato}}, \binits{G.}},
\bauthor{\bsnm{{Romanova}}, \binits{M.M.}},
\bauthor{\bsnm{{Skinner}}, \binits{S.L.}},
\bauthor{\bsnm{{Vorobyov}}, \binits{E.I.}},
\bauthor{\bsnm{{Zhu}}, \binits{Z.}}:
\bctitle{{Episodic Accretion in Young Stars}}.
In: \beditor{\bsnm{{Beuther}}, \binits{H.}},
\beditor{\bsnm{{Klessen}}, \binits{R.S.}},
\beditor{\bsnm{{Dullemond}}, \binits{C.P.}},
\beditor{\bsnm{{Henning}}, \binits{T.}} (eds.)
\bbtitle{Protostars and Planets VI},
p. \bfpage{387}
(\byear{2014}).
\doiurl{10.2458/azu_uapress_9780816531240-ch017}
\end{bchapter}
\endbibitem

\bibitem[\protect\citeauthoryear{{Wang} et~al.}{2004}]{2004ApJ...601L..83W}
\begin{barticle}
\bauthor{\bsnm{{Wang}}, \binits{H.}},
\bauthor{\bsnm{{Apai}}, \binits{D.}},
\bauthor{\bsnm{{Henning}}, \binits{T.}},
\bauthor{\bsnm{{Pascucci}}, \binits{I.}}:
\batitle{{FU Orionis: A Binary Star?}}
\bjtitle{\apjl}
\bvolume{601}(\bissue{1}),
\bfpage{83}--\blpage{86}
(\byear{2004})
\doiurl{10.1086/381705}
{\href{https://arxiv.org/abs/astro-ph/0311606}{{arXiv:astro-ph/0311606}}}
{[astro-ph]}
\end{barticle}
\endbibitem

\bibitem[\protect\citeauthoryear{{Hales} et~al.}{2015}]{2015ApJ...812..134H}
\begin{barticle}
\bauthor{\bsnm{{Hales}}, \binits{A.S.}},
\bauthor{\bsnm{{Corder}}, \binits{S.A.}},
\bauthor{\bsnm{{Dent}}, \binits{W.R.D.}},
\bauthor{\bsnm{{Andrews}}, \binits{S.M.}},
\bauthor{\bsnm{{Eisner}}, \binits{J.A.}},
\bauthor{\bsnm{{Cieza}}, \binits{L.A.}}:
\batitle{{The Early ALMA View of the FU Ori Outburst System}}.
\bjtitle{\apj}
\bvolume{812}(\bissue{2}),
\bfpage{134}
(\byear{2015})
\doiurl{10.1088/0004-637X/812/2/134}
{\href{https://arxiv.org/abs/1509.02543}{{arXiv:1509.02543}}}
{[astro-ph.SR]}
\end{barticle}
\endbibitem

\bibitem[\protect\citeauthoryear{{Liu} et~al.}{2017}]{2017A&A...602A..19L}
\begin{barticle}
\bauthor{\bsnm{{Liu}}, \binits{H.B.}},
\bauthor{\bsnm{{Vorobyov}}, \binits{E.I.}},
\bauthor{\bsnm{{Dong}}, \binits{R.}},
\bauthor{\bsnm{{Dunham}}, \binits{M.M.}},
\bauthor{\bsnm{{Takami}}, \binits{M.}},
\bauthor{\bsnm{{Galv{\'a}n-Madrid}}, \binits{R.}},
\bauthor{\bsnm{{Hashimoto}}, \binits{J.}},
\bauthor{\bsnm{{K{\'o}sp{\'a}l}}, \binits{{\'A}.}},
\bauthor{\bsnm{{Henning}}, \binits{T.}},
\bauthor{\bsnm{{Tamura}}, \binits{M.}},
\bauthor{\bsnm{{Rodr{\'\i}guez}}, \binits{L.F.}},
\bauthor{\bsnm{{Hirano}}, \binits{N.}},
\bauthor{\bsnm{{Hasegawa}}, \binits{Y.}},
\bauthor{\bsnm{{Fukagawa}}, \binits{M.}},
\bauthor{\bsnm{{Carrasco-Gonzalez}}, \binits{C.}},
\bauthor{\bsnm{{Tazzari}}, \binits{M.}}:
\batitle{{A concordant scenario to explain FU Orionis from deep centimeter and
  millimeter interferometric observations}}.
\bjtitle{\aap}
\bvolume{602},
\bfpage{19}
(\byear{2017})
\doiurl{10.1051/0004-6361/201630263}
{\href{https://arxiv.org/abs/1701.06531}{{arXiv:1701.06531}}}
{[astro-ph.SR]}
\end{barticle}
\endbibitem

\bibitem[\protect\citeauthoryear{{Liu} et~al.}{2019}]{2019ApJ...884...97L}
\begin{barticle}
\bauthor{\bsnm{{Liu}}, \binits{H.B.}},
\bauthor{\bsnm{{M{\'e}rand}}, \binits{A.}},
\bauthor{\bsnm{{Green}}, \binits{J.D.}},
\bauthor{\bsnm{{P{\'e}rez}}, \binits{S.}},
\bauthor{\bsnm{{Hales}}, \binits{A.S.}},
\bauthor{\bsnm{{Yang}}, \binits{Y.-L.}},
\bauthor{\bsnm{{Dunham}}, \binits{M.M.}},
\bauthor{\bsnm{{Hasegawa}}, \binits{Y.}},
\bauthor{\bsnm{{Henning}}, \binits{T.}},
\bauthor{\bsnm{{Galv{\'a}n-Madrid}}, \binits{R.}},
\bauthor{\bsnm{{K{\'o}sp{\'a}l}}, \binits{{\'A}.}},
\bauthor{\bsnm{{Takami}}, \binits{M.}},
\bauthor{\bsnm{{Vorobyov}}, \binits{E.I.}},
\bauthor{\bsnm{{Zhu}}, \binits{Z.}}:
\batitle{{Diagnosing 0.1-10 au Scale Morphology of the FU Ori Disk Using ALMA
  and VLTI/GRAVITY}}.
\bjtitle{\apj}
\bvolume{884}(\bissue{1}),
\bfpage{97}
(\byear{2019})
\doiurl{10.3847/1538-4357/ab391c}
{\href{https://arxiv.org/abs/1908.02981}{{arXiv:1908.02981}}}
{[astro-ph.SR]}
\end{barticle}
\endbibitem

\bibitem[\protect\citeauthoryear{{P{\'e}rez}
  et~al.}{2020}]{2020ApJ...889...59P}
\begin{barticle}
\bauthor{\bsnm{{P{\'e}rez}}, \binits{S.}},
\bauthor{\bsnm{{Hales}}, \binits{A.}},
\bauthor{\bsnm{{Liu}}, \binits{H.B.}},
\bauthor{\bsnm{{Zhu}}, \binits{Z.}},
\bauthor{\bsnm{{Casassus}}, \binits{S.}},
\bauthor{\bsnm{{Williams}}, \binits{J.}},
\bauthor{\bsnm{{Zurlo}}, \binits{A.}},
\bauthor{\bsnm{{Cuello}}, \binits{N.}},
\bauthor{\bsnm{{Cieza}}, \binits{L.}},
\bauthor{\bsnm{{Principe}}, \binits{D.}}:
\batitle{{Resolving the FU Orionis System with ALMA: Interacting Twin Disks?}}
\bjtitle{\apj}
\bvolume{889}(\bissue{1}),
\bfpage{59}
(\byear{2020})
\doiurl{10.3847/1538-4357/ab5c1b}
{\href{https://arxiv.org/abs/1911.11282}{{arXiv:1911.11282}}}
{[astro-ph.EP]}
\end{barticle}
\endbibitem

\bibitem[\protect\citeauthoryear{{Vorobyov} et~al.}{2021}]{2021A&A...647A..44V}
\begin{barticle}
\bauthor{\bsnm{{Vorobyov}}, \binits{E.I.}},
\bauthor{\bsnm{{Elbakyan}}, \binits{V.G.}},
\bauthor{\bsnm{{Liu}}, \binits{H.B.}},
\bauthor{\bsnm{{Takami}}, \binits{M.}}:
\batitle{{Distinguishing between different mechanisms of FU-Orionis-type
  luminosity outbursts}}.
\bjtitle{\aap}
\bvolume{647},
\bfpage{44}
(\byear{2021})
\doiurl{10.1051/0004-6361/202039391}
{\href{https://arxiv.org/abs/2101.01596}{{arXiv:2101.01596}}}
{[astro-ph.SR]}
\end{barticle}
\endbibitem

\bibitem[\protect\citeauthoryear{{Borchert} et~al.}{2022}]{2022MNRAS.517.4436B}
\begin{barticle}
\bauthor{\bsnm{{Borchert}}, \binits{E.M.A.}},
\bauthor{\bsnm{{Price}}, \binits{D.J.}},
\bauthor{\bsnm{{Pinte}}, \binits{C.}},
\bauthor{\bsnm{{Cuello}}, \binits{N.}}:
\batitle{{Sustained FU Orionis-type outbursts from colliding discs in stellar
  flybys}}.
\bjtitle{\mnras}
\bvolume{517}(\bissue{3}),
\bfpage{4436}--\blpage{4446}
(\byear{2022})
\doiurl{10.1093/mnras/stac2872}
{\href{https://arxiv.org/abs/2210.01143}{{arXiv:2210.01143}}}
{[astro-ph.SR]}
\end{barticle}
\endbibitem

\bibitem[\protect\citeauthoryear{{Canovas} et~al.}{2015}]{2015A&A...578L...1C}
\begin{barticle}
\bauthor{\bsnm{{Canovas}}, \binits{H.}},
\bauthor{\bsnm{{Perez}}, \binits{S.}},
\bauthor{\bsnm{{Dougados}}, \binits{C.}},
\bauthor{\bsnm{{de Boer}}, \binits{J.}},
\bauthor{\bsnm{{M{\'e}nard}}, \binits{F.}},
\bauthor{\bsnm{{Casassus}}, \binits{S.}},
\bauthor{\bsnm{{Schreiber}}, \binits{M.R.}},
\bauthor{\bsnm{{Cieza}}, \binits{L.A.}},
\bauthor{\bsnm{{Caceres}}, \binits{C.}},
\bauthor{\bsnm{{Girard}}, \binits{J.H.}}:
\batitle{{The inner environment of Z Canis Majoris: High-contrast imaging
  polarimetry with NaCo}}.
\bjtitle{\aap}
\bvolume{578},
\bfpage{1}
(\byear{2015})
\doiurl{10.1051/0004-6361/201526134}
{\href{https://arxiv.org/abs/1504.05197}{{arXiv:1504.05197}}}
{[astro-ph.SR]}
\end{barticle}
\endbibitem

\bibitem[\protect\citeauthoryear{{Liu} et~al.}{2016}]{2016SciA....2E0875L}
\begin{barticle}
\bauthor{\bsnm{{Liu}}, \binits{H.B.}},
\bauthor{\bsnm{{Takami}}, \binits{M.}},
\bauthor{\bsnm{{Kudo}}, \binits{T.}},
\bauthor{\bsnm{{Hashimoto}}, \binits{J.}},
\bauthor{\bsnm{{Dong}}, \binits{R.}},
\bauthor{\bsnm{{Vorobyov}}, \binits{E.I.}},
\bauthor{\bsnm{{Pyo}}, \binits{T.-S.}},
\bauthor{\bsnm{{Fukagawa}}, \binits{M.}},
\bauthor{\bsnm{{Tamura}}, \binits{M.}},
\bauthor{\bsnm{{Henning}}, \binits{T.}},
\bauthor{\bsnm{{Dunham}}, \binits{M.M.}},
\bauthor{\bsnm{{Karr}}, \binits{J.L.}},
\bauthor{\bsnm{{Kusakabe}}, \binits{N.}},
\bauthor{\bsnm{{Tsuribe}}, \binits{T.}}:
\batitle{{Circumstellar disks of the most vigorously accreting young stars}}.
\bjtitle{Science Advances}
\bvolume{2}(\bissue{2}),
\bfpage{1500875}
(\byear{2016})
\doiurl{10.1126/sciadv.1500875}
{\href{https://arxiv.org/abs/1602.04068}{{arXiv:1602.04068}}}
{[astro-ph.SR]}
\end{barticle}
\endbibitem

\bibitem[\protect\citeauthoryear{{Dong} et~al.}{2022}]{2022NatAs...6..331D}
\begin{barticle}
\bauthor{\bsnm{{Dong}}, \binits{R.}},
\bauthor{\bsnm{{Liu}}, \binits{H.B.}},
\bauthor{\bsnm{{Cuello}}, \binits{N.}},
\bauthor{\bsnm{{Pinte}}, \binits{C.}},
\bauthor{\bsnm{{{\'A}brah{\'a}m}}, \binits{P.}},
\bauthor{\bsnm{{Vorobyov}}, \binits{E.}},
\bauthor{\bsnm{{Hashimoto}}, \binits{J.}},
\bauthor{\bsnm{{K{\'o}sp{\'a}l}}, \binits{{\'A}.}},
\bauthor{\bsnm{{Chiang}}, \binits{E.}},
\bauthor{\bsnm{{Takami}}, \binits{M.}},
\bauthor{\bsnm{{Chen}}, \binits{L.}},
\bauthor{\bsnm{{Dunham}}, \binits{M.}},
\bauthor{\bsnm{{Fukagawa}}, \binits{M.}},
\bauthor{\bsnm{{Green}}, \binits{J.}},
\bauthor{\bsnm{{Hasegawa}}, \binits{Y.}},
\bauthor{\bsnm{{Henning}}, \binits{T.}},
\bauthor{\bsnm{{Pavlyuchenkov}}, \binits{Y.}},
\bauthor{\bsnm{{Pyo}}, \binits{T.-S.}},
\bauthor{\bsnm{{Tamura}}, \binits{M.}}:
\batitle{{A likely flyby of binary protostar Z CMa caught in action}}.
\bjtitle{Nature Astronomy}
\bvolume{6},
\bfpage{331}--\blpage{338}
(\byear{2022})
\doiurl{10.1038/s41550-021-01558-y10.48550/arXiv.2201.05617}
{\href{https://arxiv.org/abs/2201.05617}{{arXiv:2201.05617}}}
{[astro-ph.SR]}
\end{barticle}
\endbibitem

\bibitem[\protect\citeauthoryear{{Avenhaus} et~al.}{2017}]{2017AJ....154...33A}
\begin{barticle}
\bauthor{\bsnm{{Avenhaus}}, \binits{H.}},
\bauthor{\bsnm{{Quanz}}, \binits{S.P.}},
\bauthor{\bsnm{{Schmid}}, \binits{H.M.}},
\bauthor{\bsnm{{Dominik}}, \binits{C.}},
\bauthor{\bsnm{{Stolker}}, \binits{T.}},
\bauthor{\bsnm{{Ginski}}, \binits{C.}},
\bauthor{\bsnm{{de Boer}}, \binits{J.}},
\bauthor{\bsnm{{Szul{\'a}gyi}}, \binits{J.}},
\bauthor{\bsnm{{Garufi}}, \binits{A.}},
\bauthor{\bsnm{{Zurlo}}, \binits{A.}},
\bauthor{\bsnm{{Hagelberg}}, \binits{J.}},
\bauthor{\bsnm{{Benisty}}, \binits{M.}},
\bauthor{\bsnm{{Henning}}, \binits{T.}},
\bauthor{\bsnm{{M{\'e}nard}}, \binits{F.}},
\bauthor{\bsnm{{Meyer}}, \binits{M.R.}},
\bauthor{\bsnm{{Baruffolo}}, \binits{A.}},
\bauthor{\bsnm{{Bazzon}}, \binits{A.}},
\bauthor{\bsnm{{Beuzit}}, \binits{J.L.}},
\bauthor{\bsnm{{Costille}}, \binits{A.}},
\bauthor{\bsnm{{Dohlen}}, \binits{K.}},
\bauthor{\bsnm{{Girard}}, \binits{J.H.}},
\bauthor{\bsnm{{Gisler}}, \binits{D.}},
\bauthor{\bsnm{{Kasper}}, \binits{M.}},
\bauthor{\bsnm{{Mouillet}}, \binits{D.}},
\bauthor{\bsnm{{Pragt}}, \binits{J.}},
\bauthor{\bsnm{{Roelfsema}}, \binits{R.}},
\bauthor{\bsnm{{Salasnich}}, \binits{B.}},
\bauthor{\bsnm{{Sauvage}}, \binits{J.-F.}}:
\batitle{{Exploring Dust around HD 142527 down to 0.025 (4 au) Using
  SPHERE/ZIMPOL}}.
\bjtitle{\aj}
\bvolume{154}(\bissue{1}),
\bfpage{33}
(\byear{2017})
\doiurl{10.3847/1538-3881/aa7560}
{\href{https://arxiv.org/abs/1705.09680}{{arXiv:1705.09680}}}
{[astro-ph.EP]}
\end{barticle}
\endbibitem

\bibitem[\protect\citeauthoryear{{Kraus} et~al.}{2020}]{2020Sci...369.1233K}
\begin{barticle}
\bauthor{\bsnm{{Kraus}}, \binits{S.}},
\bauthor{\bsnm{{Kreplin}}, \binits{A.}},
\bauthor{\bsnm{{Young}}, \binits{A.K.}},
\bauthor{\bsnm{{Bate}}, \binits{M.R.}},
\bauthor{\bsnm{{Monnier}}, \binits{J.D.}},
\bauthor{\bsnm{{Harries}}, \binits{T.J.}},
\bauthor{\bsnm{{Avenhaus}}, \binits{H.}},
\bauthor{\bsnm{{Kluska}}, \binits{J.}},
\bauthor{\bsnm{{Laws}}, \binits{A.S.E.}},
\bauthor{\bsnm{{Rich}}, \binits{E.A.}},
\bauthor{\bsnm{{Willson}}, \binits{M.}},
\bauthor{\bsnm{{Aarnio}}, \binits{A.N.}},
\bauthor{\bsnm{{Adams}}, \binits{F.C.}},
\bauthor{\bsnm{{Andrews}}, \binits{S.M.}},
\bauthor{\bsnm{{Anugu}}, \binits{N.}},
\bauthor{\bsnm{{Bae}}, \binits{J.}},
\bauthor{\bsnm{{ten Brummelaar}}, \binits{T.}},
\bauthor{\bsnm{{Calvet}}, \binits{N.}},
\bauthor{\bsnm{{Cur{\'e}}}, \binits{M.}},
\bauthor{\bsnm{{Davies}}, \binits{C.L.}},
\bauthor{\bsnm{{Ennis}}, \binits{J.}},
\bauthor{\bsnm{{Espaillat}}, \binits{C.}},
\bauthor{\bsnm{{Gardner}}, \binits{T.}},
\bauthor{\bsnm{{Hartmann}}, \binits{L.}},
\bauthor{\bsnm{{Hinkley}}, \binits{S.}},
\bauthor{\bsnm{{Labdon}}, \binits{A.}},
\bauthor{\bsnm{{Lanthermann}}, \binits{C.}},
\bauthor{\bsnm{{LeBouquin}}, \binits{J.-B.}},
\bauthor{\bsnm{{Schaefer}}, \binits{G.H.}},
\bauthor{\bsnm{{Setterholm}}, \binits{B.R.}},
\bauthor{\bsnm{{Wilner}}, \binits{D.}},
\bauthor{\bsnm{{Zhu}}, \binits{Z.}}:
\batitle{{A triple-star system with a misaligned and warped circumstellar disk
  shaped by disk tearing}}.
\bjtitle{Science}
\bvolume{369}(\bissue{6508}),
\bfpage{1233}--\blpage{1238}
(\byear{2020})
\doiurl{10.1126/science.aba4633}
{\href{https://arxiv.org/abs/2004.01204}{{arXiv:2004.01204}}}
{[astro-ph.SR]}
\end{barticle}
\endbibitem

\bibitem[\protect\citeauthoryear{{Kennedy} et~al.}{2019}]{2019NatAs...3..230K}
\begin{barticle}
\bauthor{\bsnm{{Kennedy}}, \binits{G.M.}},
\bauthor{\bsnm{{Matr{\`a}}}, \binits{L.}},
\bauthor{\bsnm{{Facchini}}, \binits{S.}},
\bauthor{\bsnm{{Milli}}, \binits{J.}},
\bauthor{\bsnm{{Pani{\'c}}}, \binits{O.}},
\bauthor{\bsnm{{Price}}, \binits{D.}},
\bauthor{\bsnm{{Wilner}}, \binits{D.J.}},
\bauthor{\bsnm{{Wyatt}}, \binits{M.C.}},
\bauthor{\bsnm{{Yelverton}}, \binits{B.M.}}:
\batitle{{A circumbinary protoplanetary disk in a polar configuration}}.
\bjtitle{Nature Astronomy}
\bvolume{3},
\bfpage{230}--\blpage{235}
(\byear{2019})
\doiurl{10.1038/s41550-018-0667-x}
\end{barticle}
\endbibitem

\bibitem[\protect\citeauthoryear{{Marino} et~al.}{2015}]{2015ApJ...798L..44M}
\begin{barticle}
\bauthor{\bsnm{{Marino}}, \binits{S.}},
\bauthor{\bsnm{{Perez}}, \binits{S.}},
\bauthor{\bsnm{{Casassus}}, \binits{S.}}:
\batitle{{Shadows Cast by a Warp in the HD 142527 Protoplanetary Disk}}.
\bjtitle{\apjl}
\bvolume{798}(\bissue{2}),
\bfpage{44}
(\byear{2015})
\doiurl{10.1088/2041-8205/798/2/L44}
{\href{https://arxiv.org/abs/1412.4632}{{arXiv:1412.4632}}}
{[astro-ph.EP]}
\end{barticle}
\endbibitem

\bibitem[\protect\citeauthoryear{{Casassus} et~al.}{2015}]{2015ApJ...812..126C}
\begin{barticle}
\bauthor{\bsnm{{Casassus}}, \binits{S.}},
\bauthor{\bsnm{{Wright}}, \binits{C.M.}},
\bauthor{\bsnm{{Marino}}, \binits{S.}},
\bauthor{\bsnm{{Maddison}}, \binits{S.T.}},
\bauthor{\bsnm{{Wootten}}, \binits{A.}},
\bauthor{\bsnm{{Roman}}, \binits{P.}},
\bauthor{\bsnm{{P{\'e}rez}}, \binits{S.}},
\bauthor{\bsnm{{Pinilla}}, \binits{P.}},
\bauthor{\bsnm{{Wyatt}}, \binits{M.}},
\bauthor{\bsnm{{Moral}}, \binits{V.}},
\bauthor{\bsnm{{M{\'e}nard}}, \binits{F.}},
\bauthor{\bsnm{{Christiaens}}, \binits{V.}},
\bauthor{\bsnm{{Cieza}}, \binits{L.}},
\bauthor{\bsnm{{van der Plas}}, \binits{G.}}:
\batitle{{A Compact Concentration of Large Grains in the HD 142527
  Protoplanetary Dust Trap}}.
\bjtitle{\apj}
\bvolume{812}(\bissue{2}),
\bfpage{126}
(\byear{2015})
\doiurl{10.1088/0004-637X/812/2/126}
{\href{https://arxiv.org/abs/1505.07743}{{arXiv:1505.07743}}}
{[astro-ph.SR]}
\end{barticle}
\endbibitem

\bibitem[\protect\citeauthoryear{{Price} et~al.}{2018}]{2018MNRAS.477.1270P}
\begin{barticle}
\bauthor{\bsnm{{Price}}, \binits{D.J.}},
\bauthor{\bsnm{{Cuello}}, \binits{N.}},
\bauthor{\bsnm{{Pinte}}, \binits{C.}},
\bauthor{\bsnm{{Mentiplay}}, \binits{D.}},
\bauthor{\bsnm{{Casassus}}, \binits{S.}},
\bauthor{\bsnm{{Christiaens}}, \binits{V.}},
\bauthor{\bsnm{{Kennedy}}, \binits{G.M.}},
\bauthor{\bsnm{{Cuadra}}, \binits{J.}},
\bauthor{\bsnm{{Sebastian Perez}}, \binits{M.}},
\bauthor{\bsnm{{Marino}}, \binits{S.}},
\bauthor{\bsnm{{Armitage}}, \binits{P.J.}},
\bauthor{\bsnm{{Zurlo}}, \binits{A.}},
\bauthor{\bsnm{{Juhasz}}, \binits{A.}},
\bauthor{\bsnm{{Ragusa}}, \binits{E.}},
\bauthor{\bsnm{{Laibe}}, \binits{G.}},
\bauthor{\bsnm{{Lodato}}, \binits{G.}}:
\batitle{{Circumbinary, not transitional: on the spiral arms, cavity, shadows,
  fast radial flows, streamers, and horseshoe in the HD 142527 disc}}.
\bjtitle{\mnras}
\bvolume{477}(\bissue{1}),
\bfpage{1270}--\blpage{1284}
(\byear{2018})
\doiurl{10.1093/mnras/sty647}
{\href{https://arxiv.org/abs/1803.02484}{{arXiv:1803.02484}}}
{[astro-ph.SR]}
\end{barticle}
\endbibitem

\bibitem[\protect\citeauthoryear{{Bi} et~al.}{2020}]{2020ApJ...895L..18B}
\begin{barticle}
\bauthor{\bsnm{{Bi}}, \binits{J.}},
\bauthor{\bsnm{{van der Marel}}, \binits{N.}},
\bauthor{\bsnm{{Dong}}, \binits{R.}},
\bauthor{\bsnm{{Muto}}, \binits{T.}},
\bauthor{\bsnm{{Martin}}, \binits{R.G.}},
\bauthor{\bsnm{{Smallwood}}, \binits{J.L.}},
\bauthor{\bsnm{{Hashimoto}}, \binits{J.}},
\bauthor{\bsnm{{Liu}}, \binits{H.B.}},
\bauthor{\bsnm{{Nomura}}, \binits{H.}},
\bauthor{\bsnm{{Hasegawa}}, \binits{Y.}},
\bauthor{\bsnm{{Takami}}, \binits{M.}},
\bauthor{\bsnm{{Konishi}}, \binits{M.}},
\bauthor{\bsnm{{Momose}}, \binits{M.}},
\bauthor{\bsnm{{Kanagawa}}, \binits{K.D.}},
\bauthor{\bsnm{{Kataoka}}, \binits{A.}},
\bauthor{\bsnm{{Ono}}, \binits{T.}},
\bauthor{\bsnm{{Sitko}}, \binits{M.L.}},
\bauthor{\bsnm{{Takahashi}}, \binits{S.Z.}},
\bauthor{\bsnm{{Tomida}}, \binits{K.}},
\bauthor{\bsnm{{Tsukagoshi}}, \binits{T.}}:
\batitle{{GW Ori: Interactions between a Triple-star System and Its
  Circumtriple Disk in Action}}.
\bjtitle{\apjl}
\bvolume{895}(\bissue{1}),
\bfpage{18}
(\byear{2020})
\doiurl{10.3847/2041-8213/ab8eb4}
{\href{https://arxiv.org/abs/2004.03135}{{arXiv:2004.03135}}}
{[astro-ph.SR]}
\end{barticle}
\endbibitem

\bibitem[\protect\citeauthoryear{{Beck} et~al.}{2020}]{2020ApJ...902..132B}
\begin{barticle}
\bauthor{\bsnm{{Beck}}, \binits{T.L.}},
\bauthor{\bsnm{{Schaefer}}, \binits{G.H.}},
\bauthor{\bsnm{{Guilloteau}}, \binits{S.}},
\bauthor{\bsnm{{Simon}}, \binits{M.}},
\bauthor{\bsnm{{Dutrey}}, \binits{A.}},
\bauthor{\bsnm{{Di Folco}}, \binits{E.}},
\bauthor{\bsnm{{Chapillon}}, \binits{E.}}:
\batitle{{On the Nature of the T Tauri Triple System}}.
\bjtitle{\apj}
\bvolume{902}(\bissue{2}),
\bfpage{132}
(\byear{2020})
\doiurl{10.3847/1538-4357/abb5f5}
{\href{https://arxiv.org/abs/2009.03861}{{arXiv:2009.03861}}}
{[astro-ph.SR]}
\end{barticle}
\endbibitem

\bibitem[\protect\citeauthoryear{{Low} et~al.}{1984}]{Low1984}
\begin{barticle}
\bauthor{\bsnm{{Low}}, \binits{F.J.}},
\bauthor{\bsnm{{Beintema}}, \binits{D.A.}},
\bauthor{\bsnm{{Gautier}}, \binits{T.N.}},
\bauthor{\bsnm{{Gillett}}, \binits{F.C.}},
\bauthor{\bsnm{{Beichman}}, \binits{C.A.}},
\bauthor{\bsnm{{Neugebauer}}, \binits{G.}},
\bauthor{\bsnm{{Young}}, \binits{E.}},
\bauthor{\bsnm{{Aumann}}, \binits{H.H.}},
\bauthor{\bsnm{{Boggess}}, \binits{N.}},
\bauthor{\bsnm{{Emerson}}, \binits{J.P.}},
\bauthor{\bsnm{{Habing}}, \binits{H.J.}},
\bauthor{\bsnm{{Hauser}}, \binits{M.G.}},
\bauthor{\bsnm{{Houck}}, \binits{J.R.}},
\bauthor{\bsnm{{Rowan-Robinson}}, \binits{M.}},
\bauthor{\bsnm{{Soifer}}, \binits{B.T.}},
\bauthor{\bsnm{{Walker}}, \binits{R.G.}},
\bauthor{\bsnm{{Wesselius}}, \binits{P.R.}}:
\batitle{{Infrared cirrus: new components of the extended infrared emission.}}
\bjtitle{\apjl}
\bvolume{278},
\bfpage{19}--\blpage{22}
(\byear{1984})
\doiurl{10.1086/184213}
\end{barticle}
\endbibitem

\bibitem[\protect\citeauthoryear{{Miville-Desch{\^e}nes}
  et~al.}{2010}]{Miville2010}
\begin{barticle}
\bauthor{\bsnm{{Miville-Desch{\^e}nes}}, \binits{M.-A.}},
\bauthor{\bsnm{{Martin}}, \binits{P.G.}},
\bauthor{\bsnm{{Abergel}}, \binits{A.}},
\bauthor{\bsnm{{Bernard}}, \binits{J.-P.}},
\bauthor{\bsnm{{Boulanger}}, \binits{F.}},
\bauthor{\bsnm{{Lagache}}, \binits{G.}},
\bauthor{\bsnm{{Anderson}}, \binits{L.D.}},
\bauthor{\bsnm{{Andr{\'e}}}, \binits{P.}},
\bauthor{\bsnm{{Arab}}, \binits{H.}},
\bauthor{\bsnm{{Baluteau}}, \binits{J.-P.}},
\bauthor{\bsnm{{Blagrave}}, \binits{K.}},
\bauthor{\bsnm{{Bontemps}}, \binits{S.}},
\bauthor{\bsnm{{Cohen}}, \binits{M.}},
\bauthor{\bsnm{{Compiegne}}, \binits{M.}},
\bauthor{\bsnm{{Cox}}, \binits{P.}},
\bauthor{\bsnm{{Dartois}}, \binits{E.}},
\bauthor{\bsnm{{Davis}}, \binits{G.}},
\bauthor{\bsnm{{Emery}}, \binits{R.}},
\bauthor{\bsnm{{Fulton}}, \binits{T.}},
\bauthor{\bsnm{{Gry}}, \binits{C.}},
\bauthor{\bsnm{{Habart}}, \binits{E.}},
\bauthor{\bsnm{{Huang}}, \binits{M.}},
\bauthor{\bsnm{{Joblin}}, \binits{C.}},
\bauthor{\bsnm{{Jones}}, \binits{S.C.}},
\bauthor{\bsnm{{Kirk}}, \binits{J.}},
\bauthor{\bsnm{{Lim}}, \binits{T.}},
\bauthor{\bsnm{{Madden}}, \binits{S.}},
\bauthor{\bsnm{{Makiwa}}, \binits{G.}},
\bauthor{\bsnm{{Menshchikov}}, \binits{A.}},
\bauthor{\bsnm{{Molinari}}, \binits{S.}},
\bauthor{\bsnm{{Moseley}}, \binits{H.}},
\bauthor{\bsnm{{Motte}}, \binits{F.}},
\bauthor{\bsnm{{Naylor}}, \binits{D.A.}},
\bauthor{\bsnm{{Okumura}}, \binits{K.}},
\bauthor{\bsnm{{Pinheiro Gon{\c{c}}alves}}, \binits{D.}},
\bauthor{\bsnm{{Polehampton}}, \binits{E.}},
\bauthor{\bsnm{{Rod{\'o}n}}, \binits{J.A.}},
\bauthor{\bsnm{{Russeil}}, \binits{D.}},
\bauthor{\bsnm{{Saraceno}}, \binits{P.}},
\bauthor{\bsnm{{Schneider}}, \binits{N.}},
\bauthor{\bsnm{{Sidher}}, \binits{S.}},
\bauthor{\bsnm{{Spencer}}, \binits{L.}},
\bauthor{\bsnm{{Swinyard}}, \binits{B.}},
\bauthor{\bsnm{{Ward-Thompson}}, \binits{D.}},
\bauthor{\bsnm{{White}}, \binits{G.J.}},
\bauthor{\bsnm{{Zavagno}}, \binits{A.}}:
\batitle{{Herschel-SPIRE observations of the Polaris flare: Structure of the
  diffuse interstellar medium at the sub-parsec scale}}.
\bjtitle{\aap}
\bvolume{518},
\bfpage{104}
(\byear{2010})
\doiurl{10.1051/0004-6361/201014678}
{\href{https://arxiv.org/abs/1005.2746}{{arXiv:1005.2746}}}
{[astro-ph.GA]}
\end{barticle}
\endbibitem

\bibitem[\protect\citeauthoryear{{Ward-Thompson}
  et~al.}{2010}]{Ward-Thompson2010}
\begin{barticle}
\bauthor{\bsnm{{Ward-Thompson}}, \binits{D.}},
\bauthor{\bsnm{{Kirk}}, \binits{J.M.}},
\bauthor{\bsnm{{Andr{\'e}}}, \binits{P.}},
\bauthor{\bsnm{{Saraceno}}, \binits{P.}},
\bauthor{\bsnm{{Didelon}}, \binits{P.}},
\bauthor{\bsnm{{K{\"o}nyves}}, \binits{V.}},
\bauthor{\bsnm{{Schneider}}, \binits{N.}},
\bauthor{\bsnm{{Abergel}}, \binits{A.}},
\bauthor{\bsnm{{Baluteau}}, \binits{J.-P.}},
\bauthor{\bsnm{{Bernard}}, \binits{J.-P.}},
\bauthor{\bsnm{{Bontemps}}, \binits{S.}},
\bauthor{\bsnm{{Cambr{\'e}sy}}, \binits{L.}},
\bauthor{\bsnm{{Cox}}, \binits{P.}},
\bauthor{\bsnm{{di Francesco}}, \binits{J.}},
\bauthor{\bsnm{{di Giorgio}}, \binits{A.M.}},
\bauthor{\bsnm{{Griffin}}, \binits{M.}},
\bauthor{\bsnm{{Hargrave}}, \binits{P.}},
\bauthor{\bsnm{{Huang}}, \binits{M.}},
\bauthor{\bsnm{{Li}}, \binits{J.Z.}},
\bauthor{\bsnm{{Martin}}, \binits{P.}},
\bauthor{\bsnm{{Men'shchikov}}, \binits{A.}},
\bauthor{\bsnm{{Minier}}, \binits{V.}},
\bauthor{\bsnm{{Molinari}}, \binits{S.}},
\bauthor{\bsnm{{Motte}}, \binits{F.}},
\bauthor{\bsnm{{Olofsson}}, \binits{G.}},
\bauthor{\bsnm{{Pezzuto}}, \binits{S.}},
\bauthor{\bsnm{{Russeil}}, \binits{D.}},
\bauthor{\bsnm{{Sauvage}}, \binits{M.}},
\bauthor{\bsnm{{Sibthorpe}}, \binits{B.}},
\bauthor{\bsnm{{Spinoglio}}, \binits{L.}},
\bauthor{\bsnm{{Testi}}, \binits{L.}},
\bauthor{\bsnm{{White}}, \binits{G.}},
\bauthor{\bsnm{{Wilson}}, \binits{C.}},
\bauthor{\bsnm{{Woodcraft}}, \binits{A.}},
\bauthor{\bsnm{{Zavagno}}, \binits{A.}}:
\batitle{{A Herschel study of the properties of starless cores in the Polaris
  Flare dark cloud region using PACS and SPIRE}}.
\bjtitle{\aap}
\bvolume{518},
\bfpage{92}
(\byear{2010})
\doiurl{10.1051/0004-6361/201014618}
{\href{https://arxiv.org/abs/1005.2519}{{arXiv:1005.2519}}}
{[astro-ph.GA]}
\end{barticle}
\endbibitem

\bibitem[\protect\citeauthoryear{{Andr{\'e}} et~al.}{2010}]{Andre2010}
\begin{barticle}
\bauthor{\bsnm{{Andr{\'e}}}, \binits{P.}},
\bauthor{\bsnm{{Men'shchikov}}, \binits{A.}},
\bauthor{\bsnm{{Bontemps}}, \binits{S.}},
\bauthor{\bsnm{{K{\"o}nyves}}, \binits{V.}},
\bauthor{\bsnm{{Motte}}, \binits{F.}},
\bauthor{\bsnm{{Schneider}}, \binits{N.}},
\bauthor{\bsnm{{Didelon}}, \binits{P.}},
\bauthor{\bsnm{{Minier}}, \binits{V.}},
\bauthor{\bsnm{{Saraceno}}, \binits{P.}},
\bauthor{\bsnm{{Ward-Thompson}}, \binits{D.}},
\bauthor{\bsnm{{di Francesco}}, \binits{J.}},
\bauthor{\bsnm{{White}}, \binits{G.}},
\bauthor{\bsnm{{Molinari}}, \binits{S.}},
\bauthor{\bsnm{{Testi}}, \binits{L.}},
\bauthor{\bsnm{{Abergel}}, \binits{A.}},
\bauthor{\bsnm{{Griffin}}, \binits{M.}},
\bauthor{\bsnm{{Henning}}, \binits{T.}},
\bauthor{\bsnm{{Royer}}, \binits{P.}},
\bauthor{\bsnm{{Mer{\'\i}n}}, \binits{B.}},
\bauthor{\bsnm{{Vavrek}}, \binits{R.}},
\bauthor{\bsnm{{Attard}}, \binits{M.}},
\bauthor{\bsnm{{Arzoumanian}}, \binits{D.}},
\bauthor{\bsnm{{Wilson}}, \binits{C.D.}},
\bauthor{\bsnm{{Ade}}, \binits{P.}},
\bauthor{\bsnm{{Aussel}}, \binits{H.}},
\bauthor{\bsnm{{Baluteau}}, \binits{J.-P.}},
\bauthor{\bsnm{{Benedettini}}, \binits{M.}},
\bauthor{\bsnm{{Bernard}}, \binits{J.-P.}},
\bauthor{\bsnm{{Blommaert}}, \binits{J.A.D.L.}},
\bauthor{\bsnm{{Cambr{\'e}sy}}, \binits{L.}},
\bauthor{\bsnm{{Cox}}, \binits{P.}},
\bauthor{\bsnm{{di Giorgio}}, \binits{A.}},
\bauthor{\bsnm{{Hargrave}}, \binits{P.}},
\bauthor{\bsnm{{Hennemann}}, \binits{M.}},
\bauthor{\bsnm{{Huang}}, \binits{M.}},
\bauthor{\bsnm{{Kirk}}, \binits{J.}},
\bauthor{\bsnm{{Krause}}, \binits{O.}},
\bauthor{\bsnm{{Launhardt}}, \binits{R.}},
\bauthor{\bsnm{{Leeks}}, \binits{S.}},
\bauthor{\bsnm{{Le Pennec}}, \binits{J.}},
\bauthor{\bsnm{{Li}}, \binits{J.Z.}},
\bauthor{\bsnm{{Martin}}, \binits{P.G.}},
\bauthor{\bsnm{{Maury}}, \binits{A.}},
\bauthor{\bsnm{{Olofsson}}, \binits{G.}},
\bauthor{\bsnm{{Omont}}, \binits{A.}},
\bauthor{\bsnm{{Peretto}}, \binits{N.}},
\bauthor{\bsnm{{Pezzuto}}, \binits{S.}},
\bauthor{\bsnm{{Prusti}}, \binits{T.}},
\bauthor{\bsnm{{Roussel}}, \binits{H.}},
\bauthor{\bsnm{{Russeil}}, \binits{D.}},
\bauthor{\bsnm{{Sauvage}}, \binits{M.}},
\bauthor{\bsnm{{Sibthorpe}}, \binits{B.}},
\bauthor{\bsnm{{Sicilia-Aguilar}}, \binits{A.}},
\bauthor{\bsnm{{Spinoglio}}, \binits{L.}},
\bauthor{\bsnm{{Waelkens}}, \binits{C.}},
\bauthor{\bsnm{{Woodcraft}}, \binits{A.}},
\bauthor{\bsnm{{Zavagno}}, \binits{A.}}:
\batitle{{From filamentary clouds to prestellar cores to the stellar IMF:
  Initial highlights from the Herschel Gould Belt Survey}}.
\bjtitle{\aap}
\bvolume{518},
\bfpage{102}
(\byear{2010})
\doiurl{10.1051/0004-6361/201014666}
{\href{https://arxiv.org/abs/1005.2618}{{arXiv:1005.2618}}}
{[astro-ph.GA]}
\end{barticle}
\endbibitem

\bibitem[\protect\citeauthoryear{{Offner} et~al.}{2016}]{Offner2016}
\begin{barticle}
\bauthor{\bsnm{{Offner}}, \binits{S.S.R.}},
\bauthor{\bsnm{{Dunham}}, \binits{M.M.}},
\bauthor{\bsnm{{Lee}}, \binits{K.I.}},
\bauthor{\bsnm{{Arce}}, \binits{H.G.}},
\bauthor{\bsnm{{Fielding}}, \binits{D.B.}}:
\batitle{{The Turbulent Origin of Outflow and Spin Misalignment in Multiple
  Star Systems}}.
\bjtitle{\apjl}
\bvolume{827}(\bissue{1}),
\bfpage{11}
(\byear{2016})
\doiurl{10.3847/2041-8205/827/1/L11}
{\href{https://arxiv.org/abs/1606.08445}{{arXiv:1606.08445}}}
{[astro-ph.SR]}
\end{barticle}
\endbibitem

\bibitem[\protect\citeauthoryear{{Galli} et~al.}{2006}]{Galli2006}
\begin{barticle}
\bauthor{\bsnm{{Galli}}, \binits{D.}},
\bauthor{\bsnm{{Lizano}}, \binits{S.}},
\bauthor{\bsnm{{Shu}}, \binits{F.H.}},
\bauthor{\bsnm{{Allen}}, \binits{A.}}:
\batitle{{Gravitational Collapse of Magnetized Clouds. I. Ideal
  Magnetohydrodynamic Accretion Flow}}.
\bjtitle{\apj}
\bvolume{647}(\bissue{1}),
\bfpage{374}--\blpage{381}
(\byear{2006})
\doiurl{10.1086/505257}
{\href{https://arxiv.org/abs/astro-ph/0604573}{{arXiv:astro-ph/0604573}}}
{[astro-ph]}
\end{barticle}
\endbibitem

\bibitem[\protect\citeauthoryear{{Hennebelle}}{2013}]{Hennebelle2013}
\begin{barticle}
\bauthor{\bsnm{{Hennebelle}}, \binits{P.}}:
\batitle{{On the origin of non-self-gravitating filaments in the ISM}}.
\bjtitle{\aap}
\bvolume{556},
\bfpage{153}
(\byear{2013})
\doiurl{10.1051/0004-6361/201321292}
{\href{https://arxiv.org/abs/1306.5452}{{arXiv:1306.5452}}}
{[astro-ph.GA]}
\end{barticle}
\endbibitem

\bibitem[\protect\citeauthoryear{{Wang} et~al.}{2011}]{Wang2011}
\begin{barticle}
\bauthor{\bsnm{{Wang}}, \binits{K.}},
\bauthor{\bsnm{{Zhang}}, \binits{Q.}},
\bauthor{\bsnm{{Wu}}, \binits{Y.}},
\bauthor{\bsnm{{Zhang}}, \binits{H.}}:
\batitle{{Hierarchical Fragmentation and Jet-like Outflows in IRDC G28.34+0.06:
  A Growing Massive Protostar Cluster}}.
\bjtitle{\apj}
\bvolume{735}(\bissue{1}),
\bfpage{64}
(\byear{2011})
\doiurl{10.1088/0004-637X/735/1/64}
{\href{https://arxiv.org/abs/1105.4559}{{arXiv:1105.4559}}}
{[astro-ph.GA]}
\end{barticle}
\endbibitem

\bibitem[\protect\citeauthoryear{{Shu} et~al.}{1987}]{Shu1987}
\begin{barticle}
\bauthor{\bsnm{{Shu}}, \binits{F.H.}},
\bauthor{\bsnm{{Adams}}, \binits{F.C.}},
\bauthor{\bsnm{{Lizano}}, \binits{S.}}:
\batitle{{Star formation in molecular clouds: observation and theory.}}
\bjtitle{\araa}
\bvolume{25},
\bfpage{23}--\blpage{81}
(\byear{1987})
\doiurl{10.1146/annurev.aa.25.090187.000323}
\end{barticle}
\endbibitem

\bibitem[\protect\citeauthoryear{{Zhang} et~al.}{2014}]{Zhang2014}
\begin{barticle}
\bauthor{\bsnm{{Zhang}}, \binits{Q.}},
\bauthor{\bsnm{{Qiu}}, \binits{K.}},
\bauthor{\bsnm{{Girart}}, \binits{J.M.}},
\bauthor{\bsnm{{Liu}}, \binits{H.B.}},
\bauthor{\bsnm{{Tang}}, \binits{Y.-W.}},
\bauthor{\bsnm{{Koch}}, \binits{P.M.}},
\bauthor{\bsnm{{Li}}, \binits{Z.-Y.}},
\bauthor{\bsnm{{Keto}}, \binits{E.}},
\bauthor{\bsnm{{Ho}}, \binits{P.T.P.}},
\bauthor{\bsnm{{Rao}}, \binits{R.}},
\bauthor{\bsnm{{Lai}}, \binits{S.-P.}},
\bauthor{\bsnm{{Ching}}, \binits{T.-C.}},
\bauthor{\bsnm{{Frau}}, \binits{P.}},
\bauthor{\bsnm{{Chen}}, \binits{H.-H.}},
\bauthor{\bsnm{{Li}}, \binits{H.-B.}},
\bauthor{\bsnm{{Padovani}}, \binits{M.}},
\bauthor{\bsnm{{Bontemps}}, \binits{S.}},
\bauthor{\bsnm{{Csengeri}}, \binits{T.}},
\bauthor{\bsnm{{Ju{\'a}rez}}, \binits{C.}}:
\batitle{{Magnetic Fields and Massive Star Formation}}.
\bjtitle{\apj}
\bvolume{792}(\bissue{2}),
\bfpage{116}
(\byear{2014})
\doiurl{10.1088/0004-637X/792/2/116}
{\href{https://arxiv.org/abs/1407.3984}{{arXiv:1407.3984}}}
{[astro-ph.GA]}
\end{barticle}
\endbibitem

\bibitem[\protect\citeauthoryear{{Targon} et~al.}{2011}]{Targon2011}
\begin{barticle}
\bauthor{\bsnm{{Targon}}, \binits{C.G.}},
\bauthor{\bsnm{{Rodrigues}}, \binits{C.V.}},
\bauthor{\bsnm{{Cerqueira}}, \binits{A.H.}},
\bauthor{\bsnm{{Hickel}}, \binits{G.R.}}:
\batitle{{Correlating the Interstellar Magnetic Field with Protostellar Jets
  and Its Sources}}.
\bjtitle{\apj}
\bvolume{743}(\bissue{1}),
\bfpage{54}
(\byear{2011})
\doiurl{10.1088/0004-637X/743/1/54}
{\href{https://arxiv.org/abs/1109.2031}{{arXiv:1109.2031}}}
{[astro-ph.SR]}
\end{barticle}
\endbibitem

\bibitem[\protect\citeauthoryear{{Chapman} et~al.}{2013}]{Chapman2013}
\begin{barticle}
\bauthor{\bsnm{{Chapman}}, \binits{N.L.}},
\bauthor{\bsnm{{Davidson}}, \binits{J.A.}},
\bauthor{\bsnm{{Goldsmith}}, \binits{P.F.}},
\bauthor{\bsnm{{Houde}}, \binits{M.}},
\bauthor{\bsnm{{Kwon}}, \binits{W.}},
\bauthor{\bsnm{{Li}}, \binits{Z.-Y.}},
\bauthor{\bsnm{{Looney}}, \binits{L.W.}},
\bauthor{\bsnm{{Matthews}}, \binits{B.}},
\bauthor{\bsnm{{Matthews}}, \binits{T.G.}},
\bauthor{\bsnm{{Novak}}, \binits{G.}},
\bauthor{\bsnm{{Peng}}, \binits{R.}},
\bauthor{\bsnm{{Vaillancourt}}, \binits{J.E.}},
\bauthor{\bsnm{{Volgenau}}, \binits{N.H.}}:
\batitle{{Alignment between Flattened Protostellar Infall Envelopes and Ambient
  Magnetic Fields}}.
\bjtitle{\apj}
\bvolume{770}(\bissue{2}),
\bfpage{151}
(\byear{2013})
\doiurl{10.1088/0004-637X/770/2/151}
{\href{https://arxiv.org/abs/1305.2922}{{arXiv:1305.2922}}}
{[astro-ph.GA]}
\end{barticle}
\endbibitem

\bibitem[\protect\citeauthoryear{{Adams} et~al.}{1989}]{Adams1989}
\begin{barticle}
\bauthor{\bsnm{{Adams}}, \binits{F.C.}},
\bauthor{\bsnm{{Ruden}}, \binits{S.P.}},
\bauthor{\bsnm{{Shu}}, \binits{F.H.}}:
\batitle{{Eccentric Gravitational Instabilities in Nearly Keplerian Disks}}.
\bjtitle{\apj}
\bvolume{347},
\bfpage{959}
(\byear{1989})
\doiurl{10.1086/168187}
\end{barticle}
\endbibitem

\bibitem[\protect\citeauthoryear{{Kratter} et~al.}{2010}]{Kratter2010}
\begin{barticle}
\bauthor{\bsnm{{Kratter}}, \binits{K.M.}},
\bauthor{\bsnm{{Matzner}}, \binits{C.D.}},
\bauthor{\bsnm{{Krumholz}}, \binits{M.R.}},
\bauthor{\bsnm{{Klein}}, \binits{R.I.}}:
\batitle{{On the Role of Disks in the Formation of Stellar Systems: A Numerical
  Parameter Study of Rapid Accretion}}.
\bjtitle{\apj}
\bvolume{708}(\bissue{2}),
\bfpage{1585}--\blpage{1597}
(\byear{2010})
\doiurl{10.1088/0004-637X/708/2/1585}
{\href{https://arxiv.org/abs/0907.3476}{{arXiv:0907.3476}}}
{[astro-ph.SR]}
\end{barticle}
\endbibitem

\bibitem[\protect\citeauthoryear{{Dunhill} et~al.}{2015}]{Dunhill2015}
\begin{barticle}
\bauthor{\bsnm{{Dunhill}}, \binits{A.C.}},
\bauthor{\bsnm{{Cuadra}}, \binits{J.}},
\bauthor{\bsnm{{Dougados}}, \binits{C.}}:
\batitle{{Precession and accretion in circumbinary discs: the case of HD
  104237}}.
\bjtitle{\mnras}
\bvolume{448}(\bissue{4}),
\bfpage{3545}--\blpage{3554}
(\byear{2015})
\doiurl{10.1093/mnras/stv284}
{\href{https://arxiv.org/abs/1411.0687}{{arXiv:1411.0687}}}
{[astro-ph.SR]}
\end{barticle}
\endbibitem

\bibitem[\protect\citeauthoryear{{Clarke}}{2012}]{Clarke2012}
\begin{bchapter}
\bauthor{\bsnm{{Clarke}}, \binits{C.J.}}:
\bctitle{{Binary Star Formation Simulations}}.
In: \beditor{\bsnm{{Richards}}, \binits{M.T.}},
\beditor{\bsnm{{Hubeny}}, \binits{I.}} (eds.)
\bbtitle{From Interacting Binaries to Exoplanets: Essential Modeling Tools},
vol. \bseriesno{282},
pp. \bfpage{409}--\blpage{416}
(\byear{2012}).
\doiurl{10.1017/S1743921311027955}
\end{bchapter}
\endbibitem

\bibitem[\protect\citeauthoryear{{Ward}}{1997}]{Ward1997}
\begin{barticle}
\bauthor{\bsnm{{Ward}}, \binits{W.R.}}:
\batitle{{Protoplanet Migration by Nebula Tides}}.
\bjtitle{\icarus}
\bvolume{126}(\bissue{2}),
\bfpage{261}--\blpage{281}
(\byear{1997})
\doiurl{10.1006/icar.1996.5647}
\end{barticle}
\endbibitem

\bibitem[\protect\citeauthoryear{{Moe} and {Kratter}}{2018}]{Moe2018}
\begin{barticle}
\bauthor{\bsnm{{Moe}}, \binits{M.}},
\bauthor{\bsnm{{Kratter}}, \binits{K.M.}}:
\batitle{{Dynamical Formation of Close Binaries during the Pre-main-sequence
  Phase}}.
\bjtitle{\apj}
\bvolume{854}(\bissue{1}),
\bfpage{44}
(\byear{2018})
\doiurl{10.3847/1538-4357/aaa6d2}
{\href{https://arxiv.org/abs/1706.09894}{{arXiv:1706.09894}}}
{[astro-ph.SR]}
\end{barticle}
\endbibitem

\bibitem[\protect\citeauthoryear{{Arzamasskiy} et~al.}{2018}]{Arzamasskiy2018}
\begin{barticle}
\bauthor{\bsnm{{Arzamasskiy}}, \binits{L.}},
\bauthor{\bsnm{{Zhu}}, \binits{Z.}},
\bauthor{\bsnm{{Stone}}, \binits{J.M.}}:
\batitle{{Three-dimensional disc-satellite interaction: torques, migration, and
  observational signatures}}.
\bjtitle{\mnras}
\bvolume{475}(\bissue{3}),
\bfpage{3201}--\blpage{3212}
(\byear{2018})
\doiurl{10.1093/mnras/sty001}
{\href{https://arxiv.org/abs/1710.11128}{{arXiv:1710.11128}}}
{[astro-ph.EP]}
\end{barticle}
\endbibitem

\bibitem[\protect\citeauthoryear{{Larson}}{1969}]{Larson1969}
\begin{barticle}
\bauthor{\bsnm{{Larson}}, \binits{R.B.}}:
\batitle{{Numerical calculations of the dynamics of collapsing proto-star}}.
\bjtitle{\mnras}
\bvolume{145},
\bfpage{271}
(\byear{1969})
\doiurl{10.1093/mnras/145.3.271}
\end{barticle}
\endbibitem

\bibitem[\protect\citeauthoryear{{Bate} et~al.}{2002}]{Bate2002}
\begin{barticle}
\bauthor{\bsnm{{Bate}}, \binits{M.R.}},
\bauthor{\bsnm{{Bonnell}}, \binits{I.A.}},
\bauthor{\bsnm{{Bromm}}, \binits{V.}}:
\batitle{{The formation of close binary systems by dynamical interactions and
  orbital decay}}.
\bjtitle{\mnras}
\bvolume{336}(\bissue{3}),
\bfpage{705}--\blpage{713}
(\byear{2002})
\doiurl{10.1046/j.1365-8711.2002.05775.x}
{\href{https://arxiv.org/abs/astro-ph/0212403}{{arXiv:astro-ph/0212403}}}
{[astro-ph]}
\end{barticle}
\endbibitem

\bibitem[\protect\citeauthoryear{{Bate}}{2019}]{Bate2019}
\begin{barticle}
\bauthor{\bsnm{{Bate}}, \binits{M.R.}}:
\batitle{{The statistical properties of stars and their dependence on
  metallicity}}.
\bjtitle{\mnras}
\bvolume{484}(\bissue{2}),
\bfpage{2341}--\blpage{2361}
(\byear{2019})
\doiurl{10.1093/mnras/stz103}
{\href{https://arxiv.org/abs/1901.03713}{{arXiv:1901.03713}}}
{[astro-ph.SR]}
\end{barticle}
\endbibitem

\bibitem[\protect\citeauthoryear{{Tanaka} and {Ward}}{2004}]{Tanaka2004}
\begin{barticle}
\bauthor{\bsnm{{Tanaka}}, \binits{H.}},
\bauthor{\bsnm{{Ward}}, \binits{W.R.}}:
\batitle{{Three-dimensional Interaction between a Planet and an Isothermal
  Gaseous Disk. II. Eccentricity Waves and Bending Waves}}.
\bjtitle{\apj}
\bvolume{602}(\bissue{1}),
\bfpage{388}--\blpage{395}
(\byear{2004})
\doiurl{10.1086/380992}
\end{barticle}
\endbibitem

\bibitem[\protect\citeauthoryear{{Ceppi} et~al.}{2022}]{Ceppi2022}
\begin{barticle}
\bauthor{\bsnm{{Ceppi}}, \binits{S.}},
\bauthor{\bsnm{{Cuello}}, \binits{N.}},
\bauthor{\bsnm{{Lodato}}, \binits{G.}},
\bauthor{\bsnm{{Clarke}}, \binits{C.}},
\bauthor{\bsnm{{Toci}}, \binits{C.}},
\bauthor{\bsnm{{Price}}, \binits{D.J.}}:
\batitle{{Accretion rates in hierarchical triple systems with discs}}.
\bjtitle{\mnras}
\bvolume{514}(\bissue{1}),
\bfpage{906}--\blpage{919}
(\byear{2022})
\doiurl{10.1093/mnras/stac1390}
{\href{https://arxiv.org/abs/2205.08784}{{arXiv:2205.08784}}}
{[astro-ph.SR]}
\end{barticle}
\endbibitem

\bibitem[\protect\citeauthoryear{{Tokovinin} and {Moe}}{2020}]{Tokovinin2020}
\begin{barticle}
\bauthor{\bsnm{{Tokovinin}}, \binits{A.}},
\bauthor{\bsnm{{Moe}}, \binits{M.}}:
\batitle{{Formation of close binaries by disc fragmentation and migration, and
  its statistical modelling}}.
\bjtitle{\mnras}
\bvolume{491}(\bissue{4}),
\bfpage{5158}--\blpage{5171}
(\byear{2020})
\doiurl{10.1093/mnras/stz3299}
{\href{https://arxiv.org/abs/1910.01522}{{arXiv:1910.01522}}}
{[astro-ph.SR]}
\end{barticle}
\endbibitem

\bibitem[\protect\citeauthoryear{{Aizawa} et~al.}{2020}]{Aizawa2020}
\begin{barticle}
\bauthor{\bsnm{{Aizawa}}, \binits{M.}},
\bauthor{\bsnm{{Suto}}, \binits{Y.}},
\bauthor{\bsnm{{Oya}}, \binits{Y.}},
\bauthor{\bsnm{{Ikeda}}, \binits{S.}},
\bauthor{\bsnm{{Nakazato}}, \binits{T.}}:
\batitle{{Search for Alignment of Disk Orientations in Nearby Star-forming
  Regions: Lupus, Taurus, Upper Scorpius, {\ensuremath{\rho}} Ophiuchi, and
  Orion}}.
\bjtitle{\apj}
\bvolume{899}(\bissue{1}),
\bfpage{55}
(\byear{2020})
\doiurl{10.3847/1538-4357/aba43d}
{\href{https://arxiv.org/abs/2007.03393}{{arXiv:2007.03393}}}
{[astro-ph.SR]}
\end{barticle}
\endbibitem

\bibitem[\protect\citeauthoryear{{Bally}}{2016}]{Bally2016}
\begin{barticle}
\bauthor{\bsnm{{Bally}}, \binits{J.}}:
\batitle{{Protostellar Outflows}}.
\bjtitle{\araa}
\bvolume{54},
\bfpage{491}--\blpage{528}
(\byear{2016})
\doiurl{10.1146/annurev-astro-081915-023341}
\end{barticle}
\endbibitem

\bibitem[\protect\citeauthoryear{{Kumar} et~al.}{2011}]{Kumar2011}
\begin{barticle}
\bauthor{\bsnm{{Kumar}}, \binits{M.S.N.}},
\bauthor{\bsnm{{Sharma}}, \binits{S.}},
\bauthor{\bsnm{{Davis}}, \binits{C.J.}},
\bauthor{\bsnm{{Borissova}}, \binits{J.}},
\bauthor{\bsnm{{Grave}}, \binits{J.M.C.}}:
\batitle{{H$_{2}$ flows in the Corona Australis cloud and their driving
  sources}}.
\bjtitle{\aap}
\bvolume{533},
\bfpage{137}
(\byear{2011})
\doiurl{10.1051/0004-6361/201116860}
{\href{https://arxiv.org/abs/1108.3799}{{arXiv:1108.3799}}}
{[astro-ph.SR]}
\end{barticle}
\endbibitem

\bibitem[\protect\citeauthoryear{{Lee} et~al.}{2016}]{Lee2016}
\begin{barticle}
\bauthor{\bsnm{{Lee}}, \binits{K.I.}},
\bauthor{\bsnm{{Dunham}}, \binits{M.M.}},
\bauthor{\bsnm{{Myers}}, \binits{P.C.}},
\bauthor{\bsnm{{Arce}}, \binits{H.G.}},
\bauthor{\bsnm{{Bourke}}, \binits{T.L.}},
\bauthor{\bsnm{{Goodman}}, \binits{A.A.}},
\bauthor{\bsnm{{J{\o}rgensen}}, \binits{J.K.}},
\bauthor{\bsnm{{Kristensen}}, \binits{L.E.}},
\bauthor{\bsnm{{Offner}}, \binits{S.S.R.}},
\bauthor{\bsnm{{Pineda}}, \binits{J.E.}},
\bauthor{\bsnm{{Tobin}}, \binits{J.J.}},
\bauthor{\bsnm{{Vorobyov}}, \binits{E.I.}}:
\batitle{{Misalignment of Outflow Axes in the Proto-multiple Systems in
  Perseus}}.
\bjtitle{\apjl}
\bvolume{820}(\bissue{1}),
\bfpage{2}
(\byear{2016})
\doiurl{10.3847/2041-8205/820/1/L2}
{\href{https://arxiv.org/abs/1602.07397}{{arXiv:1602.07397}}}
{[astro-ph.SR]}
\end{barticle}
\endbibitem

\bibitem[\protect\citeauthoryear{{Baug} et~al.}{2020}]{Baug2020}
\begin{barticle}
\bauthor{\bsnm{{Baug}}, \binits{T.}},
\bauthor{\bsnm{{Wang}}, \binits{K.}},
\bauthor{\bsnm{{Liu}}, \binits{T.}},
\bauthor{\bsnm{{Tang}}, \binits{M.}},
\bauthor{\bsnm{{Zhang}}, \binits{Q.}},
\bauthor{\bsnm{{Li}}, \binits{D.}},
\bauthor{\bsnm{{Eswaraiah}}, \binits{C.}},
\bauthor{\bsnm{{Liu}}, \binits{S.-Y.}},
\bauthor{\bsnm{{Tej}}, \binits{A.}},
\bauthor{\bsnm{{Goldsmith}}, \binits{P.F.}},
\bauthor{\bsnm{{Bronfman}}, \binits{L.}},
\bauthor{\bsnm{{Qin}}, \binits{S.-L.}},
\bauthor{\bsnm{{T{\'o}th}}, \binits{V.L.}},
\bauthor{\bsnm{{Li}}, \binits{P.-S.}},
\bauthor{\bsnm{{Kim}}, \binits{K.-T.}}:
\batitle{{ALMA Observations Reveal No Preferred Outflow-filament and
  Outflow-magnetic Field Orientations in Protoclusters}}.
\bjtitle{\apj}
\bvolume{890}(\bissue{1}),
\bfpage{44}
(\byear{2020})
\doiurl{10.3847/1538-4357/ab66b6}
{\href{https://arxiv.org/abs/1912.13240}{{arXiv:1912.13240}}}
{[astro-ph.GA]}
\end{barticle}
\endbibitem

\bibitem[\protect\citeauthoryear{{Tokovinin}}{2017}]{Tokovinin2017}
\begin{barticle}
\bauthor{\bsnm{{Tokovinin}}, \binits{A.}}:
\batitle{{Orbit Alignment in Triple Stars}}.
\bjtitle{\apj}
\bvolume{844}(\bissue{2}),
\bfpage{103}
(\byear{2017})
\doiurl{10.3847/1538-4357/aa7746}
{\href{https://arxiv.org/abs/1706.00748}{{arXiv:1706.00748}}}
{[astro-ph.SR]}
\end{barticle}
\endbibitem

\bibitem[\protect\citeauthoryear{{Fisher}}{2004}]{Fisher2004}
\begin{barticle}
\bauthor{\bsnm{{Fisher}}, \binits{R.T.}}:
\batitle{{A Turbulent Interstellar Medium Origin of the Binary Period
  Distribution}}.
\bjtitle{\apj}
\bvolume{600}(\bissue{2}),
\bfpage{769}--\blpage{780}
(\byear{2004})
\doiurl{10.1086/380111}
{\href{https://arxiv.org/abs/astro-ph/0303280}{{arXiv:astro-ph/0303280}}}
{[astro-ph]}
\end{barticle}
\endbibitem

\bibitem[\protect\citeauthoryear{{Goodwin} et~al.}{2004}]{Goodwin2004}
\begin{barticle}
\bauthor{\bsnm{{Goodwin}}, \binits{S.P.}},
\bauthor{\bsnm{{Whitworth}}, \binits{A.P.}},
\bauthor{\bsnm{{Ward-Thompson}}, \binits{D.}}:
\batitle{{Simulating star formation in molecular cores. II. The effects of
  different levels of turbulence}}.
\bjtitle{\aap}
\bvolume{423},
\bfpage{169}--\blpage{182}
(\byear{2004})
\doiurl{10.1051/0004-6361:20040285}
{\href{https://arxiv.org/abs/astro-ph/0405117}{{arXiv:astro-ph/0405117}}}
{[astro-ph]}
\end{barticle}
\endbibitem

\bibitem[\protect\citeauthoryear{{Offner} et~al.}{2010}]{Offner2010}
\begin{barticle}
\bauthor{\bsnm{{Offner}}, \binits{S.S.R.}},
\bauthor{\bsnm{{Kratter}}, \binits{K.M.}},
\bauthor{\bsnm{{Matzner}}, \binits{C.D.}},
\bauthor{\bsnm{{Krumholz}}, \binits{M.R.}},
\bauthor{\bsnm{{Klein}}, \binits{R.I.}}:
\batitle{{The Formation of Low-mass Binary Star Systems Via Turbulent
  Fragmentation}}.
\bjtitle{\apj}
\bvolume{725}(\bissue{2}),
\bfpage{1485}--\blpage{1494}
(\byear{2010})
\doiurl{10.1088/0004-637X/725/2/1485}
{\href{https://arxiv.org/abs/1010.3702}{{arXiv:1010.3702}}}
{[astro-ph.SR]}
\end{barticle}
\endbibitem

\bibitem[\protect\citeauthoryear{{Czekala} et~al.}{2019}]{Czekala2019}
\begin{barticle}
\bauthor{\bsnm{{Czekala}}, \binits{I.}},
\bauthor{\bsnm{{Chiang}}, \binits{E.}},
\bauthor{\bsnm{{Andrews}}, \binits{S.M.}},
\bauthor{\bsnm{{Jensen}}, \binits{E.L.N.}},
\bauthor{\bsnm{{Torres}}, \binits{G.}},
\bauthor{\bsnm{{Wilner}}, \binits{D.J.}},
\bauthor{\bsnm{{Stassun}}, \binits{K.G.}},
\bauthor{\bsnm{{Macintosh}}, \binits{B.}}:
\batitle{{The Degree of Alignment between Circumbinary Disks and Their Binary
  Hosts}}.
\bjtitle{\apj}
\bvolume{883}(\bissue{1}),
\bfpage{22}
(\byear{2019})
\doiurl{10.3847/1538-4357/ab287b}
{\href{https://arxiv.org/abs/1906.03269}{{arXiv:1906.03269}}}
{[astro-ph.EP]}
\end{barticle}
\endbibitem

\bibitem[\protect\citeauthoryear{{Bohn} et~al.}{2022}]{Bohn2022}
\begin{barticle}
\bauthor{\bsnm{{Bohn}}, \binits{A.J.}},
\bauthor{\bsnm{{Benisty}}, \binits{M.}},
\bauthor{\bsnm{{Perraut}}, \binits{K.}},
\bauthor{\bsnm{{van der Marel}}, \binits{N.}},
\bauthor{\bsnm{{W{\"o}lfer}}, \binits{L.}},
\bauthor{\bsnm{{van Dishoeck}}, \binits{E.F.}},
\bauthor{\bsnm{{Facchini}}, \binits{S.}},
\bauthor{\bsnm{{Manara}}, \binits{C.F.}},
\bauthor{\bsnm{{Teague}}, \binits{R.}},
\bauthor{\bsnm{{Francis}}, \binits{L.}},
\bauthor{\bsnm{{Berger}}, \binits{J.-P.}},
\bauthor{\bsnm{{Garcia-Lopez}}, \binits{R.}},
\bauthor{\bsnm{{Ginski}}, \binits{C.}},
\bauthor{\bsnm{{Henning}}, \binits{T.}},
\bauthor{\bsnm{{Kenworthy}}, \binits{M.}},
\bauthor{\bsnm{{Kraus}}, \binits{S.}},
\bauthor{\bsnm{{M{\'e}nard}}, \binits{F.}},
\bauthor{\bsnm{{M{\'e}rand}}, \binits{A.}},
\bauthor{\bsnm{{P{\'e}rez}}, \binits{L.M.}}:
\batitle{{Probing inner and outer disk misalignments in transition disks.
  Constraints from VLTI/GRAVITY and ALMA observations}}.
\bjtitle{\aap}
\bvolume{658},
\bfpage{183}
(\byear{2022})
\doiurl{10.1051/0004-6361/202142070}
{\href{https://arxiv.org/abs/2112.00123}{{arXiv:2112.00123}}}
{[astro-ph.EP]}
\end{barticle}
\endbibitem

\bibitem[\protect\citeauthoryear{{D'Orazi} et~al.}{2019}]{D'Orazi2019}
\begin{barticle}
\bauthor{\bsnm{{D'Orazi}}, \binits{V.}},
\bauthor{\bsnm{{Gratton}}, \binits{R.}},
\bauthor{\bsnm{{Desidera}}, \binits{S.}},
\bauthor{\bsnm{{Avenhaus}}, \binits{H.}},
\bauthor{\bsnm{{Mesa}}, \binits{D.}},
\bauthor{\bsnm{{Stolker}}, \binits{T.}},
\bauthor{\bsnm{{Giro}}, \binits{E.}},
\bauthor{\bsnm{{Benatti}}, \binits{S.}},
\bauthor{\bsnm{{Jang-Condell}}, \binits{H.}},
\bauthor{\bsnm{{Rigliaco}}, \binits{E.}},
\bauthor{\bsnm{{Sissa}}, \binits{E.}},
\bauthor{\bsnm{{Scatolin}}, \binits{T.}},
\bauthor{\bsnm{{Benisty}}, \binits{M.}},
\bauthor{\bsnm{{Bhowmik}}, \binits{T.}},
\bauthor{\bsnm{{Boccaletti}}, \binits{A.}},
\bauthor{\bsnm{{Bonnefoy}}, \binits{M.}},
\bauthor{\bsnm{{Brandner}}, \binits{W.}},
\bauthor{\bsnm{{Buenzli}}, \binits{E.}},
\bauthor{\bsnm{{Chauvin}}, \binits{G.}},
\bauthor{\bsnm{{Daemgen}}, \binits{S.}},
\bauthor{\bsnm{{Damasso}}, \binits{M.}},
\bauthor{\bsnm{{Feldt}}, \binits{M.}},
\bauthor{\bsnm{{Galicher}}, \binits{R.}},
\bauthor{\bsnm{{Girard}}, \binits{J.}},
\bauthor{\bsnm{{Janson}}, \binits{M.}},
\bauthor{\bsnm{{Hagelberg}}, \binits{J.}},
\bauthor{\bsnm{{Mouillet}}, \binits{D.}},
\bauthor{\bsnm{{Kral}}, \binits{Q.}},
\bauthor{\bsnm{{Lannier}}, \binits{J.}},
\bauthor{\bsnm{{Lagrange}}, \binits{A.-M.}},
\bauthor{\bsnm{{Langlois}}, \binits{M.}},
\bauthor{\bsnm{{Maire}}, \binits{A.-L.}},
\bauthor{\bsnm{{Menard}}, \binits{F.}},
\bauthor{\bsnm{{Moeller-Nilsson}}, \binits{O.}},
\bauthor{\bsnm{{Perrot}}, \binits{C.}},
\bauthor{\bsnm{{Peretti}}, \binits{S.}},
\bauthor{\bsnm{{Rabou}}, \binits{P.}},
\bauthor{\bsnm{{Ramos}}, \binits{J.}},
\bauthor{\bsnm{{Rodet}}, \binits{L.}},
\bauthor{\bsnm{{Roelfsema}}, \binits{R.}},
\bauthor{\bsnm{{Roux}}, \binits{A.}},
\bauthor{\bsnm{{Salter}}, \binits{G.}},
\bauthor{\bsnm{{Schlieder}}, \binits{J.E.}},
\bauthor{\bsnm{{Schmidt}}, \binits{T.}},
\bauthor{\bsnm{{Szulagyi}}, \binits{J.}},
\bauthor{\bsnm{{Thalmann}}, \binits{C.}},
\bauthor{\bsnm{{Thebault}}, \binits{P.}},
\bauthor{\bsnm{{van der Plas}}, \binits{G.}},
\bauthor{\bsnm{{Vigan}}, \binits{A.}},
\bauthor{\bsnm{{Zurlo}}, \binits{A.}}:
\batitle{{Mapping of shadows cast on a protoplanetary disk by a close binary
  system}}.
\bjtitle{Nature Astronomy}
\bvolume{3},
\bfpage{167}--\blpage{172}
(\byear{2019})
\doiurl{10.1038/s41550-018-0626-6}
{\href{https://arxiv.org/abs/1811.10621}{{arXiv:1811.10621}}}
{[astro-ph.SR]}
\end{barticle}
\endbibitem

\bibitem[\protect\citeauthoryear{{Martinez-Brunner}
  et~al.}{2022}]{Martinez2022}
\begin{barticle}
\bauthor{\bsnm{{Martinez-Brunner}}, \binits{R.}},
\bauthor{\bsnm{{Casassus}}, \binits{S.}},
\bauthor{\bsnm{{P{\'e}rez}}, \binits{S.}},
\bauthor{\bsnm{{Hales}}, \binits{A.}},
\bauthor{\bsnm{{Weber}}, \binits{P.}},
\bauthor{\bsnm{{C{\'a}rcamo}}, \binits{M.}},
\bauthor{\bsnm{{Arce-Tord}}, \binits{C.}},
\bauthor{\bsnm{{Cieza}}, \binits{L.}},
\bauthor{\bsnm{{Garufi}}, \binits{A.}},
\bauthor{\bsnm{{Marino}}, \binits{S.}},
\bauthor{\bsnm{{Zurlo}}, \binits{A.}}:
\batitle{{High-resolution ALMA observations of V4046 Sgr: a circumbinary disc
  with a thin ring}}.
\bjtitle{\mnras}
\bvolume{510}(\bissue{1}),
\bfpage{1248}--\blpage{1257}
(\byear{2022})
\doiurl{10.1093/mnras/stab3440}
{\href{https://arxiv.org/abs/2111.12668}{{arXiv:2111.12668}}}
{[astro-ph.EP]}
\end{barticle}
\endbibitem

\bibitem[\protect\citeauthoryear{{Gillen} et~al.}{2014}]{Gillen2014}
\begin{barticle}
\bauthor{\bsnm{{Gillen}}, \binits{E.}},
\bauthor{\bsnm{{Aigrain}}, \binits{S.}},
\bauthor{\bsnm{{McQuillan}}, \binits{A.}},
\bauthor{\bsnm{{Bouvier}}, \binits{J.}},
\bauthor{\bsnm{{Hodgkin}}, \binits{S.}},
\bauthor{\bsnm{{Alencar}}, \binits{S.H.P.}},
\bauthor{\bsnm{{Terquem}}, \binits{C.}},
\bauthor{\bsnm{{Southworth}}, \binits{J.}},
\bauthor{\bsnm{{Gibson}}, \binits{N.P.}},
\bauthor{\bsnm{{Cody}}, \binits{A.}},
\bauthor{\bsnm{{Lendl}}, \binits{M.}},
\bauthor{\bsnm{{Morales-Calder{\'o}n}}, \binits{M.}},
\bauthor{\bsnm{{Favata}}, \binits{F.}},
\bauthor{\bsnm{{Stauffer}}, \binits{J.}},
\bauthor{\bsnm{{Micela}}, \binits{G.}}:
\batitle{{CoRoT 223992193: A new, low-mass, pre-main sequence eclipsing binary
  with evidence of a circumbinary disk}}.
\bjtitle{\aap}
\bvolume{562},
\bfpage{50}
(\byear{2014})
\doiurl{10.1051/0004-6361/201322493}
{\href{https://arxiv.org/abs/1311.3990}{{arXiv:1311.3990}}}
{[astro-ph.SR]}
\end{barticle}
\endbibitem

\bibitem[\protect\citeauthoryear{{Terquem} et~al.}{2015}]{Terquem2015}
\begin{barticle}
\bauthor{\bsnm{{Terquem}}, \binits{C.}},
\bauthor{\bsnm{{S{\o}rensen-Clark}}, \binits{P.M.}},
\bauthor{\bsnm{{Bouvier}}, \binits{J.}}:
\batitle{{A circumbinary disc model for the variability of the eclipsing binary
  CoRoT 223992193}}.
\bjtitle{\mnras}
\bvolume{454}(\bissue{4}),
\bfpage{3472}--\blpage{3479}
(\byear{2015})
\doiurl{10.1093/mnras/stv2258}
{\href{https://arxiv.org/abs/1509.08509}{{arXiv:1509.08509}}}
{[astro-ph.SR]}
\end{barticle}
\endbibitem

\bibitem[\protect\citeauthoryear{{Gillen} et~al.}{2017}]{Gillen2017}
\begin{barticle}
\bauthor{\bsnm{{Gillen}}, \binits{E.}},
\bauthor{\bsnm{{Aigrain}}, \binits{S.}},
\bauthor{\bsnm{{Terquem}}, \binits{C.}},
\bauthor{\bsnm{{Bouvier}}, \binits{J.}},
\bauthor{\bsnm{{Alencar}}, \binits{S.H.P.}},
\bauthor{\bsnm{{Gandolfi}}, \binits{D.}},
\bauthor{\bsnm{{Stauffer}}, \binits{J.}},
\bauthor{\bsnm{{Cody}}, \binits{A.}},
\bauthor{\bsnm{{Venuti}}, \binits{L.}},
\bauthor{\bsnm{{Almeida}}, \binits{P.V.}},
\bauthor{\bsnm{{Micela}}, \binits{G.}},
\bauthor{\bsnm{{Favata}}, \binits{F.}},
\bauthor{\bsnm{{Deeg}}, \binits{H.J.}}:
\batitle{{CoRoT 223992193: Investigating the variability in a low-mass,
  pre-main sequence eclipsing binary with evidence of a circumbinary disk}}.
\bjtitle{\aap}
\bvolume{599},
\bfpage{27}
(\byear{2017})
\doiurl{10.1051/0004-6361/201628483}
{\href{https://arxiv.org/abs/1611.05461}{{arXiv:1611.05461}}}
{[astro-ph.SR]}
\end{barticle}
\endbibitem

\bibitem[\protect\citeauthoryear{{Kennedy}}{2015}]{Kennedy2015}
\begin{barticle}
\bauthor{\bsnm{{Kennedy}}, \binits{G.M.}}:
\batitle{{Nature or nurture of coplanar Tatooines: the aligned circumbinary
  Kuiper Belt analogue around HD 131511.}}
\bjtitle{\mnras}
\bvolume{447},
\bfpage{75}--\blpage{79}
(\byear{2015})
\doiurl{10.1093/mnrasl/slu190}
{\href{https://arxiv.org/abs/1412.0674}{{arXiv:1412.0674}}}
{[astro-ph.EP]}
\end{barticle}
\endbibitem

\bibitem[\protect\citeauthoryear{{Alencar} et~al.}{2003}]{Alencar2003}
\begin{barticle}
\bauthor{\bsnm{{Alencar}}, \binits{S.H.P.}},
\bauthor{\bsnm{{Melo}}, \binits{C.H.F.}},
\bauthor{\bsnm{{Dullemond}}, \binits{C.P.}},
\bauthor{\bsnm{{Andersen}}, \binits{J.}},
\bauthor{\bsnm{{Batalha}}, \binits{C.}},
\bauthor{\bsnm{{Vaz}}, \binits{L.P.R.}},
\bauthor{\bsnm{{Mathieu}}, \binits{R.D.}}:
\batitle{{The pre-main sequence spectroscopic binary AK Scorpii revisited}}.
\bjtitle{\aap}
\bvolume{409},
\bfpage{1037}--\blpage{1053}
(\byear{2003})
\doiurl{10.1051/0004-6361:20031229}
\end{barticle}
\endbibitem

\bibitem[\protect\citeauthoryear{{Anthonioz} et~al.}{2015}]{Anthonioz2015}
\begin{barticle}
\bauthor{\bsnm{{Anthonioz}}, \binits{F.}},
\bauthor{\bsnm{{M{\'e}nard}}, \binits{F.}},
\bauthor{\bsnm{{Pinte}}, \binits{C.}},
\bauthor{\bsnm{{Le Bouquin}}, \binits{J.-B.}},
\bauthor{\bsnm{{Benisty}}, \binits{M.}},
\bauthor{\bsnm{{Thi}}, \binits{W.-F.}},
\bauthor{\bsnm{{Absil}}, \binits{O.}},
\bauthor{\bsnm{{Duch{\^e}ne}}, \binits{G.}},
\bauthor{\bsnm{{Augereau}}, \binits{J.-C.}},
\bauthor{\bsnm{{Berger}}, \binits{J.-P.}},
\bauthor{\bsnm{{Casassus}}, \binits{S.}},
\bauthor{\bsnm{{Duvert}}, \binits{G.}},
\bauthor{\bsnm{{Lazareff}}, \binits{B.}},
\bauthor{\bsnm{{Malbet}}, \binits{F.}},
\bauthor{\bsnm{{Millan-Gabet}}, \binits{R.}},
\bauthor{\bsnm{{Schreiber}}, \binits{M.R.}},
\bauthor{\bsnm{{Traub}}, \binits{W.}},
\bauthor{\bsnm{{Zins}}, \binits{G.}}:
\batitle{{The VLTI/PIONIER near-infrared interferometric survey of southern T
  Tauri stars. I. First results}}.
\bjtitle{\aap}
\bvolume{574},
\bfpage{41}
(\byear{2015})
\doiurl{10.1051/0004-6361/201424520}
{\href{https://arxiv.org/abs/1412.1052}{{arXiv:1412.1052}}}
{[astro-ph.SR]}
\end{barticle}
\endbibitem

\bibitem[\protect\citeauthoryear{{Janson} et~al.}{2016}]{Janson2016}
\begin{barticle}
\bauthor{\bsnm{{Janson}}, \binits{M.}},
\bauthor{\bsnm{{Thalmann}}, \binits{C.}},
\bauthor{\bsnm{{Boccaletti}}, \binits{A.}},
\bauthor{\bsnm{{Maire}}, \binits{A.-L.}},
\bauthor{\bsnm{{Zurlo}}, \binits{A.}},
\bauthor{\bsnm{{Marzari}}, \binits{F.}},
\bauthor{\bsnm{{Meyer}}, \binits{M.R.}},
\bauthor{\bsnm{{Carson}}, \binits{J.C.}},
\bauthor{\bsnm{{Augereau}}, \binits{J.-C.}},
\bauthor{\bsnm{{Garufi}}, \binits{A.}},
\bauthor{\bsnm{{Henning}}, \binits{T.}},
\bauthor{\bsnm{{Desidera}}, \binits{S.}},
\bauthor{\bsnm{{Asensio-Torres}}, \binits{R.}},
\bauthor{\bsnm{{Pohl}}, \binits{A.}}:
\batitle{{Detection of Sharp Symmetric Features in the Circumbinary Disk around
  AK Sco}}.
\bjtitle{\apjl}
\bvolume{816}(\bissue{1}),
\bfpage{1}
(\byear{2016})
\doiurl{10.3847/2041-8205/816/1/L1}
{\href{https://arxiv.org/abs/1512.04552}{{arXiv:1512.04552}}}
{[astro-ph.SR]}
\end{barticle}
\endbibitem

\bibitem[\protect\citeauthoryear{{Czekala} et~al.}{2016}]{Czekala2016}
\begin{barticle}
\bauthor{\bsnm{{Czekala}}, \binits{I.}},
\bauthor{\bsnm{{Andrews}}, \binits{S.M.}},
\bauthor{\bsnm{{Torres}}, \binits{G.}},
\bauthor{\bsnm{{Jensen}}, \binits{E.L.N.}},
\bauthor{\bsnm{{Stassun}}, \binits{K.G.}},
\bauthor{\bsnm{{Wilner}}, \binits{D.J.}},
\bauthor{\bsnm{{Latham}}, \binits{D.W.}}:
\batitle{{A Disk-based Dynamical Constraint on the Mass of the Young Binary DQ
  Tau}}.
\bjtitle{\apj}
\bvolume{818}(\bissue{2}),
\bfpage{156}
(\byear{2016})
\doiurl{10.3847/0004-637X/818/2/156}
{\href{https://arxiv.org/abs/1601.03806}{{arXiv:1601.03806}}}
{[astro-ph.SR]}
\end{barticle}
\endbibitem

\bibitem[\protect\citeauthoryear{{Guilloteau} et~al.}{2011}]{Guilloteau2011}
\begin{barticle}
\bauthor{\bsnm{{Guilloteau}}, \binits{S.}},
\bauthor{\bsnm{{Dutrey}}, \binits{A.}},
\bauthor{\bsnm{{Pi{\'e}tu}}, \binits{V.}},
\bauthor{\bsnm{{Boehler}}, \binits{Y.}}:
\batitle{{A dual-frequency sub-arcsecond study of proto-planetary disks at mm
  wavelengths: first evidence for radial variations of the dust properties}}.
\bjtitle{\aap}
\bvolume{529},
\bfpage{105}
(\byear{2011})
\doiurl{10.1051/0004-6361/201015209}
{\href{https://arxiv.org/abs/1103.1296}{{arXiv:1103.1296}}}
{[astro-ph.GA]}
\end{barticle}
\endbibitem

\bibitem[\protect\citeauthoryear{{Long} et~al.}{2018}]{Long2018}
\begin{barticle}
\bauthor{\bsnm{{Long}}, \binits{F.}},
\bauthor{\bsnm{{Pinilla}}, \binits{P.}},
\bauthor{\bsnm{{Herczeg}}, \binits{G.J.}},
\bauthor{\bsnm{{Harsono}}, \binits{D.}},
\bauthor{\bsnm{{Dipierro}}, \binits{G.}},
\bauthor{\bsnm{{Pascucci}}, \binits{I.}},
\bauthor{\bsnm{{Hendler}}, \binits{N.}},
\bauthor{\bsnm{{Tazzari}}, \binits{M.}},
\bauthor{\bsnm{{Ragusa}}, \binits{E.}},
\bauthor{\bsnm{{Salyk}}, \binits{C.}},
\bauthor{\bsnm{{Edwards}}, \binits{S.}},
\bauthor{\bsnm{{Lodato}}, \binits{G.}},
\bauthor{\bsnm{{van de Plas}}, \binits{G.}},
\bauthor{\bsnm{{Johnstone}}, \binits{D.}},
\bauthor{\bsnm{{Liu}}, \binits{Y.}},
\bauthor{\bsnm{{Boehler}}, \binits{Y.}},
\bauthor{\bsnm{{Cabrit}}, \binits{S.}},
\bauthor{\bsnm{{Manara}}, \binits{C.F.}},
\bauthor{\bsnm{{Menard}}, \binits{F.}},
\bauthor{\bsnm{{Mulders}}, \binits{G.D.}},
\bauthor{\bsnm{{Nisini}}, \binits{B.}},
\bauthor{\bsnm{{Fischer}}, \binits{W.J.}},
\bauthor{\bsnm{{Rigliaco}}, \binits{E.}},
\bauthor{\bsnm{{Banzatti}}, \binits{A.}},
\bauthor{\bsnm{{Avenhaus}}, \binits{H.}},
\bauthor{\bsnm{{Gully-Santiago}}, \binits{M.}}:
\batitle{{Gaps and Rings in an ALMA Survey of Disks in the Taurus Star-forming
  Region}}.
\bjtitle{\apj}
\bvolume{869}(\bissue{1}),
\bfpage{17}
(\byear{2018})
\doiurl{10.3847/1538-4357/aae8e1}
{\href{https://arxiv.org/abs/1810.06044}{{arXiv:1810.06044}}}
{[astro-ph.SR]}
\end{barticle}
\endbibitem

\bibitem[\protect\citeauthoryear{{Andrews} et~al.}{2010}]{Andrews2010}
\begin{barticle}
\bauthor{\bsnm{{Andrews}}, \binits{S.M.}},
\bauthor{\bsnm{{Czekala}}, \binits{I.}},
\bauthor{\bsnm{{Wilner}}, \binits{D.J.}},
\bauthor{\bsnm{{Espaillat}}, \binits{C.}},
\bauthor{\bsnm{{Dullemond}}, \binits{C.P.}},
\bauthor{\bsnm{{Hughes}}, \binits{A.M.}}:
\batitle{{Truncated Disks in TW Hya Association Multiple Star Systems}}.
\bjtitle{\apj}
\bvolume{710}(\bissue{1}),
\bfpage{462}--\blpage{469}
(\byear{2010})
\doiurl{10.1088/0004-637X/710/1/462}
{\href{https://arxiv.org/abs/0912.3537}{{arXiv:0912.3537}}}
{[astro-ph.SR]}
\end{barticle}
\endbibitem

\bibitem[\protect\citeauthoryear{{Kellogg} et~al.}{2017}]{Kellogg2017}
\begin{barticle}
\bauthor{\bsnm{{Kellogg}}, \binits{K.}},
\bauthor{\bsnm{{Prato}}, \binits{L.}},
\bauthor{\bsnm{{Torres}}, \binits{G.}},
\bauthor{\bsnm{{Schaefer}}, \binits{G.H.}},
\bauthor{\bsnm{{Avilez}}, \binits{I.}},
\bauthor{\bsnm{{Ru{\'\i}z-Rodr{\'\i}guez}}, \binits{D.}},
\bauthor{\bsnm{{Wasserman}}, \binits{L.H.}},
\bauthor{\bsnm{{Bonanos}}, \binits{A.Z.}},
\bauthor{\bsnm{{Guenther}}, \binits{E.W.}},
\bauthor{\bsnm{{Neuh{\"a}user}}, \binits{R.}},
\bauthor{\bsnm{{Levine}}, \binits{S.E.}},
\bauthor{\bsnm{{Bosh}}, \binits{A.S.}},
\bauthor{\bsnm{{Morzinski}}, \binits{K.M.}},
\bauthor{\bsnm{{Close}}, \binits{L.}},
\bauthor{\bsnm{{Bailey}}, \binits{V.}},
\bauthor{\bsnm{{Hinz}}, \binits{P.}},
\bauthor{\bsnm{{Males}}, \binits{J.R.}}:
\batitle{{The TWA 3 Young Triple System: Orbits, Disks, Evolution}}.
\bjtitle{\apj}
\bvolume{844}(\bissue{2}),
\bfpage{168}
(\byear{2017})
\doiurl{10.3847/1538-4357/aa7c60}
{\href{https://arxiv.org/abs/1707.00591}{{arXiv:1707.00591}}}
{[astro-ph.SR]}
\end{barticle}
\endbibitem

\bibitem[\protect\citeauthoryear{{Czekala} et~al.}{2021}]{Czekala2021}
\begin{barticle}
\bauthor{\bsnm{{Czekala}}, \binits{I.}},
\bauthor{\bsnm{{Ribas}}, \binits{{\'A}.}},
\bauthor{\bsnm{{Cuello}}, \binits{N.}},
\bauthor{\bsnm{{Chiang}}, \binits{E.}},
\bauthor{\bsnm{{Mac{\'\i}as}}, \binits{E.}},
\bauthor{\bsnm{{Duch{\^e}ne}}, \binits{G.}},
\bauthor{\bsnm{{Andrews}}, \binits{S.M.}},
\bauthor{\bsnm{{Espaillat}}, \binits{C.C.}}:
\batitle{{A Coplanar Circumbinary Protoplanetary Disk in the TWA 3 Triple M
  Dwarf System}}.
\bjtitle{\apj}
\bvolume{912}(\bissue{1}),
\bfpage{6}
(\byear{2021})
\doiurl{10.3847/1538-4357/abebe3}
{\href{https://arxiv.org/abs/2102.11875}{{arXiv:2102.11875}}}
{[astro-ph.EP]}
\end{barticle}
\endbibitem

\bibitem[\protect\citeauthoryear{{Herbst} et~al.}{2002}]{Herbst2002}
\begin{barticle}
\bauthor{\bsnm{{Herbst}}, \binits{W.}},
\bauthor{\bsnm{{Hamilton}}, \binits{C.M.}},
\bauthor{\bsnm{{Vrba}}, \binits{F.J.}},
\bauthor{\bsnm{{Ibrahimov}}, \binits{M.A.}},
\bauthor{\bsnm{{Bailer-Jones}}, \binits{C.A.L.}},
\bauthor{\bsnm{{Mundt}}, \binits{R.}},
\bauthor{\bsnm{{Lamm}}, \binits{M.}},
\bauthor{\bsnm{{Mazeh}}, \binits{T.}},
\bauthor{\bsnm{{Webster}}, \binits{Z.T.}},
\bauthor{\bsnm{{Haisch}}, \binits{K.E.}},
\bauthor{\bsnm{{Williams}}, \binits{E.C.}},
\bauthor{\bsnm{{Rhodes}}, \binits{A.H.}},
\bauthor{\bsnm{{Balonek}}, \binits{T.J.}},
\bauthor{\bsnm{{Scholz}}, \binits{A.}},
\bauthor{\bsnm{{Riffeser}}, \binits{A.}}:
\batitle{{Fine Structure in the Circumstellar Environment of a Young,
  Solar-like Star: The Unique Eclipses of KH 15D}}.
\bjtitle{\pasp}
\bvolume{114}(\bissue{801}),
\bfpage{1167}--\blpage{1172}
(\byear{2002})
\doiurl{10.1086/344205}
{\href{https://arxiv.org/abs/astro-ph/0208203}{{arXiv:astro-ph/0208203}}}
{[astro-ph]}
\end{barticle}
\endbibitem

\bibitem[\protect\citeauthoryear{{Capelo} et~al.}{2012}]{Capelo2012}
\begin{barticle}
\bauthor{\bsnm{{Capelo}}, \binits{H.L.}},
\bauthor{\bsnm{{Herbst}}, \binits{W.}},
\bauthor{\bsnm{{Leggett}}, \binits{S.K.}},
\bauthor{\bsnm{{Hamilton}}, \binits{C.M.}},
\bauthor{\bsnm{{Johnson}}, \binits{J.A.}}:
\batitle{{Locating the Trailing Edge of the Circumbinary Ring in the KH 15D
  System}}.
\bjtitle{\apjl}
\bvolume{757}(\bissue{1}),
\bfpage{18}
(\byear{2012})
\doiurl{10.1088/2041-8205/757/1/L18}
{\href{https://arxiv.org/abs/1208.5497}{{arXiv:1208.5497}}}
{[astro-ph.SR]}
\end{barticle}
\endbibitem

\bibitem[\protect\citeauthoryear{{Johnson} et~al.}{2004}]{Johnson2004}
\begin{barticle}
\bauthor{\bsnm{{Johnson}}, \binits{J.A.}},
\bauthor{\bsnm{{Marcy}}, \binits{G.W.}},
\bauthor{\bsnm{{Hamilton}}, \binits{C.M.}},
\bauthor{\bsnm{{Herbst}}, \binits{W.}},
\bauthor{\bsnm{{Johns-Krull}}, \binits{C.M.}}:
\batitle{{KH 15D: A Spectroscopic Binary}}.
\bjtitle{\aj}
\bvolume{128}(\bissue{3}),
\bfpage{1265}--\blpage{1272}
(\byear{2004})
\doiurl{10.1086/422735}
{\href{https://arxiv.org/abs/astro-ph/0403099}{{arXiv:astro-ph/0403099}}}
{[astro-ph]}
\end{barticle}
\endbibitem

\bibitem[\protect\citeauthoryear{{Poon} et~al.}{2021}]{Poon2021}
\begin{barticle}
\bauthor{\bsnm{{Poon}}, \binits{M.}},
\bauthor{\bsnm{{Zanazzi}}, \binits{J.J.}},
\bauthor{\bsnm{{Zhu}}, \binits{W.}}:
\batitle{{Constraining the circumbinary disc tilt in the KH 15D system}}.
\bjtitle{\mnras}
\bvolume{503}(\bissue{2}),
\bfpage{1599}--\blpage{1614}
(\byear{2021})
\doiurl{10.1093/mnras/stab575}
{\href{https://arxiv.org/abs/2009.14204}{{arXiv:2009.14204}}}
{[astro-ph.EP]}
\end{barticle}
\endbibitem

\bibitem[\protect\citeauthoryear{{Welty}}{1995}]{Welty1995}
\begin{barticle}
\bauthor{\bsnm{{Welty}}, \binits{A.D.}}:
\batitle{{Discovery of a Pre-Main-Sequence Spectrscopic Binary: V773 Tauri}}.
\bjtitle{\aj}
\bvolume{110},
\bfpage{776}
(\byear{1995})
\doiurl{10.1086/117562}
\end{barticle}
\endbibitem

\bibitem[\protect\citeauthoryear{{Boden} et~al.}{2007}]{Boden2007}
\begin{barticle}
\bauthor{\bsnm{{Boden}}, \binits{A.F.}},
\bauthor{\bsnm{{Torres}}, \binits{G.}},
\bauthor{\bsnm{{Sargent}}, \binits{A.I.}},
\bauthor{\bsnm{{Akeson}}, \binits{R.L.}},
\bauthor{\bsnm{{Carpenter}}, \binits{J.M.}},
\bauthor{\bsnm{{Boboltz}}, \binits{D.A.}},
\bauthor{\bsnm{{Massi}}, \binits{M.}},
\bauthor{\bsnm{{Ghez}}, \binits{A.M.}},
\bauthor{\bsnm{{Latham}}, \binits{D.W.}},
\bauthor{\bsnm{{Johnston}}, \binits{K.J.}},
\bauthor{\bsnm{{Menten}}, \binits{K.M.}},
\bauthor{\bsnm{{Ros}}, \binits{E.}}:
\batitle{{Dynamical Masses for Pre-Main-Sequence Stars: A Preliminary Physical
  Orbit for V773 Tau A}}.
\bjtitle{\apj}
\bvolume{670}(\bissue{2}),
\bfpage{1214}--\blpage{1224}
(\byear{2007})
\doiurl{10.1086/521296}
{\href{https://arxiv.org/abs/0706.2376}{{arXiv:0706.2376}}}
{[astro-ph]}
\end{barticle}
\endbibitem

\bibitem[\protect\citeauthoryear{{Kenworthy} et~al.}{2022}]{Kenworthy2022}
\begin{barticle}
\bauthor{\bsnm{{Kenworthy}}, \binits{M.A.}},
\bauthor{\bsnm{{Gonz{\'a}lez Picos}}, \binits{D.}},
\bauthor{\bsnm{{Elizondo}}, \binits{E.}},
\bauthor{\bsnm{{Martin}}, \binits{R.G.}},
\bauthor{\bsnm{{van Dam}}, \binits{D.M.}},
\bauthor{\bsnm{{Rodriguez}}, \binits{J.E.}},
\bauthor{\bsnm{{Kennedy}}, \binits{G.M.}},
\bauthor{\bsnm{{Ginski}}, \binits{C.}},
\bauthor{\bsnm{{Mugrauer}}, \binits{M.}},
\bauthor{\bsnm{{Vogt}}, \binits{N.}},
\bauthor{\bsnm{{Adam}}, \binits{C.}},
\bauthor{\bsnm{{Oelkers}}, \binits{R.J.}}:
\batitle{{Eclipse of the V773 Tau B circumbinary disc}}.
\bjtitle{\aap}
\bvolume{666},
\bfpage{61}
(\byear{2022})
\doiurl{10.1051/0004-6361/202243441}
{\href{https://arxiv.org/abs/2207.05575}{{arXiv:2207.05575}}}
{[astro-ph.SR]}
\end{barticle}
\endbibitem

\bibitem[\protect\citeauthoryear{{Sissa} et~al.}{2019}]{Sissa2019}
\begin{barticle}
\bauthor{\bsnm{{Sissa}}, \binits{E.}},
\bauthor{\bsnm{{Gratton}}, \binits{R.}},
\bauthor{\bsnm{{Alcal{\`a}}}, \binits{J.M.}},
\bauthor{\bsnm{{Desidera}}, \binits{S.}},
\bauthor{\bsnm{{Messina}}, \binits{S.}},
\bauthor{\bsnm{{Mesa}}, \binits{D.}},
\bauthor{\bsnm{{D'Orazi}}, \binits{V.}},
\bauthor{\bsnm{{Rigliaco}}, \binits{E.}}:
\batitle{{The origin of R CrA variability. A complex triple system hosting a
  disk}}.
\bjtitle{\aap}
\bvolume{630},
\bfpage{132}
(\byear{2019})
\doiurl{10.1051/0004-6361/201936101}
{\href{https://arxiv.org/abs/1908.07256}{{arXiv:1908.07256}}}
{[astro-ph.SR]}
\end{barticle}
\endbibitem

\bibitem[\protect\citeauthoryear{{Mesa} et~al.}{2019}]{Mesa2019}
\begin{barticle}
\bauthor{\bsnm{{Mesa}}, \binits{D.}},
\bauthor{\bsnm{{Bonnefoy}}, \binits{M.}},
\bauthor{\bsnm{{Gratton}}, \binits{R.}},
\bauthor{\bsnm{{Van Der Plas}}, \binits{G.}},
\bauthor{\bsnm{{D'Orazi}}, \binits{V.}},
\bauthor{\bsnm{{Sissa}}, \binits{E.}},
\bauthor{\bsnm{{Zurlo}}, \binits{A.}},
\bauthor{\bsnm{{Rigliaco}}, \binits{E.}},
\bauthor{\bsnm{{Schmidt}}, \binits{T.}},
\bauthor{\bsnm{{Langlois}}, \binits{M.}},
\bauthor{\bsnm{{Vigan}}, \binits{A.}},
\bauthor{\bsnm{{Ubeira Gabellini}}, \binits{M.G.}},
\bauthor{\bsnm{{Desidera}}, \binits{S.}},
\bauthor{\bsnm{{Antoniucci}}, \binits{S.}},
\bauthor{\bsnm{{Barbieri}}, \binits{M.}},
\bauthor{\bsnm{{Benisty}}, \binits{M.}},
\bauthor{\bsnm{{Boccaletti}}, \binits{A.}},
\bauthor{\bsnm{{Claudi}}, \binits{R.}},
\bauthor{\bsnm{{Fedele}}, \binits{D.}},
\bauthor{\bsnm{{Gasparri}}, \binits{D.}},
\bauthor{\bsnm{{Henning}}, \binits{T.}},
\bauthor{\bsnm{{Kasper}}, \binits{M.}},
\bauthor{\bsnm{{Lagrange}}, \binits{A.-M.}},
\bauthor{\bsnm{{Lazzoni}}, \binits{C.}},
\bauthor{\bsnm{{Lodato}}, \binits{G.}},
\bauthor{\bsnm{{Maire}}, \binits{A.-L.}},
\bauthor{\bsnm{{Manara}}, \binits{C.F.}},
\bauthor{\bsnm{{Meyer}}, \binits{M.}},
\bauthor{\bsnm{{Reggiani}}, \binits{M.}},
\bauthor{\bsnm{{Samland}}, \binits{M.}},
\bauthor{\bsnm{{Van den Ancker}}, \binits{M.}},
\bauthor{\bsnm{{Chauvin}}, \binits{G.}},
\bauthor{\bsnm{{Cheetham}}, \binits{A.}},
\bauthor{\bsnm{{Feldt}}, \binits{M.}},
\bauthor{\bsnm{{Hugot}}, \binits{E.}},
\bauthor{\bsnm{{Janson}}, \binits{M.}},
\bauthor{\bsnm{{Ligi}}, \binits{R.}},
\bauthor{\bsnm{{M{\"o}ller-Nilsson}}, \binits{O.}},
\bauthor{\bsnm{{Petit}}, \binits{C.}},
\bauthor{\bsnm{{Rickman}}, \binits{E.L.}},
\bauthor{\bsnm{{Rigal}}, \binits{F.}},
\bauthor{\bsnm{{Wildi}}, \binits{F.}}:
\batitle{{Exploring the R CrA environment with SPHERE. Discovery of a new
  stellar companion}}.
\bjtitle{\aap}
\bvolume{624},
\bfpage{4}
(\byear{2019})
\doiurl{10.1051/0004-6361/201834682}
{\href{https://arxiv.org/abs/1902.02536}{{arXiv:1902.02536}}}
{[astro-ph.SR]}
\end{barticle}
\endbibitem

\bibitem[\protect\citeauthoryear{{Czekala} et~al.}{2017}]{Czekala2017}
\begin{barticle}
\bauthor{\bsnm{{Czekala}}, \binits{I.}},
\bauthor{\bsnm{{Andrews}}, \binits{S.M.}},
\bauthor{\bsnm{{Torres}}, \binits{G.}},
\bauthor{\bsnm{{Rodriguez}}, \binits{J.E.}},
\bauthor{\bsnm{{Jensen}}, \binits{E.L.N.}},
\bauthor{\bsnm{{Stassun}}, \binits{K.G.}},
\bauthor{\bsnm{{Latham}}, \binits{D.W.}},
\bauthor{\bsnm{{Wilner}}, \binits{D.J.}},
\bauthor{\bsnm{{Gully-Santiago}}, \binits{M.A.}},
\bauthor{\bsnm{{Grankin}}, \binits{K.N.}},
\bauthor{\bsnm{{Lund}}, \binits{M.B.}},
\bauthor{\bsnm{{Kuhn}}, \binits{R.B.}},
\bauthor{\bsnm{{Stevens}}, \binits{D.J.}},
\bauthor{\bsnm{{Siverd}}, \binits{R.J.}},
\bauthor{\bsnm{{James}}, \binits{D.}},
\bauthor{\bsnm{{Gaudi}}, \binits{B.S.}},
\bauthor{\bsnm{{Shappee}}, \binits{B.J.}},
\bauthor{\bsnm{{Holoien}}, \binits{T.W.-S.}}:
\batitle{{The Architecture of the GW Ori Young Triple-star System and Its Disk:
  Dynamical Masses, Mutual Inclinations, and Recurrent Eclipses}}.
\bjtitle{\apj}
\bvolume{851}(\bissue{2}),
\bfpage{132}
(\byear{2017})
\doiurl{10.3847/1538-4357/aa9be7}
{\href{https://arxiv.org/abs/1710.03153}{{arXiv:1710.03153}}}
{[astro-ph.EP]}
\end{barticle}
\endbibitem

\bibitem[\protect\citeauthoryear{{Kennedy} et~al.}{2012}]{Kennedy2012}
\begin{barticle}
\bauthor{\bsnm{{Kennedy}}, \binits{G.M.}},
\bauthor{\bsnm{{Wyatt}}, \binits{M.C.}},
\bauthor{\bsnm{{Sibthorpe}}, \binits{B.}},
\bauthor{\bsnm{{Duch{\^e}ne}}, \binits{G.}},
\bauthor{\bsnm{{Kalas}}, \binits{P.}},
\bauthor{\bsnm{{Matthews}}, \binits{B.C.}},
\bauthor{\bsnm{{Greaves}}, \binits{J.S.}},
\bauthor{\bsnm{{Su}}, \binits{K.Y.L.}},
\bauthor{\bsnm{{Fitzgerald}}, \binits{M.P.}}:
\batitle{{99 Herculis: host to a circumbinary polar-ring debris disc}}.
\bjtitle{\mnras}
\bvolume{421}(\bissue{3}),
\bfpage{2264}--\blpage{2276}
(\byear{2012})
\doiurl{10.1111/j.1365-2966.2012.20448.x}
{\href{https://arxiv.org/abs/1201.1911}{{arXiv:1201.1911}}}
{[astro-ph.EP]}
\end{barticle}
\endbibitem

\bibitem[\protect\citeauthoryear{{Monnier} et~al.}{2006}]{Monnier2006}
\begin{barticle}
\bauthor{\bsnm{{Monnier}}, \binits{J.D.}},
\bauthor{\bsnm{{Berger}}, \binits{J.-P.}},
\bauthor{\bsnm{{Millan-Gabet}}, \binits{R.}},
\bauthor{\bsnm{{Traub}}, \binits{W.A.}},
\bauthor{\bsnm{{Schloerb}}, \binits{F.P.}},
\bauthor{\bsnm{{Pedretti}}, \binits{E.}},
\bauthor{\bsnm{{Benisty}}, \binits{M.}},
\bauthor{\bsnm{{Carleton}}, \binits{N.P.}},
\bauthor{\bsnm{{Haguenauer}}, \binits{P.}},
\bauthor{\bsnm{{Kern}}, \binits{P.}},
\bauthor{\bsnm{{Labeye}}, \binits{P.}},
\bauthor{\bsnm{{Lacasse}}, \binits{M.G.}},
\bauthor{\bsnm{{Malbet}}, \binits{F.}},
\bauthor{\bsnm{{Perraut}}, \binits{K.}},
\bauthor{\bsnm{{Pearlman}}, \binits{M.}},
\bauthor{\bsnm{{Zhao}}, \binits{M.}}:
\batitle{{Few Skewed Disks Found in First Closure-Phase Survey of Herbig Ae/Be
  Stars}}.
\bjtitle{\apj}
\bvolume{647}(\bissue{1}),
\bfpage{444}--\blpage{463}
(\byear{2006})
\doiurl{10.1086/505340}
{\href{https://arxiv.org/abs/astro-ph/0606052}{{arXiv:astro-ph/0606052}}}
{[astro-ph]}
\end{barticle}
\endbibitem

\bibitem[\protect\citeauthoryear{{Okamoto} et~al.}{2009}]{Okamoto2009}
\begin{barticle}
\bauthor{\bsnm{{Okamoto}}, \binits{Y.K.}},
\bauthor{\bsnm{{Kataza}}, \binits{H.}},
\bauthor{\bsnm{{Honda}}, \binits{M.}},
\bauthor{\bsnm{{Fujiwara}}, \binits{H.}},
\bauthor{\bsnm{{Momose}}, \binits{M.}},
\bauthor{\bsnm{{Ohashi}}, \binits{N.}},
\bauthor{\bsnm{{Fujiyoshi}}, \binits{T.}},
\bauthor{\bsnm{{Sakon}}, \binits{I.}},
\bauthor{\bsnm{{Sako}}, \binits{S.}},
\bauthor{\bsnm{{Yamashita}}, \binits{T.}},
\bauthor{\bsnm{{Miyata}}, \binits{T.}},
\bauthor{\bsnm{{Onaka}}, \binits{T.}}:
\batitle{{Direct Detection of a Flared Disk Around a Young Massive Star
  HD200775 and its 10 to 1000 AU Scale Properties}}.
\bjtitle{\apj}
\bvolume{706}(\bissue{1}),
\bfpage{665}--\blpage{675}
(\byear{2009})
\doiurl{10.1088/0004-637X/706/1/665}
{\href{https://arxiv.org/abs/0910.4328}{{arXiv:0910.4328}}}
{[astro-ph.GA]}
\end{barticle}
\endbibitem

\bibitem[\protect\citeauthoryear{{Benisty} et~al.}{2013}]{Benisty2013}
\begin{barticle}
\bauthor{\bsnm{{Benisty}}, \binits{M.}},
\bauthor{\bsnm{{Perraut}}, \binits{K.}},
\bauthor{\bsnm{{Mourard}}, \binits{D.}},
\bauthor{\bsnm{{Stee}}, \binits{P.}},
\bauthor{\bsnm{{Lima}}, \binits{G.H.R.A.}},
\bauthor{\bsnm{{Le Bouquin}}, \binits{J.B.}},
\bauthor{\bsnm{{Borges Fernandes}}, \binits{M.}},
\bauthor{\bsnm{{Chesneau}}, \binits{O.}},
\bauthor{\bsnm{{Nardetto}}, \binits{N.}},
\bauthor{\bsnm{{Tallon-Bosc}}, \binits{I.}},
\bauthor{\bsnm{{McAlister}}, \binits{H.}},
\bauthor{\bsnm{{Ten Brummelaar}}, \binits{T.}},
\bauthor{\bsnm{{Ridgway}}, \binits{S.}},
\bauthor{\bsnm{{Sturmann}}, \binits{J.}},
\bauthor{\bsnm{{Sturmann}}, \binits{L.}},
\bauthor{\bsnm{{Turner}}, \binits{N.}},
\bauthor{\bsnm{{Farrington}}, \binits{C.}},
\bauthor{\bsnm{{Goldfinger}}, \binits{P.J.}}:
\batitle{{Enhanced H$_{{\ensuremath{\alpha}}}$ activity at periastron in the
  young and massive spectroscopic binary HD 200775}}.
\bjtitle{\aap}
\bvolume{555},
\bfpage{113}
(\byear{2013})
\doiurl{10.1051/0004-6361/201219893}
{\href{https://arxiv.org/abs/1306.0390}{{arXiv:1306.0390}}}
{[astro-ph.SR]}
\end{barticle}
\endbibitem

\bibitem[\protect\citeauthoryear{{Aly} et~al.}{2018}]{Aly2018}
\begin{barticle}
\bauthor{\bsnm{{Aly}}, \binits{H.}},
\bauthor{\bsnm{{Lodato}}, \binits{G.}},
\bauthor{\bsnm{{Cazzoletti}}, \binits{P.}}:
\batitle{{On the secular evolution of GG Tau A circumbinary disc: a misaligned
  disc scenario}}.
\bjtitle{\mnras}
\bvolume{480}(\bissue{4}),
\bfpage{4738}--\blpage{4745}
(\byear{2018})
\doiurl{10.1093/mnras/sty2179}
{\href{https://arxiv.org/abs/1809.06383}{{arXiv:1809.06383}}}
{[astro-ph.SR]}
\end{barticle}
\endbibitem

\bibitem[\protect\citeauthoryear{{Guilloteau} et~al.}{1999}]{Guilloteau1999}
\begin{barticle}
\bauthor{\bsnm{{Guilloteau}}, \binits{S.}},
\bauthor{\bsnm{{Dutrey}}, \binits{A.}},
\bauthor{\bsnm{{Simon}}, \binits{M.}}:
\batitle{{GG Tauri: the ring world}}.
\bjtitle{\aap}
\bvolume{348},
\bfpage{570}--\blpage{578}
(\byear{1999})
\end{barticle}
\endbibitem

\bibitem[\protect\citeauthoryear{{Rigliaco} et~al.}{2023}]{Rigliaco2023}
\begin{barticle}
\bauthor{\bsnm{{Rigliaco}}, \binits{E.}},
\bauthor{\bsnm{{Gratton}}, \binits{R.}},
\bauthor{\bsnm{{Ceppi}}, \binits{S.}},
\bauthor{\bsnm{{Ginski}}, \binits{C.}},
\bauthor{\bsnm{{Hogerheijde}}, \binits{M.}},
\bauthor{\bsnm{{Benisty}}, \binits{M.}},
\bauthor{\bsnm{{Birnstiel}}, \binits{T.}},
\bauthor{\bsnm{{Dima}}, \binits{M.}},
\bauthor{\bsnm{{Facchini}}, \binits{S.}},
\bauthor{\bsnm{{Garufi}}, \binits{A.}},
\bauthor{\bsnm{{Bae}}, \binits{J.}},
\bauthor{\bsnm{{Langlois}}, \binits{M.}},
\bauthor{\bsnm{{Lodato}}, \binits{G.}},
\bauthor{\bsnm{{Mamajek}}, \binits{E.}},
\bauthor{\bsnm{{Manara}}, \binits{C.F.}},
\bauthor{\bsnm{{M{\'e}nard}}, \binits{F.}},
\bauthor{\bsnm{{Ribas}}, \binits{A.}},
\bauthor{\bsnm{{Zurlo}}, \binits{A.}}:
\batitle{{Disk Evolution Study Through Imaging of Nearby Young Stars
  (DESTINYS): Characterization of the young star T CrA and its circumstellar
  environment}}.
\bjtitle{\aap}
\bvolume{671},
\bfpage{82}
(\byear{2023})
\doiurl{10.1051/0004-6361/202245192}
{\href{https://arxiv.org/abs/2301.01486}{{arXiv:2301.01486}}}
{[astro-ph.SR]}
\end{barticle}
\endbibitem

\bibitem[\protect\citeauthoryear{{Claudi} et~al.}{2019}]{Claudi2019}
\begin{barticle}
\bauthor{\bsnm{{Claudi}}, \binits{R.}},
\bauthor{\bsnm{{Maire}}, \binits{A.-L.}},
\bauthor{\bsnm{{Mesa}}, \binits{D.}},
\bauthor{\bsnm{{Cheetham}}, \binits{A.}},
\bauthor{\bsnm{{Fontanive}}, \binits{C.}},
\bauthor{\bsnm{{Gratton}}, \binits{R.}},
\bauthor{\bsnm{{Zurlo}}, \binits{A.}},
\bauthor{\bsnm{{Avenhaus}}, \binits{H.}},
\bauthor{\bsnm{{Bhowmik}}, \binits{T.}},
\bauthor{\bsnm{{Biller}}, \binits{B.}},
\bauthor{\bsnm{{Boccaletti}}, \binits{A.}},
\bauthor{\bsnm{{Bonavita}}, \binits{M.}},
\bauthor{\bsnm{{Bonnefoy}}, \binits{M.}},
\bauthor{\bsnm{{Cascone}}, \binits{E.}},
\bauthor{\bsnm{{Chauvin}}, \binits{G.}},
\bauthor{\bsnm{{Delboulb{\'e}}}, \binits{A.}},
\bauthor{\bsnm{{Desidera}}, \binits{S.}},
\bauthor{\bsnm{{D'Orazi}}, \binits{V.}},
\bauthor{\bsnm{{Feautrier}}, \binits{P.}},
\bauthor{\bsnm{{Feldt}}, \binits{M.}},
\bauthor{\bsnm{{Flammini Dotti}}, \binits{F.}},
\bauthor{\bsnm{{Girard}}, \binits{J.H.}},
\bauthor{\bsnm{{Giro}}, \binits{E.}},
\bauthor{\bsnm{{Janson}}, \binits{M.}},
\bauthor{\bsnm{{Hagelberg}}, \binits{J.}},
\bauthor{\bsnm{{Keppler}}, \binits{M.}},
\bauthor{\bsnm{{Kopytova}}, \binits{T.}},
\bauthor{\bsnm{{Lacour}}, \binits{S.}},
\bauthor{\bsnm{{Lagrange}}, \binits{A.-M.}},
\bauthor{\bsnm{{Langlois}}, \binits{M.}},
\bauthor{\bsnm{{Lannier}}, \binits{J.}},
\bauthor{\bsnm{{Le Coroller}}, \binits{H.}},
\bauthor{\bsnm{{Menard}}, \binits{F.}},
\bauthor{\bsnm{{Messina}}, \binits{S.}},
\bauthor{\bsnm{{Meyer}}, \binits{M.}},
\bauthor{\bsnm{{Millward}}, \binits{M.}},
\bauthor{\bsnm{{Olofsson}}, \binits{J.}},
\bauthor{\bsnm{{Pavlov}}, \binits{A.}},
\bauthor{\bsnm{{Peretti}}, \binits{S.}},
\bauthor{\bsnm{{Perrot}}, \binits{C.}},
\bauthor{\bsnm{{Pinte}}, \binits{C.}},
\bauthor{\bsnm{{Pragt}}, \binits{J.}},
\bauthor{\bsnm{{Ramos}}, \binits{J.}},
\bauthor{\bsnm{{Rochat}}, \binits{S.}},
\bauthor{\bsnm{{Rodet}}, \binits{L.}},
\bauthor{\bsnm{{Roelfsema}}, \binits{R.}},
\bauthor{\bsnm{{Rouan}}, \binits{D.}},
\bauthor{\bsnm{{Salter}}, \binits{G.}},
\bauthor{\bsnm{{Schmidt}}, \binits{T.}},
\bauthor{\bsnm{{Sissa}}, \binits{E.}},
\bauthor{\bsnm{{Thebault}}, \binits{P.}},
\bauthor{\bsnm{{Udry}}, \binits{S.}},
\bauthor{\bsnm{{Vigan}}, \binits{A.}}:
\batitle{{SPHERE dynamical and spectroscopic characterization of HD 142527B}}.
\bjtitle{\aap}
\bvolume{622},
\bfpage{96}
(\byear{2019})
\doiurl{10.1051/0004-6361/201833990}
{\href{https://arxiv.org/abs/1812.07814}{{arXiv:1812.07814}}}
{[astro-ph.SR]}
\end{barticle}
\endbibitem

\bibitem[\protect\citeauthoryear{{Hunziker} et~al.}{2021}]{Hunziker2021}
\begin{barticle}
\bauthor{\bsnm{{Hunziker}}, \binits{S.}},
\bauthor{\bsnm{{Schmid}}, \binits{H.M.}},
\bauthor{\bsnm{{Ma}}, \binits{J.}},
\bauthor{\bsnm{{Menard}}, \binits{F.}},
\bauthor{\bsnm{{Avenhaus}}, \binits{H.}},
\bauthor{\bsnm{{Boccaletti}}, \binits{A.}},
\bauthor{\bsnm{{Beuzit}}, \binits{J.L.}},
\bauthor{\bsnm{{Chauvin}}, \binits{G.}},
\bauthor{\bsnm{{Dohlen}}, \binits{K.}},
\bauthor{\bsnm{{Dominik}}, \binits{C.}},
\bauthor{\bsnm{{Engler}}, \binits{N.}},
\bauthor{\bsnm{{Ginski}}, \binits{C.}},
\bauthor{\bsnm{{Gratton}}, \binits{R.}},
\bauthor{\bsnm{{Henning}}, \binits{T.}},
\bauthor{\bsnm{{Langlois}}, \binits{M.}},
\bauthor{\bsnm{{Milli}}, \binits{J.}},
\bauthor{\bsnm{{Mouillet}}, \binits{D.}},
\bauthor{\bsnm{{Tschudi}}, \binits{C.}},
\bauthor{\bsnm{{van Holstein}}, \binits{R.G.}},
\bauthor{\bsnm{{Vigan}}, \binits{A.}}:
\batitle{{HD 142527: quantitative disk polarimetry with SPHERE}}.
\bjtitle{\aap}
\bvolume{648},
\bfpage{110}
(\byear{2021})
\doiurl{10.1051/0004-6361/202040166}
{\href{https://arxiv.org/abs/2103.08462}{{arXiv:2103.08462}}}
{[astro-ph.EP]}
\end{barticle}
\endbibitem

\bibitem[\protect\citeauthoryear{{Fern{\'a}ndez-L{\'o}pez}
  et~al.}{2017}]{Fernandez2017}
\begin{barticle}
\bauthor{\bsnm{{Fern{\'a}ndez-L{\'o}pez}}, \binits{M.}},
\bauthor{\bsnm{{Zapata}}, \binits{L.A.}},
\bauthor{\bsnm{{Gabbasov}}, \binits{R.}}:
\batitle{{Strongly Misaligned Triple System in SR 24 Revealed by ALMA}}.
\bjtitle{\apj}
\bvolume{845}(\bissue{1}),
\bfpage{10}
(\byear{2017})
\doiurl{10.3847/1538-4357/aa7d51}
{\href{https://arxiv.org/abs/1707.01128}{{arXiv:1707.01128}}}
{[astro-ph.SR]}
\end{barticle}
\endbibitem

\bibitem[\protect\citeauthoryear{{Schaefer} et~al.}{2018}]{Schaefer2018}
\begin{barticle}
\bauthor{\bsnm{{Schaefer}}, \binits{G.H.}},
\bauthor{\bsnm{{Prato}}, \binits{L.}},
\bauthor{\bsnm{{Simon}}, \binits{M.}}:
\batitle{{Orbital Motion of Young Binaries in Ophiuchus and Upper
  Centaurus-Lupus}}.
\bjtitle{\aj}
\bvolume{155}(\bissue{3}),
\bfpage{109}
(\byear{2018})
\doiurl{10.3847/1538-3881/aaa59a}
{\href{https://arxiv.org/abs/1802.02747}{{arXiv:1802.02747}}}
{[astro-ph.SR]}
\end{barticle}
\endbibitem

\bibitem[\protect\citeauthoryear{{Cazzoletti} et~al.}{2017}]{Cazzoletti2017}
\begin{barticle}
\bauthor{\bsnm{{Cazzoletti}}, \binits{P.}},
\bauthor{\bsnm{{Ricci}}, \binits{L.}},
\bauthor{\bsnm{{Birnstiel}}, \binits{T.}},
\bauthor{\bsnm{{Lodato}}, \binits{G.}}:
\batitle{{Testing dust trapping in the circumbinary disk around GG Tauri A}}.
\bjtitle{\aap}
\bvolume{599},
\bfpage{102}
(\byear{2017})
\doiurl{10.1051/0004-6361/201629721}
{\href{https://arxiv.org/abs/1610.08381}{{arXiv:1610.08381}}}
{[astro-ph.SR]}
\end{barticle}
\endbibitem

\bibitem[\protect\citeauthoryear{{Brinch} et~al.}{2016}]{Brinch2016}
\begin{barticle}
\bauthor{\bsnm{{Brinch}}, \binits{C.}},
\bauthor{\bsnm{{J{\o}rgensen}}, \binits{J.K.}},
\bauthor{\bsnm{{Hogerheijde}}, \binits{M.R.}},
\bauthor{\bsnm{{Nelson}}, \binits{R.P.}},
\bauthor{\bsnm{{Gressel}}, \binits{O.}}:
\batitle{{Misaligned Disks in the Binary Protostar IRS 43}}.
\bjtitle{\apjl}
\bvolume{830}(\bissue{1}),
\bfpage{16}
(\byear{2016})
\doiurl{10.3847/2041-8205/830/1/L16}
{\href{https://arxiv.org/abs/1610.03626}{{arXiv:1610.03626}}}
{[astro-ph.SR]}
\end{barticle}
\endbibitem

\bibitem[\protect\citeauthoryear{{Raghavan} et~al.}{2010}]{Raghavan2010}
\begin{barticle}
\bauthor{\bsnm{{Raghavan}}, \binits{D.}},
\bauthor{\bsnm{{McAlister}}, \binits{H.A.}},
\bauthor{\bsnm{{Henry}}, \binits{T.J.}},
\bauthor{\bsnm{{Latham}}, \binits{D.W.}},
\bauthor{\bsnm{{Marcy}}, \binits{G.W.}},
\bauthor{\bsnm{{Mason}}, \binits{B.D.}},
\bauthor{\bsnm{{Gies}}, \binits{D.R.}},
\bauthor{\bsnm{{White}}, \binits{R.J.}},
\bauthor{\bsnm{{ten Brummelaar}}, \binits{T.A.}}:
\batitle{{A Survey of Stellar Families: Multiplicity of Solar-type Stars}}.
\bjtitle{\apjs}
\bvolume{190}(\bissue{1}),
\bfpage{1}--\blpage{42}
(\byear{2010})
\doiurl{10.1088/0067-0049/190/1/1}
{\href{https://arxiv.org/abs/1007.0414}{{arXiv:1007.0414}}}
{[astro-ph.SR]}
\end{barticle}
\endbibitem

\bibitem[\protect\citeauthoryear{{Lidov}}{1962}]{Lidov1962}
\begin{barticle}
\bauthor{\bsnm{{Lidov}}, \binits{M.L.}}:
\batitle{{The evolution of orbits of artificial satellites of planets under the
  action of gravitational perturbations of external bodies}}.
\bjtitle{\planss}
\bvolume{9}(\bissue{10}),
\bfpage{719}--\blpage{759}
(\byear{1962})
\doiurl{10.1016/0032-0633(62)90129-0}
\end{barticle}
\endbibitem

\bibitem[\protect\citeauthoryear{{Winn} and {Fabrycky}}{2015}]{Winn2015}
\begin{barticle}
\bauthor{\bsnm{{Winn}}, \binits{J.N.}},
\bauthor{\bsnm{{Fabrycky}}, \binits{D.C.}}:
\batitle{{The Occurrence and Architecture of Exoplanetary Systems}}.
\bjtitle{\araa}
\bvolume{53},
\bfpage{409}--\blpage{447}
(\byear{2015})
\doiurl{10.1146/annurev-astro-082214-122246}
{\href{https://arxiv.org/abs/1410.4199}{{arXiv:1410.4199}}}
{[astro-ph.EP]}
\end{barticle}
\endbibitem

\bibitem[\protect\citeauthoryear{{Orosz} et~al.}{2019}]{Orosz2019}
\begin{barticle}
\bauthor{\bsnm{{Orosz}}, \binits{J.A.}},
\bauthor{\bsnm{{Welsh}}, \binits{W.F.}},
\bauthor{\bsnm{{Haghighipour}}, \binits{N.}},
\bauthor{\bsnm{{Quarles}}, \binits{B.}},
\bauthor{\bsnm{{Short}}, \binits{D.R.}},
\bauthor{\bsnm{{Mills}}, \binits{S.M.}},
\bauthor{\bsnm{{Satyal}}, \binits{S.}},
\bauthor{\bsnm{{Torres}}, \binits{G.}},
\bauthor{\bsnm{{Agol}}, \binits{E.}},
\bauthor{\bsnm{{Fabrycky}}, \binits{D.C.}},
\bauthor{\bsnm{{Jontof-Hutter}}, \binits{D.}},
\bauthor{\bsnm{{Windmiller}}, \binits{G.}},
\bauthor{\bsnm{{M{\"u}ller}}, \binits{T.W.A.}},
\bauthor{\bsnm{{Hinse}}, \binits{T.C.}},
\bauthor{\bsnm{{Cochran}}, \binits{W.D.}},
\bauthor{\bsnm{{Endl}}, \binits{M.}},
\bauthor{\bsnm{{Ford}}, \binits{E.B.}},
\bauthor{\bsnm{{Mazeh}}, \binits{T.}},
\bauthor{\bsnm{{Lissauer}}, \binits{J.J.}}:
\batitle{{Discovery of a Third Transiting Planet in the Kepler-47 Circumbinary
  System}}.
\bjtitle{\aj}
\bvolume{157}(\bissue{5}),
\bfpage{174}
(\byear{2019})
\doiurl{10.3847/1538-3881/ab0ca0}
{\href{https://arxiv.org/abs/1904.07255}{{arXiv:1904.07255}}}
{[astro-ph.EP]}
\end{barticle}
\endbibitem

\bibitem[\protect\citeauthoryear{{Kostov} et~al.}{2020}]{Kostov2020}
\begin{barticle}
\bauthor{\bsnm{{Kostov}}, \binits{V.B.}},
\bauthor{\bsnm{{Orosz}}, \binits{J.A.}},
\bauthor{\bsnm{{Feinstein}}, \binits{A.D.}},
\bauthor{\bsnm{{Welsh}}, \binits{W.F.}},
\bauthor{\bsnm{{Cukier}}, \binits{W.}},
\bauthor{\bsnm{{Haghighipour}}, \binits{N.}},
\bauthor{\bsnm{{Quarles}}, \binits{B.}},
\bauthor{\bsnm{{Martin}}, \binits{D.V.}},
\bauthor{\bsnm{{Montet}}, \binits{B.T.}},
\bauthor{\bsnm{{Torres}}, \binits{G.}},
\bauthor{\bsnm{{Triaud}}, \binits{A.H.M.J.}},
\bauthor{\bsnm{{Barclay}}, \binits{T.}},
\bauthor{\bsnm{{Boyd}}, \binits{P.}},
\bauthor{\bsnm{{Briceno}}, \binits{C.}},
\bauthor{\bsnm{{Cameron}}, \binits{A.C.}},
\bauthor{\bsnm{{Correia}}, \binits{A.C.M.}},
\bauthor{\bsnm{{Gilbert}}, \binits{E.A.}},
\bauthor{\bsnm{{Gill}}, \binits{S.}},
\bauthor{\bsnm{{Gillon}}, \binits{M.}},
\bauthor{\bsnm{{Haqq-Misra}}, \binits{J.}},
\bauthor{\bsnm{{Hellier}}, \binits{C.}},
\bauthor{\bsnm{{Dressing}}, \binits{C.}},
\bauthor{\bsnm{{Fabrycky}}, \binits{D.C.}},
\bauthor{\bsnm{{Furesz}}, \binits{G.}},
\bauthor{\bsnm{{Jenkins}}, \binits{J.M.}},
\bauthor{\bsnm{{Kane}}, \binits{S.R.}},
\bauthor{\bsnm{{Kopparapu}}, \binits{R.}},
\bauthor{\bsnm{{Hod{\v{z}}i{\'c}}}, \binits{V.K.}},
\bauthor{\bsnm{{Latham}}, \binits{D.W.}},
\bauthor{\bsnm{{Law}}, \binits{N.}},
\bauthor{\bsnm{{Levine}}, \binits{A.M.}},
\bauthor{\bsnm{{Li}}, \binits{G.}},
\bauthor{\bsnm{{Lintott}}, \binits{C.}},
\bauthor{\bsnm{{Lissauer}}, \binits{J.J.}},
\bauthor{\bsnm{{Mann}}, \binits{A.W.}},
\bauthor{\bsnm{{Mazeh}}, \binits{T.}},
\bauthor{\bsnm{{Mardling}}, \binits{R.}},
\bauthor{\bsnm{{Maxted}}, \binits{P.F.L.}},
\bauthor{\bsnm{{Eisner}}, \binits{N.}},
\bauthor{\bsnm{{Pepe}}, \binits{F.}},
\bauthor{\bsnm{{Pepper}}, \binits{J.}},
\bauthor{\bsnm{{Pollacco}}, \binits{D.}},
\bauthor{\bsnm{{Quinn}}, \binits{S.N.}},
\bauthor{\bsnm{{Quintana}}, \binits{E.V.}},
\bauthor{\bsnm{{Rowe}}, \binits{J.F.}},
\bauthor{\bsnm{{Ricker}}, \binits{G.}},
\bauthor{\bsnm{{Rose}}, \binits{M.E.}},
\bauthor{\bsnm{{Seager}}, \binits{S.}},
\bauthor{\bsnm{{Santerne}}, \binits{A.}},
\bauthor{\bsnm{{S{\'e}gransan}}, \binits{D.}},
\bauthor{\bsnm{{Short}}, \binits{D.R.}},
\bauthor{\bsnm{{Smith}}, \binits{J.C.}},
\bauthor{\bsnm{{Standing}}, \binits{M.R.}},
\bauthor{\bsnm{{Tokovinin}}, \binits{A.}},
\bauthor{\bsnm{{Trifonov}}, \binits{T.}},
\bauthor{\bsnm{{Turner}}, \binits{O.}},
\bauthor{\bsnm{{Twicken}}, \binits{J.D.}},
\bauthor{\bsnm{{Udry}}, \binits{S.}},
\bauthor{\bsnm{{Vanderspek}}, \binits{R.}},
\bauthor{\bsnm{{Winn}}, \binits{J.N.}},
\bauthor{\bsnm{{Wolf}}, \binits{E.T.}},
\bauthor{\bsnm{{Ziegler}}, \binits{C.}},
\bauthor{\bsnm{{Ansorge}}, \binits{P.}},
\bauthor{\bsnm{{Barnet}}, \binits{F.}},
\bauthor{\bsnm{{Bergeron}}, \binits{J.}},
\bauthor{\bsnm{{Huten}}, \binits{M.}},
\bauthor{\bsnm{{Pappa}}, \binits{G.}},
\bauthor{\bsnm{{van der Straeten}}, \binits{T.}}:
\batitle{{TOI-1338: TESS' First Transiting Circumbinary Planet}}.
\bjtitle{\aj}
\bvolume{159}(\bissue{6}),
\bfpage{253}
(\byear{2020})
\doiurl{10.3847/1538-3881/ab8a48}
{\href{https://arxiv.org/abs/2004.07783}{{arXiv:2004.07783}}}
{[astro-ph.EP]}
\end{barticle}
\endbibitem

\bibitem[\protect\citeauthoryear{{Doolin} and {Blundell}}{2011}]{Doolin2011}
\begin{barticle}
\bauthor{\bsnm{{Doolin}}, \binits{S.}},
\bauthor{\bsnm{{Blundell}}, \binits{K.M.}}:
\batitle{{The dynamics and stability of circumbinary orbits}}.
\bjtitle{\mnras}
\bvolume{418}(\bissue{4}),
\bfpage{2656}--\blpage{2668}
(\byear{2011})
\doiurl{10.1111/j.1365-2966.2011.19657.x}
{\href{https://arxiv.org/abs/1108.4144}{{arXiv:1108.4144}}}
{[astro-ph.SR]}
\end{barticle}
\endbibitem

\bibitem[\protect\citeauthoryear{{Verrier} and {Evans}}{2009}]{Verrier2009}
\begin{barticle}
\bauthor{\bsnm{{Verrier}}, \binits{P.E.}},
\bauthor{\bsnm{{Evans}}, \binits{N.W.}}:
\batitle{{High-inclination planets and asteroids in multistellar systems}}.
\bjtitle{\mnras}
\bvolume{394}(\bissue{4}),
\bfpage{1721}--\blpage{1726}
(\byear{2009})
\doiurl{10.1111/j.1365-2966.2009.14446.x}
{\href{https://arxiv.org/abs/0812.4528}{{arXiv:0812.4528}}}
{[astro-ph]}
\end{barticle}
\endbibitem

\bibitem[\protect\citeauthoryear{{Martin} and {Lubow}}{2018}]{Martin2018}
\begin{barticle}
\bauthor{\bsnm{{Martin}}, \binits{R.G.}},
\bauthor{\bsnm{{Lubow}}, \binits{S.H.}}:
\batitle{{Polar alignment of a protoplanetary disc around an eccentric binary -
  II. Effect of binary and disc parameters}}.
\bjtitle{\mnras}
\bvolume{479}(\bissue{1}),
\bfpage{1297}--\blpage{1308}
(\byear{2018})
\doiurl{10.1093/mnras/sty1648}
{\href{https://arxiv.org/abs/1806.08388}{{arXiv:1806.08388}}}
{[astro-ph.EP]}
\end{barticle}
\endbibitem

\bibitem[\protect\citeauthoryear{{Ceppi} et~al.}{2023}]{Ceppi2023}
\begin{barticle}
\bauthor{\bsnm{{Ceppi}}, \binits{S.}},
\bauthor{\bsnm{{Longarini}}, \binits{C.}},
\bauthor{\bsnm{{Lodato}}, \binits{G.}},
\bauthor{\bsnm{{Cuello}}, \binits{N.}},
\bauthor{\bsnm{{Lubow}}, \binits{S.H.}}:
\batitle{{Precession and polar alignment of accretion discs in triple (or
  multiple) stellar systems}}.
\bjtitle{\mnras}
\bvolume{520}(\bissue{4}),
\bfpage{5817}--\blpage{5827}
(\byear{2023})
\doiurl{10.1093/mnras/stad444}
{\href{https://arxiv.org/abs/2302.03411}{{arXiv:2302.03411}}}
{[astro-ph.EP]}
\end{barticle}
\endbibitem

\bibitem[\protect\citeauthoryear{{Chen} et~al.}{2020}]{Chen2020}
\begin{barticle}
\bauthor{\bsnm{{Chen}}, \binits{C.}},
\bauthor{\bsnm{{Lubow}}, \binits{S.H.}},
\bauthor{\bsnm{{Martin}}, \binits{R.G.}}:
\batitle{{Polar planets around highly eccentric binaries are the most stable}}.
\bjtitle{\mnras}
\bvolume{494}(\bissue{4}),
\bfpage{4645}--\blpage{4655}
(\byear{2020})
\doiurl{10.1093/mnras/staa1037}
{\href{https://arxiv.org/abs/2004.07230}{{arXiv:2004.07230}}}
{[astro-ph.EP]}
\end{barticle}
\endbibitem

\bibitem[\protect\citeauthoryear{{Cuello} and
  {Giuppone}}{2019}]{CuelloGiuppone2019}
\begin{barticle}
\bauthor{\bsnm{{Cuello}}, \binits{N.}},
\bauthor{\bsnm{{Giuppone}}, \binits{C.A.}}:
\batitle{{Planet formation and stability in polar circumbinary discs}}.
\bjtitle{\aap}
\bvolume{628},
\bfpage{119}
(\byear{2019})
\doiurl{10.1051/0004-6361/201833976}
{\href{https://arxiv.org/abs/1906.10579}{{arXiv:1906.10579}}}
{[astro-ph.EP]}
\end{barticle}
\endbibitem

\bibitem[\protect\citeauthoryear{{Kozai}}{1962}]{1962AJ.....67..591K}
\begin{barticle}
\bauthor{\bsnm{{Kozai}}, \binits{Y.}}:
\batitle{{Secular perturbations of asteroids with high inclination and
  eccentricity}}.
\bjtitle{\aj}
\bvolume{67},
\bfpage{591}--\blpage{598}
(\byear{1962})
\doiurl{10.1086/108790}
\end{barticle}
\endbibitem

\bibitem[\protect\citeauthoryear{{Saha} et~al.}{2020}]{Saha2020}
\begin{barticle}
\bauthor{\bsnm{{Saha}}, \binits{P.}},
\bauthor{\bsnm{{Gopinathan}}, \binits{M.}},
\bauthor{\bsnm{{Kamath}}, \binits{U.}},
\bauthor{\bsnm{{Lee}}, \binits{C.W.}},
\bauthor{\bsnm{{Manoj}}, \binits{P.}},
\bauthor{\bsnm{{Mathew}}, \binits{B.}},
\bauthor{\bsnm{{Sharma}}, \binits{E.}}:
\batitle{{A census of young stellar population associated with the Herbig Be
  star HD 200775}}.
\bjtitle{\mnras}
\bvolume{494}(\bissue{4}),
\bfpage{5851}--\blpage{5871}
(\byear{2020})
\doiurl{10.1093/mnras/staa1053}
{\href{https://arxiv.org/abs/2005.00519}{{arXiv:2005.00519}}}
{[astro-ph.SR]}
\end{barticle}
\endbibitem

\end{thebibliography}


\end{document}